\documentclass[journal=jpcafh,manuscript=article]{achemso}
\usepackage{helvet,times,mathptm,epsfig,amsmath}
\usepackage{graphicx}
\usepackage{achemso,natmove}
\usepackage{epsfig}
\newcommand\beq{\begin{equation}}
\newcommand\eeq{\end{equation}}
\newcommand\bea{\begin{eqnarray}}
\newcommand\eea{\end{eqnarray}}

\fontsize{1in} {1.2in}\selectfont

\title{The Correlated Electronic States of a few Polycyclic Aromatic 
Hydrocarbons: A Computational Study}
\author{Geetanjali Giri}
\affiliation[]{Solid State and Structural Chemistry Unit, Indian Institute of Science, Bangalore 560 012, India}
\author{Y. Anusooya Pati}
\affiliation[]{Solid State and Structural Chemistry Unit, Indian Institute of Science, Bangalore 560 012, India}
\author{S. Ramasesha}
\email{ramasesh@sscu.iisc.ernet.in}
\affiliation[]{Solid State and Structural Chemistry Unit, Indian Institute of Science, Bangalore 560 012, India}
\date{\today}

\begin{document}

\begin{abstract}

{\noindent{\small In recent years Polycyclic Aromatic Hydrocarbons (PAHs) have 
been studied for their electronic properties as they are viewed as nanodots of 
graphene. They have also been of interest as functional molecules for 
applications such as light emitting diodes and solar cells. Since last few 
years varying structural and chemical properties corresponding to the size 
and geometry of these molecules have been studied both theoretically and 
experimentally. In this paper, we carry out a systematic study of the 
electronic states of several PAHs using the Pariser-Parr-Pople model 
which incorporates long-range electron correlations. In all the molecules 
studied by us, we 
find that the 2A state is below the 1B state and hence none of them will be 
fluorescent in the gaseous phase. The singlet-triplet gap is more than one-half 
of the singlet-singlet gap in all cases and hence none of these PAHs can be 
candidates for improved solar cell efficiencies in a singlet fission. We 
discuss in detail the properties of the electronic states which include 
bond orders and spin densities (in triplets) of these systems. }
}
\end{abstract}

\section{Introduction}
Electronic structure of $\pi-$ conjugated carbon systems is of enduring 
interest, although there have been  paradigm shifts in the underlying reasons. Early 
interest revolved around aromaticity \cite{salem,soos,gomes} and later with the advent of 
the era of electronic polymers, the interest has shifted to the study of relative 
energy level ordering of the low energy excited states. With recent focus on 
graphene, there has been a resurgence of interest in condensed aromatic ring systems. 
Thus, polycyclic aromatic hydrocarbons (PAHs) which were studied mainly for 
environmental impact are now of interest from the electronic structure view 
point as these molecules serve as model nanodots of graphene \cite{nanodot,gqd-jmcrev,
chen-jmc-2018}. 
The effect of shape on the optical properties of nanodots of graphene, have 
also been studied {\cite{nanoflake-opt-jcp}. The role of size, shape and
hetero atom doping on the magnetic and electronic properties of graphene quantum
dots have been studied theoretically using density functional methods 
\cite{yamijala-jpc-2013, mahasin-jmc-2014, krgeeta-jmc-2016}. Indeed, 
it has also been reported that alkali metal doped picene exhibits superconductivity 
below 18K \cite{mitsu-nature-2010} and K-doped 1,2:8,9-dibenzopentacene shows 
superconductivity above 30 K \cite{xue-sci-rep-2012}, Sm-doped phenacenes 
show very low $T_c \approx 5$K \cite{gian-sm-acene-chemcom-2014,nakagawa-jpc-2016}, which adds 
further dimension to the study of PAHs. The spectra of PAHs  have also been 
studied extensively, since they have been detected in the interstellar 
medium \cite{allam-science-1987}.

Experimental studies have mainly focused on the vibrational spectra of PAH 
molecules. Allamandola et al have assigned the unidentified infrared bands 
(UIBs) of interstellar particles to Raman spectra of partially hydrogenated,
positively charged PAHs\cite{allam-astroj-1985}. They have studied the infrared
fluorescence spectrum of chrysene molecule.  The observation of diffused 
interstellar absorption bands (DIBs) is attributed to the presence of larger
PAHs molecule.\cite{hende-asr-1997} Salama et al have obtained the spectra
of several neutral and ionic PAH molecules under the astrophysical laboratory 
conditions and compared this with the astronomical 
spectra \cite{salama-astroj-2011}. These astronomical bands can be explained 
with electronic absorptions and/or dynamical methods such as electronic 
relaxations and intramolecular vibrational
relaxations. Palewska et al have obtained the high resolution fluorescence and 
phosphorescence spectra of tetrahelicene and hexahelicene in n-heptane matrix 
at room temperature\cite{palewska}. They find the singlet-singlet transition 
to be at 3.1 eV whereas the singlet-triplet transition to be at 2.5 eV. The 
electronic absorption spectrum of chrysene molecule  in boric acid glass 
matrix is reported by Hussain \cite{hussain-spec-chimi-acta-2007}. They compared their 
excitation 
energies obtained from experiments with those obtained from semiempirical AM1 
method. They found a reasonably good agreement in the calculated bond lengths 
and bond angles with those obtained from the X-ray diffraction.

Electronic absorption spectra in complex PAH molecules were obtained within 
semiempirical approach where restricted configuration interactions were 
used\cite{canuto-astropj-1991}. Ping Du et al used semi empirical quantum 
mechanical intermediate neglect of differential overlap (INDO) method to 
compare optical spectra of naphthalene and its ions with experimental 
data \cite{ping-chemphys-1993}. Infrared spectra of neutral and cationic 
PAHs are obtained within the DFT method and  the neutral molecules showed a 
good agreement with the experimental spectra\cite{lang-jpc-1996}. The absolute 
absorption cross-section of several PAH molecules and their cations were
obtained within time-dependent density functional 
theory\cite{malloci-anda-2004}. They find good agreement with the 
experimental excitation energies for neutral PAH molecules. This is not 
surprising as TDDFT method with adiabatic local density approximation is 
equivalent to single CI method \cite{27} which has been shown to give good 
optical gaps due to cancellation of errors. However, with ions, this 
cancellation does not occur. Using time dependent DFT, Hammonds et al
have obtained the electronic spectra of  protonated and partially hydrogenated
PAHs and found similarity in the absorption peaks of diffuse interstellar 
bands \cite{hammond-pccp-2009}. Size and geometry dependence of 
larger PAH molecules on optical spectra has been studied by Cocchi et al 
within ZINDO model, using only single excitation configuration interactions 
(SCI)\cite{cocchi-jpc-2014}. The role of aliphatic side groups on the 
vibrational spectra of PAHs molecule is studied using the DFT method 
\cite{sadjadi-astropj-2015}. 

A feature common to all conjugated carbon systems is that the $\pi-$electrons 
experience strong correlations. The role of electron correlation on the 
ordering of the excited state energy levels in PAHs has been studied by 
Mazumdar et al using Pariser-Parr-Pople model hamiltonian \cite{sumit}. 
They find that the effect of electron-electron interactions is weaker in 
PAHs than in polyenes, but stronger than in the quasi-1D systems like 
poly-(paraphenylene) and poly-(paraphenylenevinylene). Basak et al have 
carried out a detailed study of the linear absorption spectra of diamond 
shaped graphene quantum dots (where pyrene is the smallest unit), with 
quadruple CI and MRSDCI basis using the correlated PPP model Hamiltonian.  
They have compared their excitation gaps with screened and standard 
PPP parameters for carbon \cite{basak-prb92-2015}. They have also studied the
two-photon absorption and photoinduced absorption of these molecules. They 
find that the effect of electron-electron correlation gets enhanced with the
increase in the size of graphene quantum dots. \cite{basak-prb98-2018}
Motivated by this large number of 
experimental and theoretical data on PAHs, the current study focuses on the 
effect of strong electron-electron correlations on one-photon and 
two-photon absorptions in PAHs. So far, a thorough study of this class of 
molecules inclusive of strong long range electron-electron correlations  
has not been reported to the best of our knowledge. 

In this paper, we have compared electronic spectra of various PAHs (see Fig. 1)
such as Fluoranthene, Pyrene (both 16 conjugated carbon atoms), Benzanthracene, 
Chrysene, Helicene, Triphenylene (18 conjugated carbon atoms), 
Benzopyrene (20 conjugated carbon atoms) and Picene (22 conjugated carbon atoms) 
using numerically exact diagrammatic valence bond (DVB) method as well as Density 
matrix renormalization group method (DMRG) for larger molecules. We have 
analyzed the bond orders and spin densities in triplet states based on the 
ring orientations of these systems. We find that none of the 
molecules belonging to this class, studied by us, fulfill the energy 
criterion for singlet fission. Details of methods used in this study are 
discussed in next section followed by a description of singlet and triplet 
states of all molecules considered, with their corresponding properties. We 
have summarized our findings in the last section.

\begin{figure}[]
\begin{center}
\includegraphics[width=0.28\textwidth]{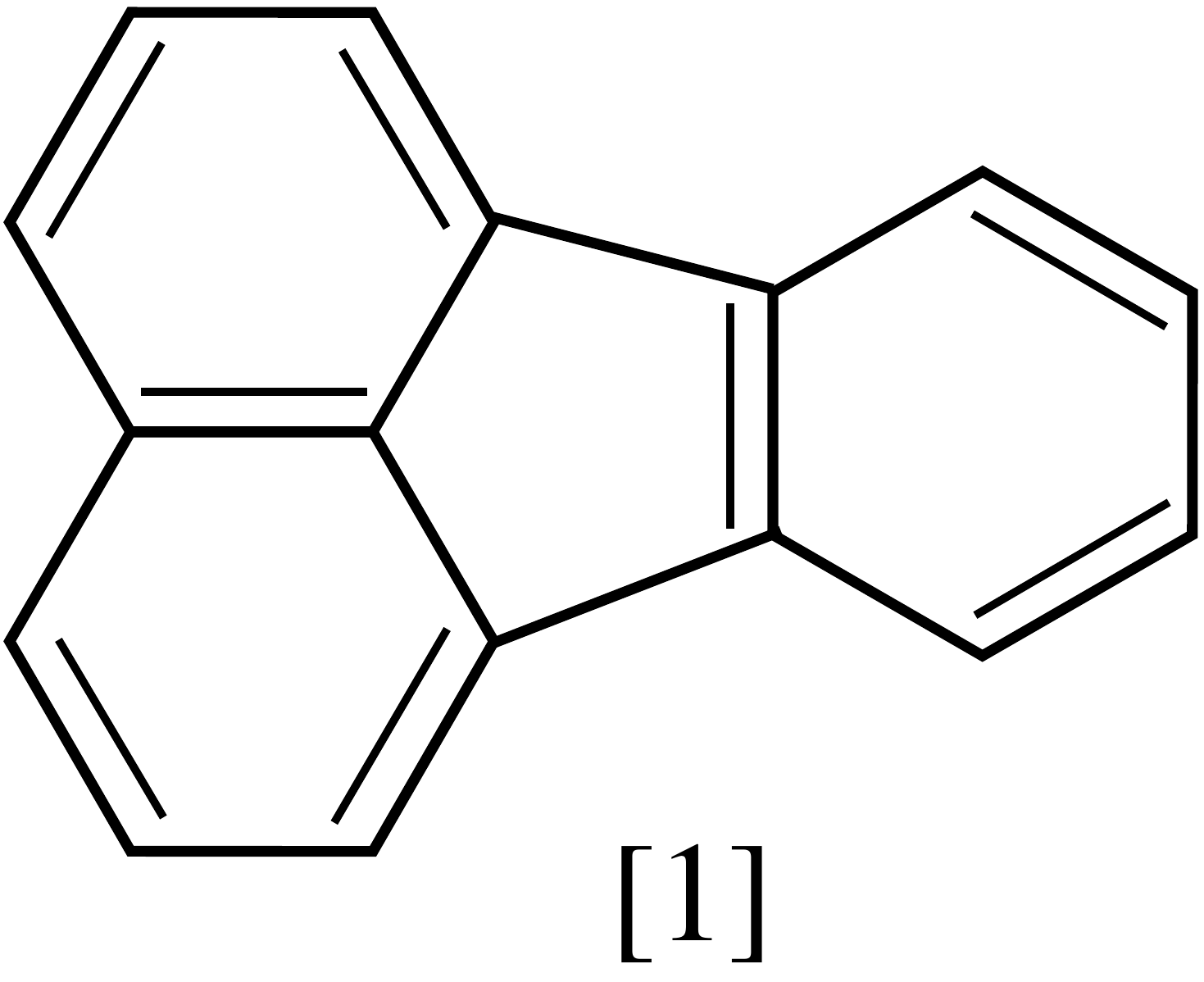}%
\hspace{1cm}
\label{fig:1} 
\includegraphics[width=0.18\textwidth]{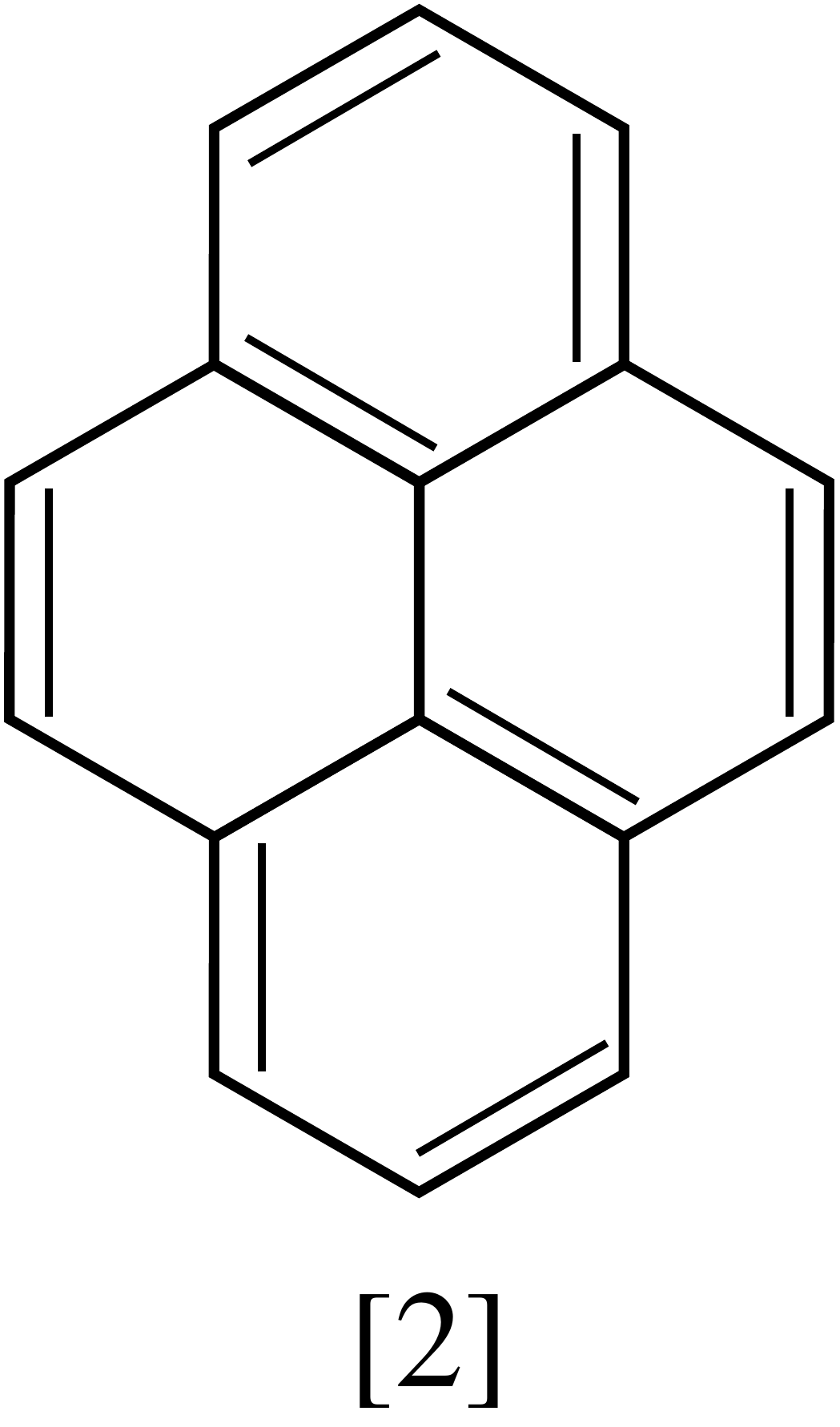}%
\hspace{1cm}
\label{fig:2}
\includegraphics[width=0.30\textwidth]{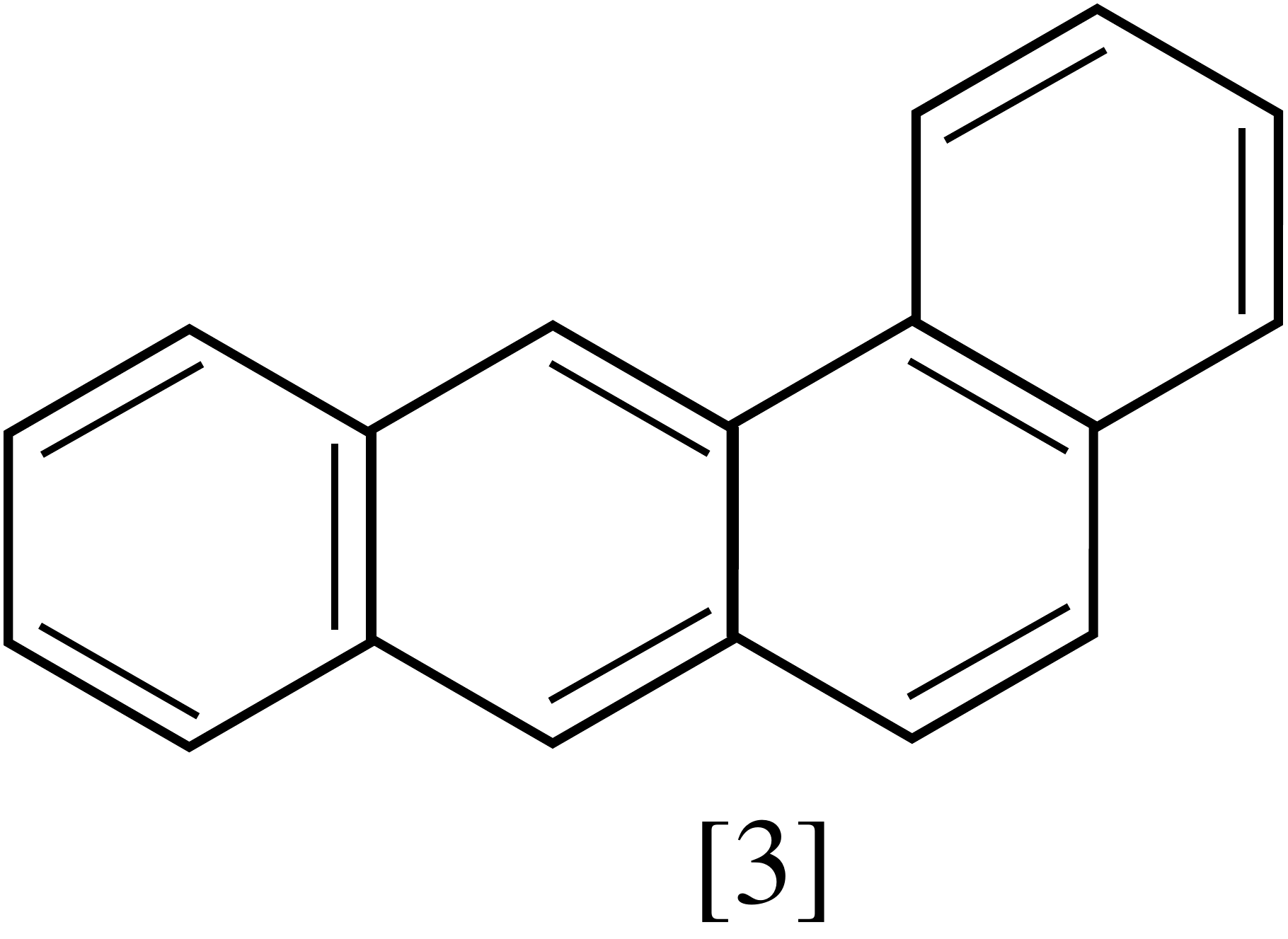}%
\label{fig:3} 
\vspace{1.0cm}
\hspace{1cm}
\includegraphics[width=0.30\textwidth]{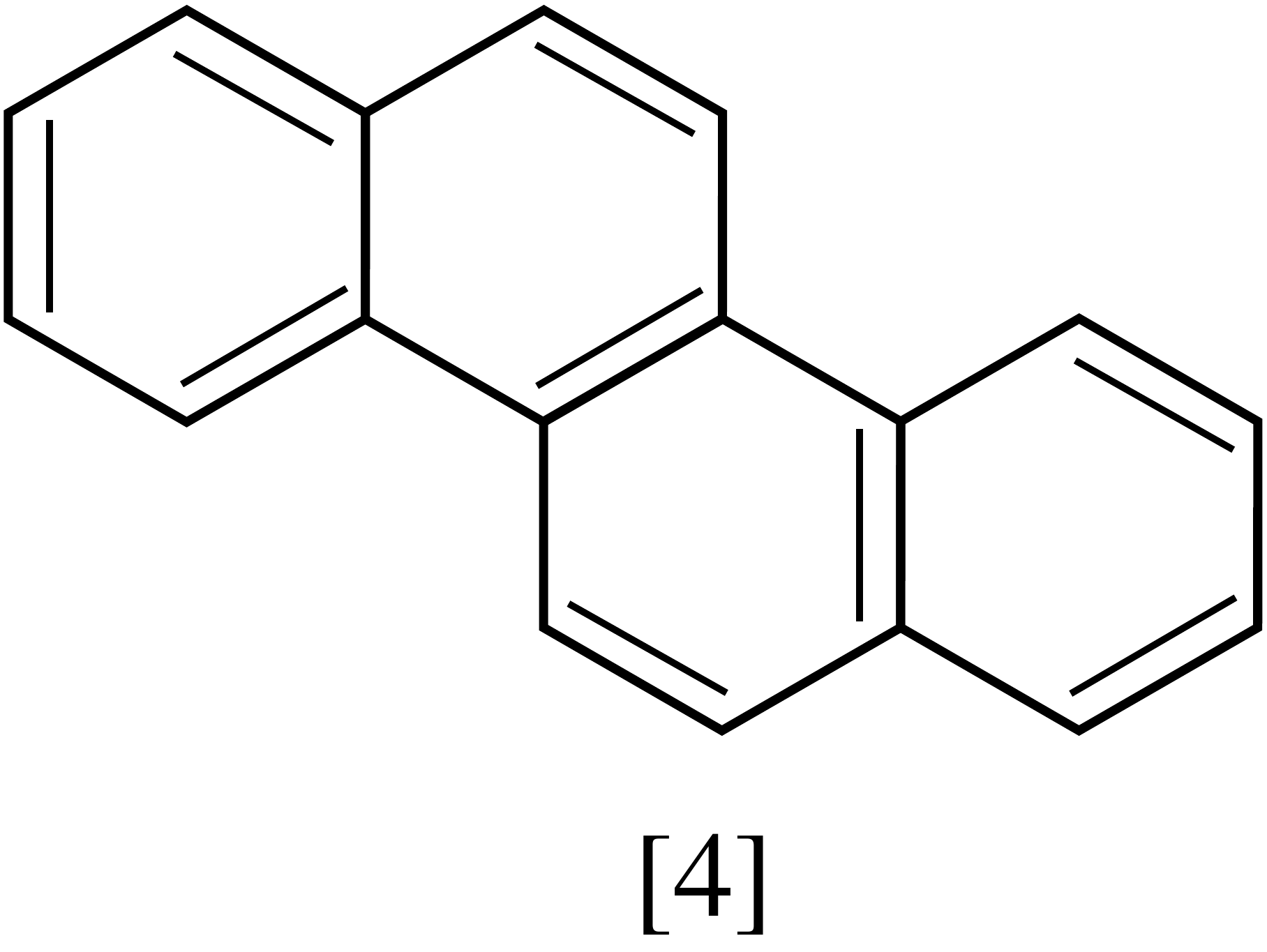}%
\hspace{2cm}
\label{fig:4} 
\includegraphics[width=0.25\textwidth]{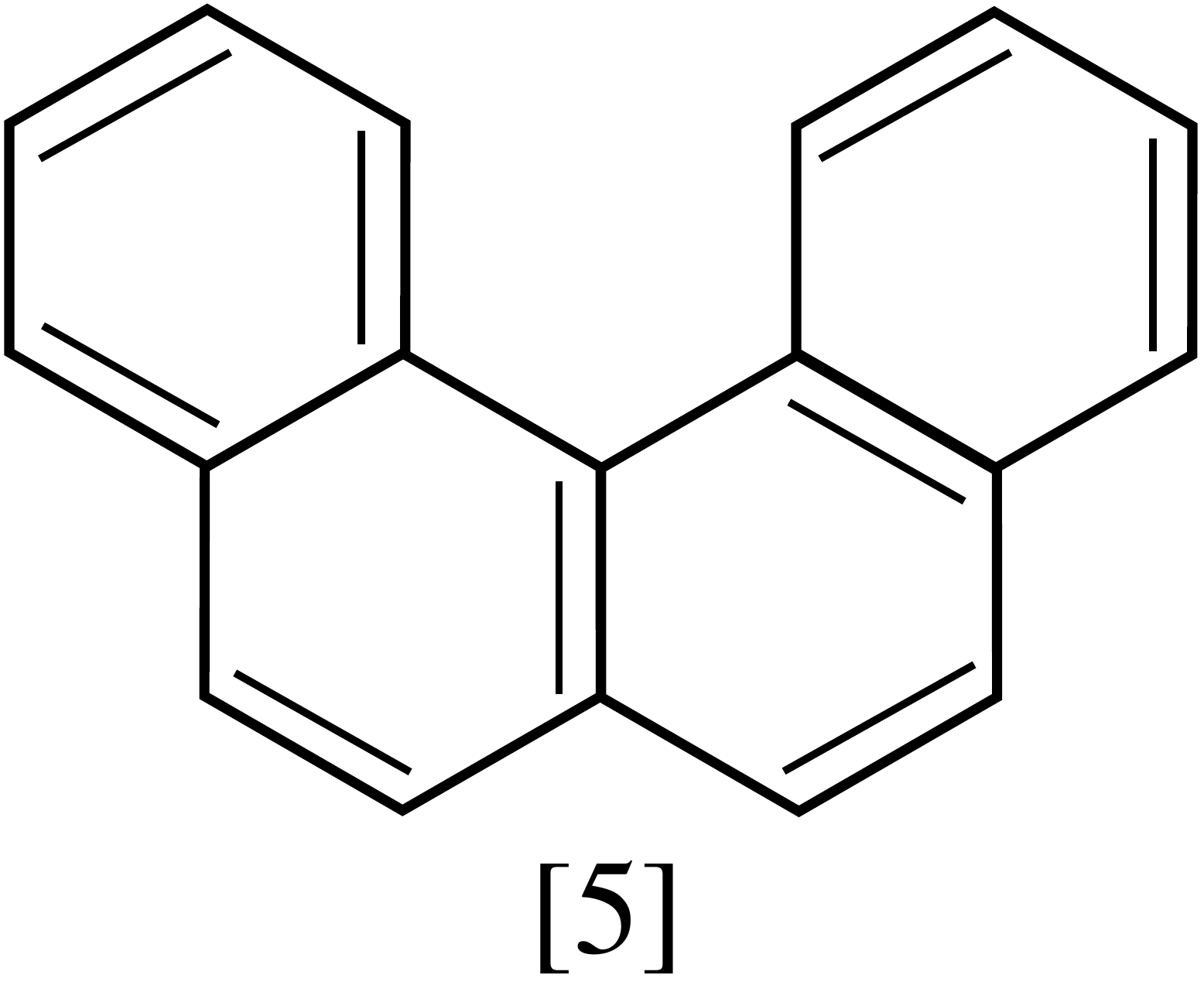}
\hspace{1cm}
\label{fig:5}
\includegraphics[width=0.22\textwidth]{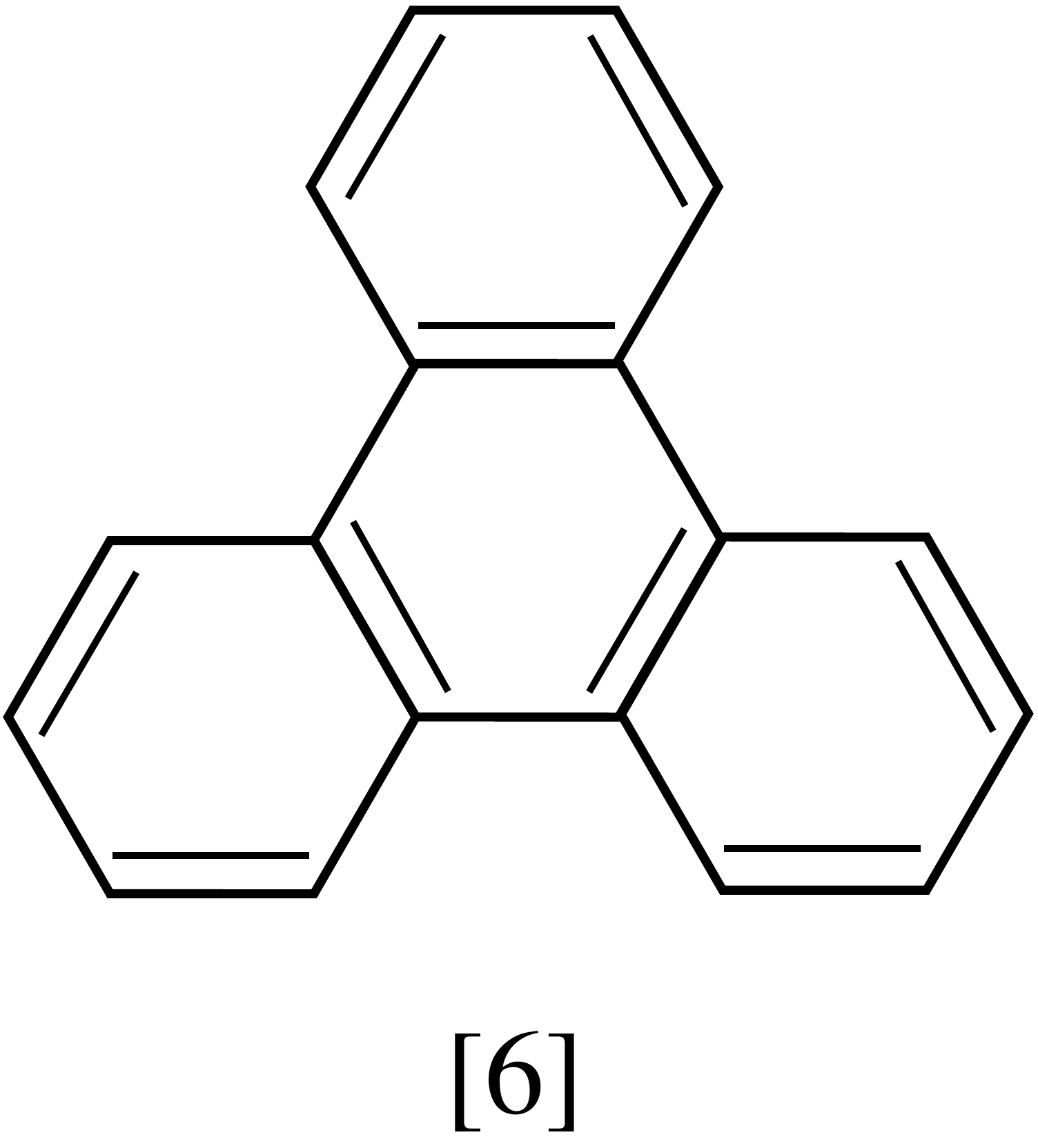}%
\hspace{1cm}
\label{fig:6} 
\vspace{1.0cm}
\includegraphics[width=0.22\textwidth]{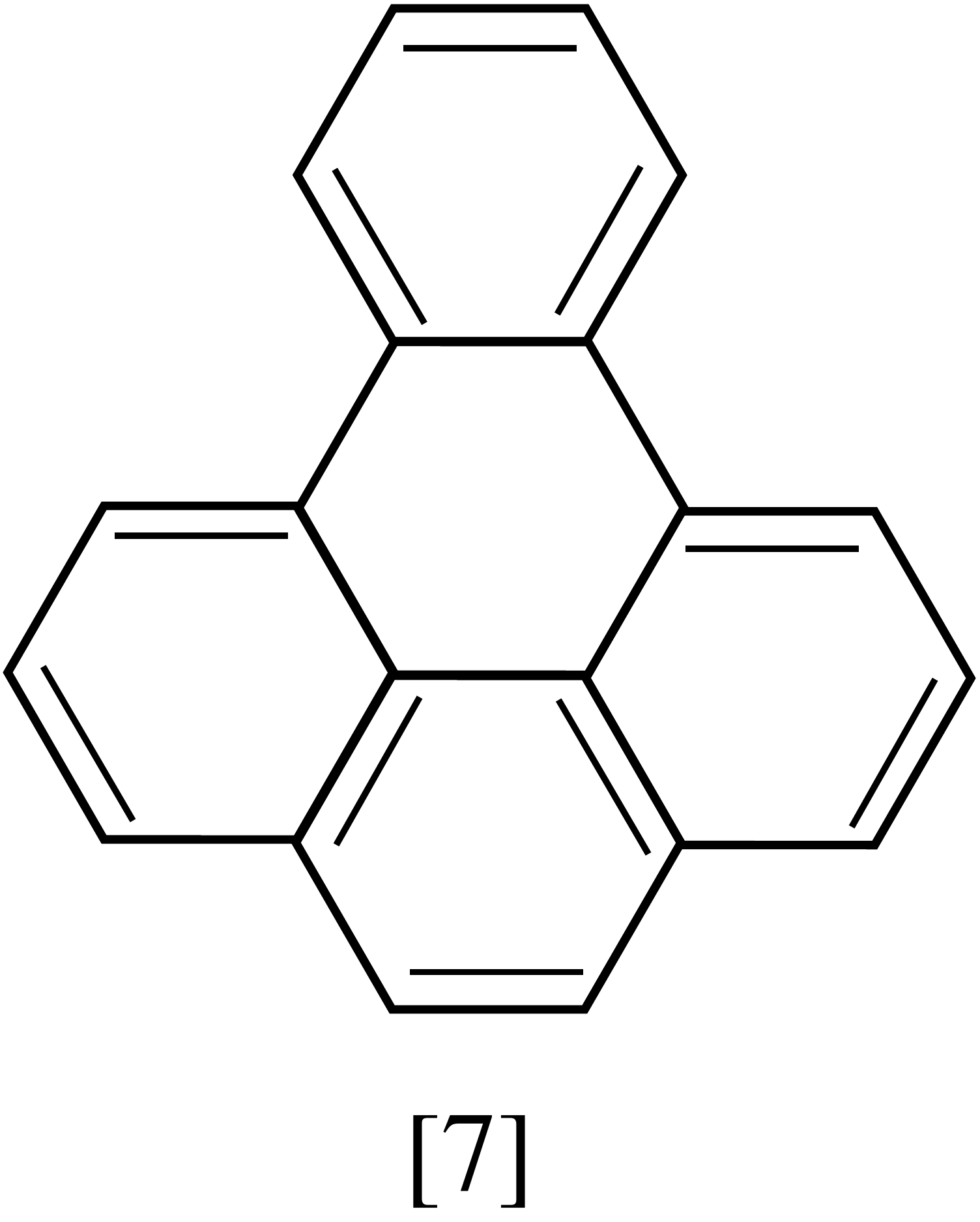}%
\hspace{1cm}
\label{fig:7}
\includegraphics[width=0.35\textwidth]{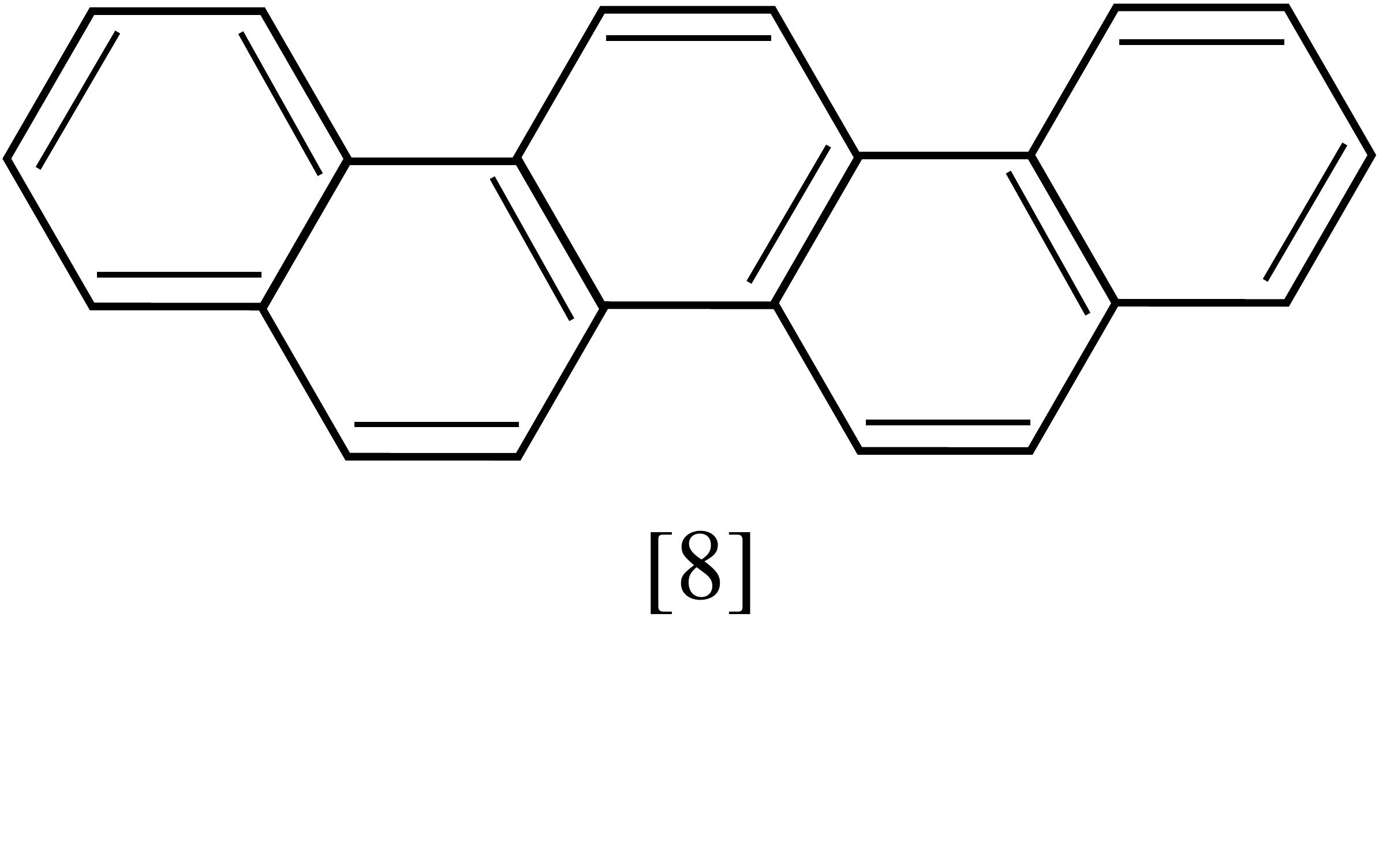}%
\label{fig:8}
\end{center}
\caption{Schematic representation of PAH molecules that have been studied. 
1) Fluoranthene, 2) Pyrene, 3) Benzanthracene, 4) Chrysene, 5) Helicene, 6) Triphenylene, 
7) Benzopyrene and 8) Picene.}
\end{figure}
 
\section{\label{sec1}Methodology} We have considered all molecules to be geometrically 
planar. Ignoring hydrogen atoms in a molecule leaves us with the carbon skeleton. Carbon 
atoms in $sp^2$ hybridization in aromatic hydrocarbons provides three sigma bonds with 
neighboring atoms resulting in a rigid framework of the system whereas $2p_z$ orbital 
gives conjugated network perpendicular to the plane of the molecule. This conjugated 
framework in aromatic systems gives rise to many interesting properties owing to the 
delocalized nature of orbitals. In our studies we focus on $\pi$-electron network 
to explore various properties of the PAHs as this network is energetically well 
separated from the underlying $\sigma-$network. This approximation is the $\sigma-\pi$ 
separability approximation \cite{32}. In the present study, we have used 
Pariser-Parr-Pople (PPP) model Hamiltonian to describe the $\pi-$electrons \cite{33,34}. 
PPP model is obtained from many-body Hamiltonian considering only $\pi$-electrons and 
invoking zero differential overlap (ZDO) approximation. The PPP Hamiltonian is given 
below:
\begin{align}
\hat{H}_{PPP}=\sum_{<ij>,\sigma} t_{0} (\hat{c}^\dagger_{i,\sigma} \hat{c}_{j,\sigma} + H.C.) + \sum_{i} \epsilon_{i}\hat{n}_i
\mspace{48mu}
\notag\\
 + \sum_{i}\dfrac{U}{2}\hat{n}_i(\hat{n}_i-1) + \sum_{i>j} V_{ij}(\hat{n}_i - z_i)(\hat{n}_j - z_j)
\end{align}
Here first term is the noninteracting part of the PPP Hamiltonian in which $t_0$ is 
nearest-neighbour hopping term between two bonded orbitals `i' and `j'. 
$\hat{c}^\dagger_{i,\sigma}$ ($\hat{c}_{i,\sigma}$) creates (annihilates) an electron 
of spin $\sigma$ in $2p_z$ orbital of carbon atom at site i. $\epsilon_i$ is the 
site energy of `i'-th carbon atom. The second term corresponds to interacting part 
of the Hamiltonian which consists of on-site Hubbard repulsion term for two electrons 
occupying the same orbital. In the third term $V_{ij}$s are electron-electron repulsion 
integrals between electrons in two separate orbitals i and j. $V_{ij}$s are 
interpolated between U and $\dfrac{e^2}{r}$ as $r_{ij} \rightarrow \infty$ using the Ohno interpolation scheme \cite{35}, 
\begin{align}
V_{ij} = 14.397 \left[ \left\{ \dfrac{28.794}{(U_i + U_j)} \right\}^2 + r_{ij}^2\right]^{-\dfrac{1}{2}}
\end{align}
The operator $\hat{n_{i}}$ is the number operator, the C-C bond length is 
fixed at 1.4 $\AA$ for nearest neighbour carbon atoms. We use the standard PPP
model parameters for carbon, namely, $t_0= -2.4$ eV, $U= 11.26$ eV and site 
energies $\epsilon_i=0.0$ eV for unsubstituted carbon site. $z_i$ is the number of 
electrons at site `i' which leaves the site neutral and correspond to local chemical 
potential. For carbon atoms in conjugation, $z_i$ values are set to 1.

      The PPP model Hamiltonian conserves total spin. So, to obtain the 
eigenstates of PAHs within PPP model we have used the diagrammatic valence bond 
(DVB) method because it generates spin-adapted basis functions contrary to 
slater basis functions which conserves only the z-component of total spin. 
To reduce the computational cost further, we have exploited the combination 
of $C_2$ and electron-hole (e-h) symmetry and factorized the Hilbert space 
into symmetry adapted bases \cite{36,37}. Since $C_2$ and eh symmetry operators
commute, we have an Abelian group with 4 one-dimensional representation. 
Except benzanthracene, all molecules studied by us have $C_2$ symmetry - 
either along X-axis or along Y-axis. Since these molecules are assumed to be 
planar, inversion symmetry and $C_2$ along Z-axis are equivalent. So, we 
have also used the term  $C_2$ for inversion symmetry. The space with character 
+1 under $C_2$ and +1 under e-h symmetry corresponds to the $A^+$ space and 
contains covalent VB diagrams, i.e. VB diagram in which every site is 
neutral. The ground state of the system is found in this space. The 
optically connected excited state (to the ground state) is found in the 
$B^-$ space, where the character under both $C_2$ and e-h symmetry is -1. 
This space does not contain any covalent VB diagram and hence is also called 
the ionic space. The e-h symmetry exists only in bipartite lattices where the 
transfer term is nonzero only between atoms in different sublattices. Among 
the PAHs we have studied only Fluoranthene does not belong to this class. 
We use the symmetry adapted VB basis to set-up the Hamiltonian matrix \cite{38}. 
A few low-lying states in each subspace are obtained from the Rettrup 
algorithm\cite{39}. With the computational resources we have, we can do 
exact calculations for 
the singlets when both $C_2$ and eh symmetries exist, for a system with 20 
carbon atoms and 20 electrons (neutral system). The dimensionalities of the 
subspaces for all the PAH systems we have studied is given in Table 1. 
Wherever we are unable to carry out exact calculations, we have resorted to 
the DMRG calculations. Thus triplets of benzopyrene and both 
singlets and triplets of picene were obtained within the DMRG approach. 
After obtaining the low-lying eigenstates, to fully characterize the state, 
we have computed the transition dipole between ground and excited states in 
appropriate spaces. We have also computed bond orders to understand the 
nature of equilibrium geometry in all eigenstates. In triplet states we 
have computed spin densities which can be verified from ESR studies.

\begin{table}[ht!]
\begin{tabular}{|l|c|c|c|c|c|}
\hline
Molecule & $N_e=N$   & \multicolumn{2}{c|}{Singlets} & \multicolumn{2}{c|}{Triplets} \\  \hline
& \multicolumn{1}{c|}{} & $\bf{A^{cov}}$   & $\bf{B^{ionic}}$   
& \multicolumn{1}{c|}{$\bf{B^{cov}}$}     & $\bf{A^{ionic}}$     \\  \hline

Pyrene           & 16            &  ~~~~~~~~8,698,485 &  ~~~~~~~~8,695,320 &  ~~~~16,692,888  &  ~~~~~~16,689,604 \\ \hline
Benzanthracene   & 18            & ~~~~224,573,349 &~~~~224,568,487 & ~~451,003,761  & ~~~~450,991,827 \\ \hline
Chrysene         & 18            & ~~~~112,313,738 &~~~~112,303,369 & ~~225,524,745  & ~~~~225,510,840 \\ \hline
Helicene         & 18            & ~~~~224,573,349 &~~~~224,568,487 & ~~451,003,761  & ~~~~450,991,827 \\ \hline
Triphenylene     & 18            & ~~~~112,313,738 &~~~~112,303,369 & ~~225,524,745  & ~~~~225,510,840 \\ \hline
BenzoPyrene         & 20            &~ 1,481,162,738 & ~1,481,122,588 & - &      -                  \\ \hline
\end{tabular}
\caption{Dimensionalities of the symmetry adapted spaces for PAHs 
studied in this paper. The total number of bases in singlet and triplet spaces 
for neutral systems with N = 16 are 34,763,300 and 66,745,536; with N = 18 
are 449,141,836 and 901,995,588; with N = 20 are 5,924,217,936 and  
12,342,120,700;  with N = 22 are 79,483,257,303 and 170,724,392,916, 
respectively.}

\label{tabular:sym-basis}
\end{table}

\section{Results and Discussion}
All the molecules studied, with the exception of fluoranthene, have electron-hole 
symmetry. Thus the states in these molecules (except fluoranthene) can be classified 
as 'covalent' ('+') or 'ionic' ('-'). The ground state is in the covalent subspace, 
so are the 2-photon states. Because of the non-alternancy, only fluoranthene will have 
non-zero dipole moment (4.22 Debye) in the ground state. The excitations to 
low-lying states in all molecules are shown in Fig. 2. In cases where the energy gap 
between levels in a molecule are small, we have ensured the linear independency of 
eigen states by computing the overlap of nearly degenerate eigen states. One common 
feature we note is that in all the molecules, the two-photon excitation
is below the one-photon excitation, except in the case of fluoranthene, where the lowest excited
singlet is in the B space and thus has a nonzero transition dipole. The lowest two-photon state
in fluoranthene is about 0.2eV above the lowest one-photon state.  In all other molecules,
the lowest two-photon state is about 0.7 eV below the lowest one-photon state. 

\begin{figure}
\includegraphics[width=0.45\textwidth]{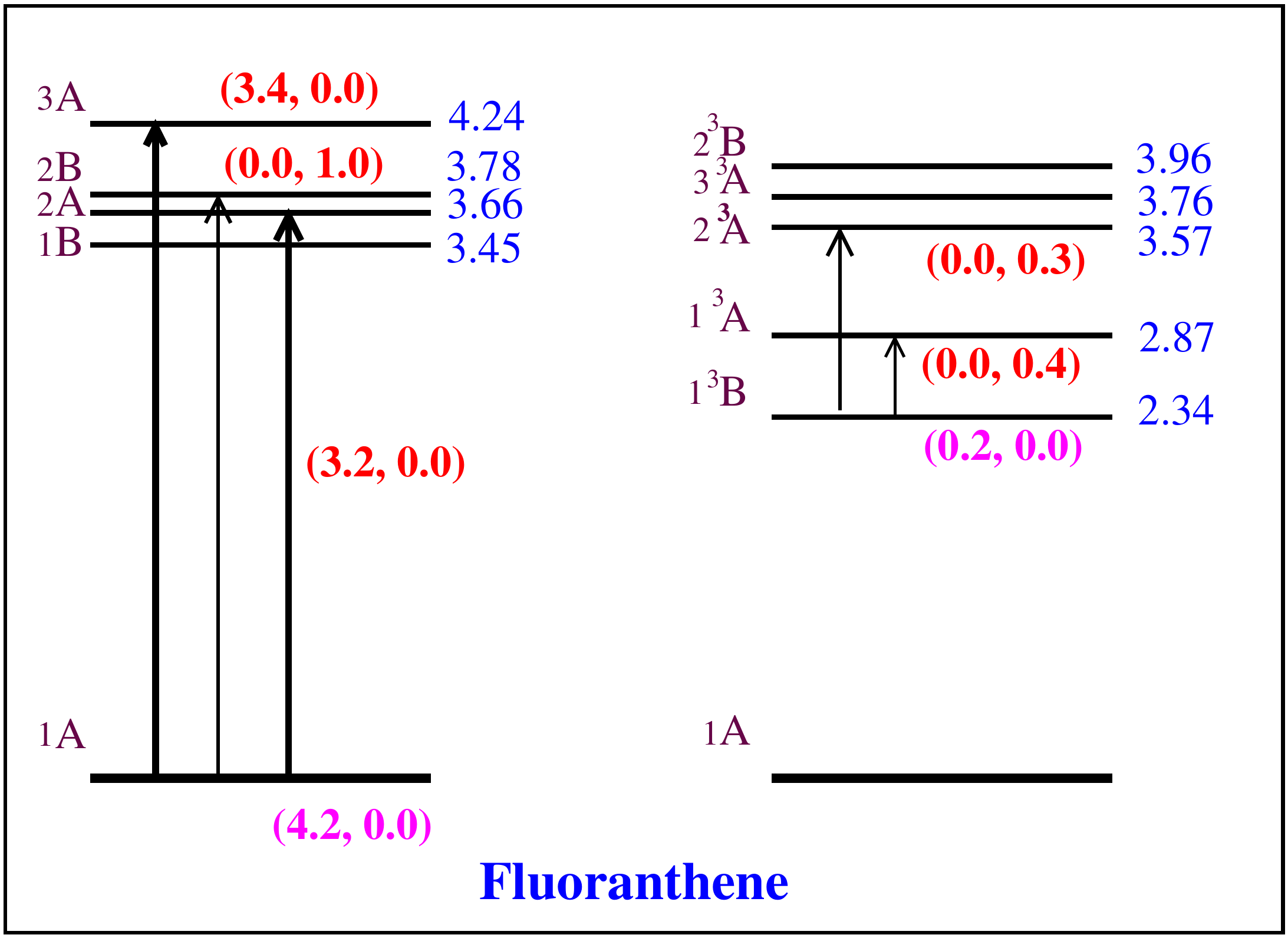}
\hspace{1cm}
\includegraphics[width=0.42\textwidth]{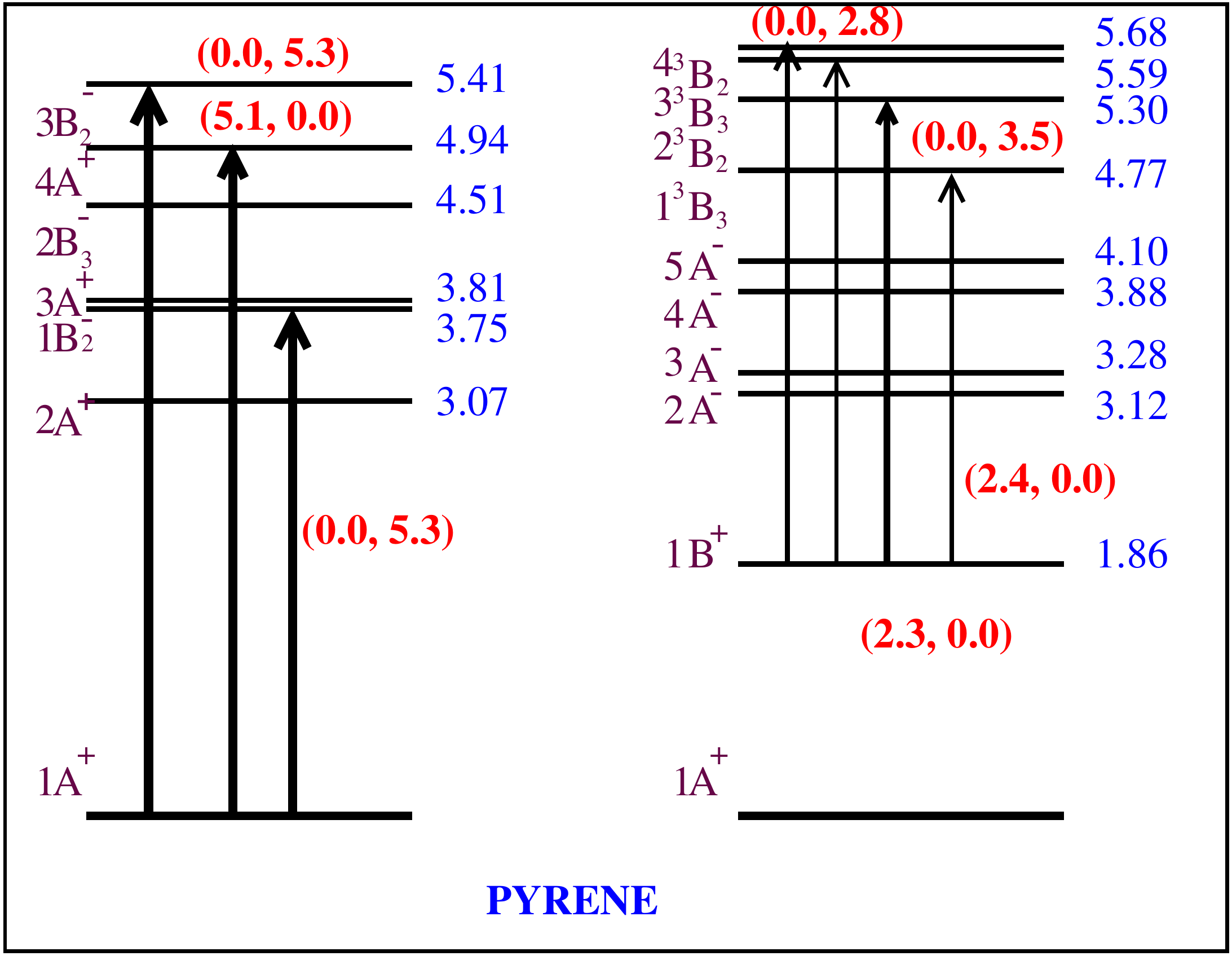}\\
\vspace*{1cm}
\includegraphics[width=0.45\textwidth]{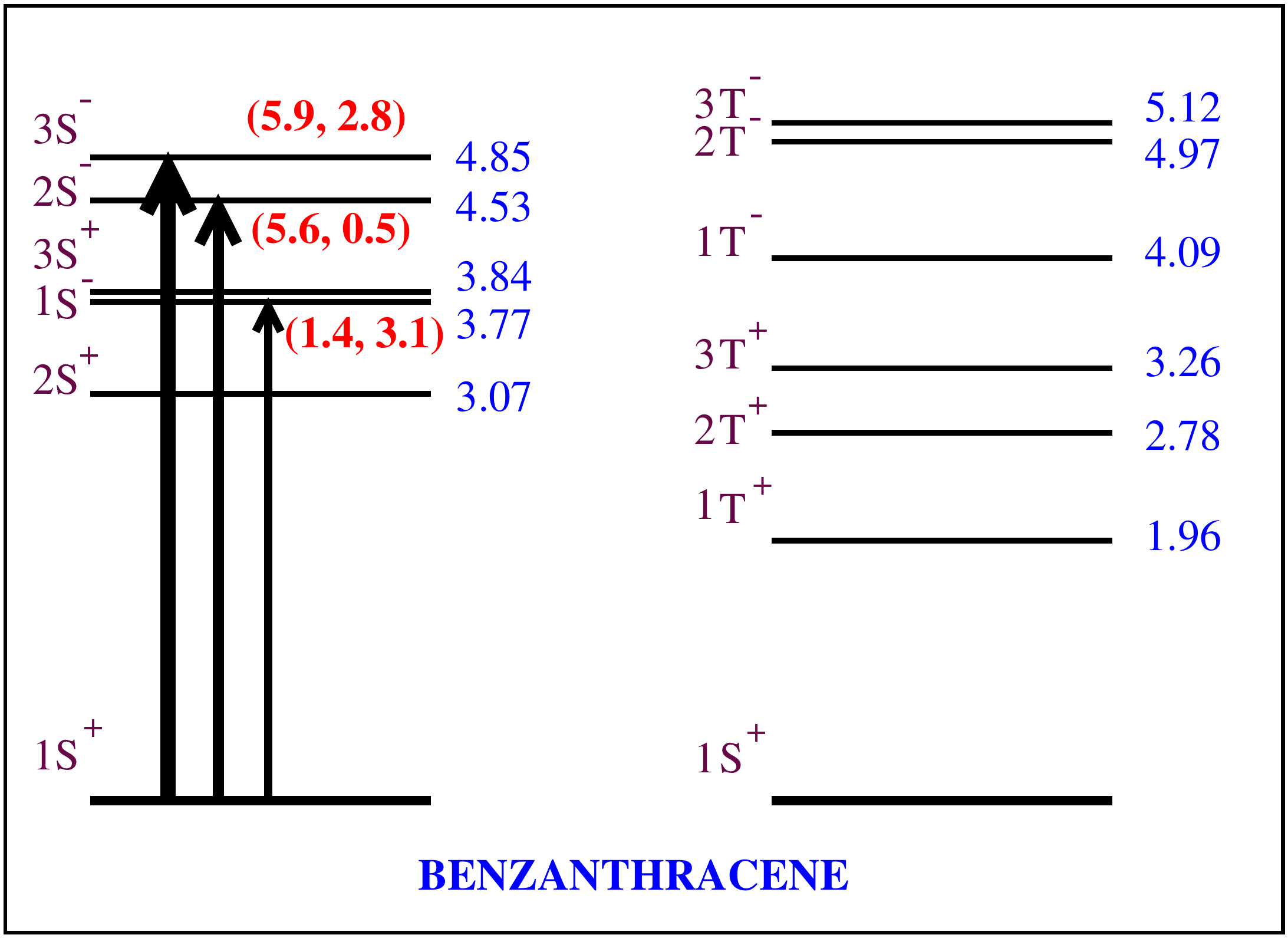}
\hspace{1cm}
\includegraphics[width=0.42\textwidth]{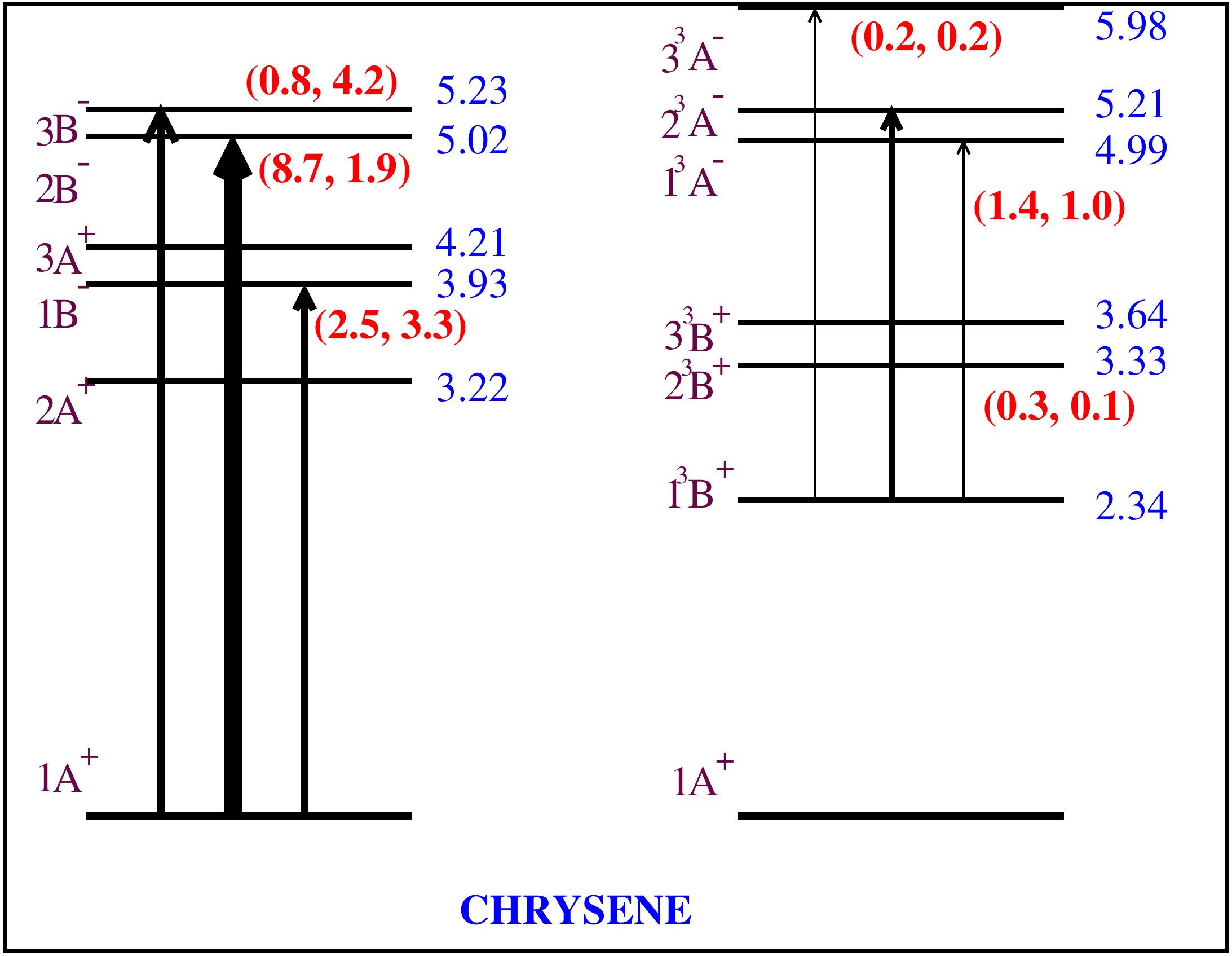}
\caption{Low-lying excitation gaps for PAH molecules. Molecules are 
labelled in each box. The states shown to the left are singlets and 
those to the right are triplets. In the case of Fluoranthene, the symmetry 
labels were assigned from analyzing the eigenstates and confirmed from the 
polarization of the transition dipole. The transition dipoles are in Debye 
and energies are in eV. The spacing between labels is scaled to the magnitude 
of the gaps. The X and Y components of the transition dipoles are given in 
parenthesis and the thickness of the arrow indicates the strength of 
the transition. State labels with '+' and '-' superscripts correspond to 
'covalent' and 'ionic' subspaces.}
\end{figure}

\begin{figure}
\begin{center}
\noindent
\includegraphics[width=0.45\textwidth]{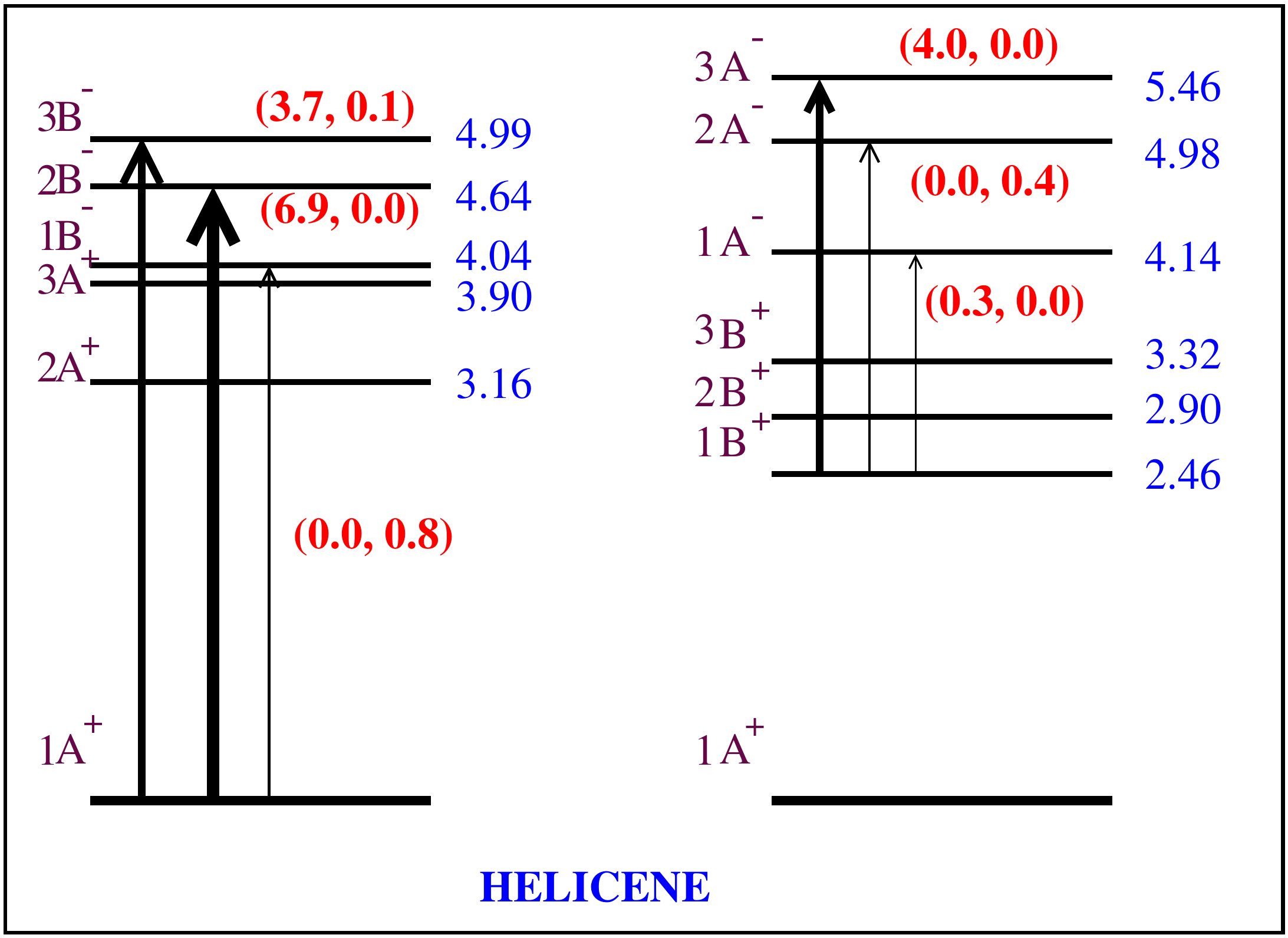}
\hspace{1cm}
\includegraphics[width=0.45\textwidth]{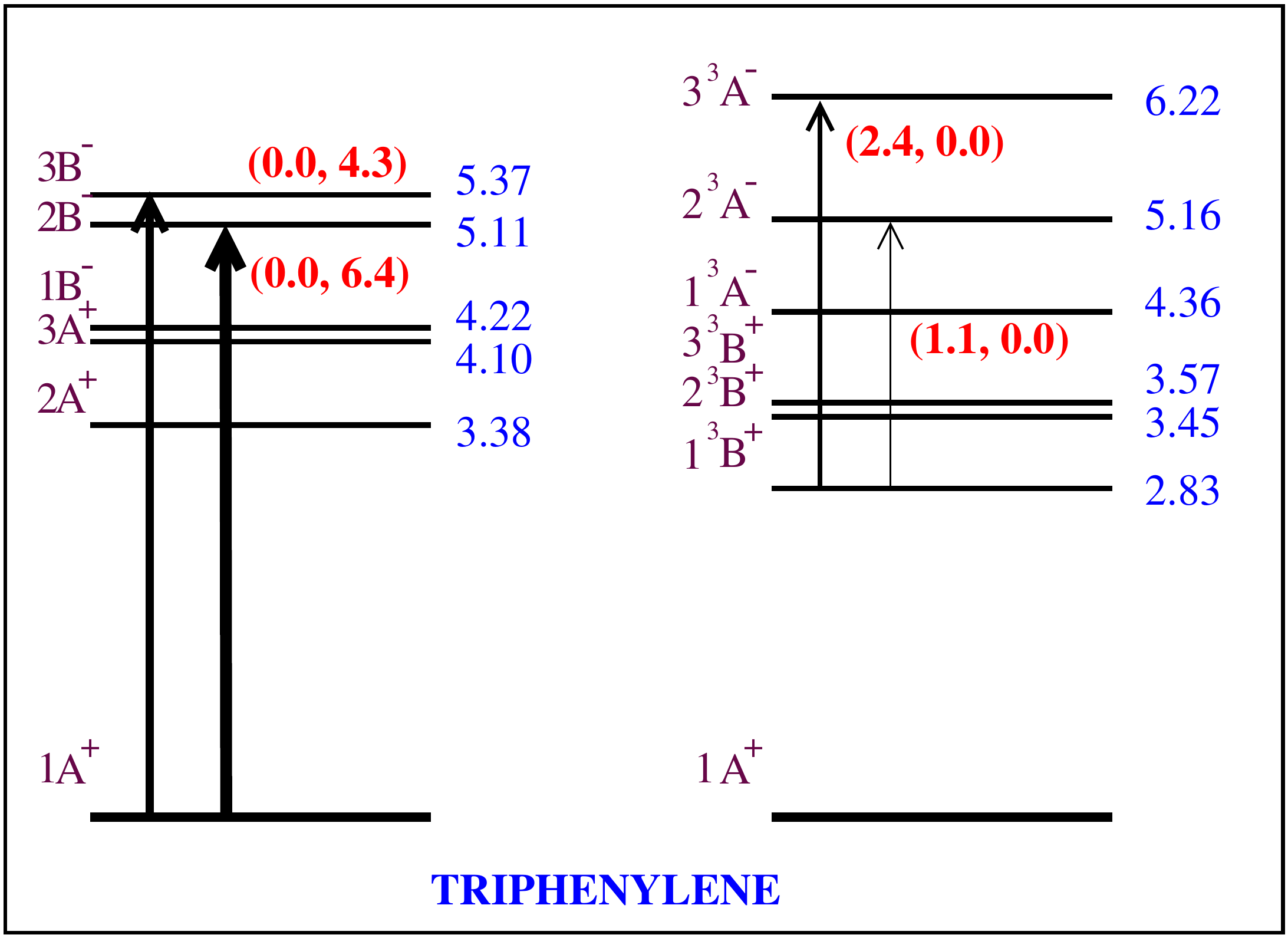}\\
\vspace*{1cm}
\includegraphics[width=0.45\textwidth]{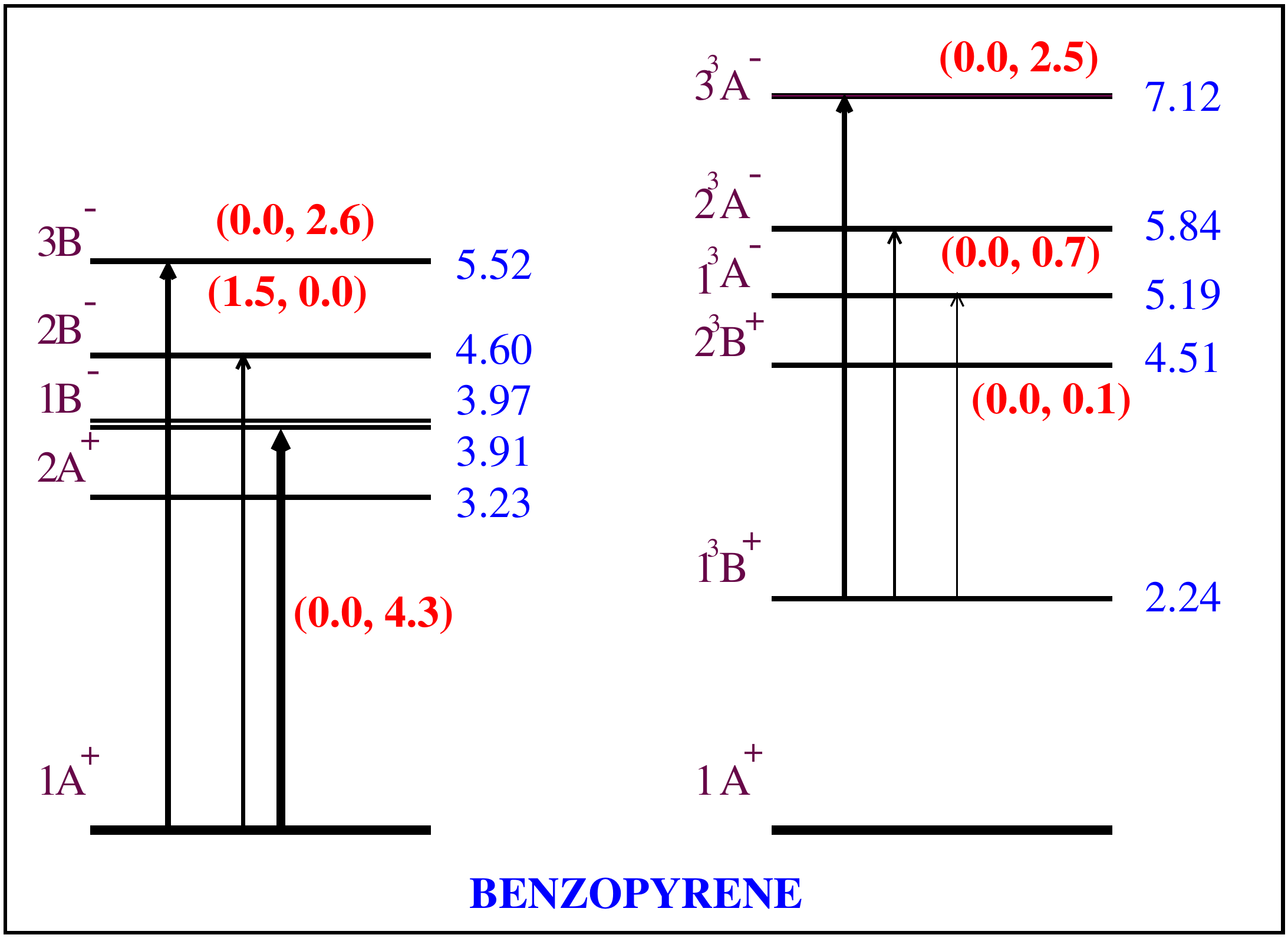}
\hspace{1cm}
\includegraphics[width=0.45\textwidth]{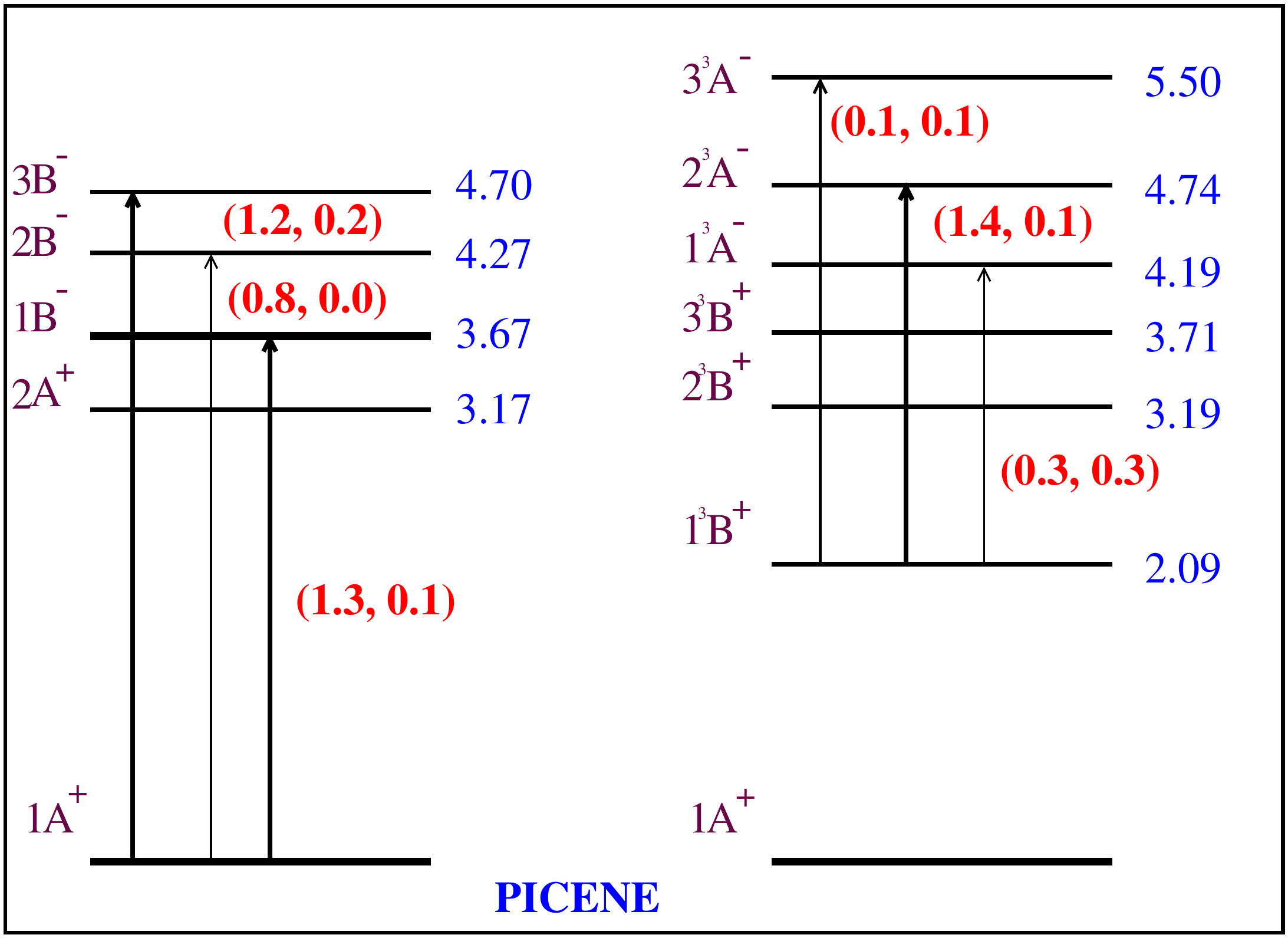} \\

\vspace*{1cm}
{Figure 2 (contd.): Low-lying excitation gaps for the PAH molecules. }
\end{center}
\end{figure}

Hence except fluoranthene, none of the other molecules will be fluorescent 
by Kasha rule \cite{40}. It is also reasonable to assume that the ($2A - 1B$) 
gap is a measure of the strength of electron correlation. We find it lowest 
in fluoranthene, implying that the effective electron correlations are weakest
in this system. Of the alternant systems, the weakest effective correlation is 
in picene and all the other systems show comparable effective correlation. 
The strongest optical absorption to 1Bu state is in benzopyrene and the 
weakest in chrysene. In many of these systems, we do observe strong 
excitations to higher singlet states. In Table 2, we give a comparison of 
experimental gaps with the theoretical 
gaps.\cite{clarence-spectra-1959,pyrene-fluro-jacs-1993} We find that our 
theoretical gaps compare very well with the experimental gaps. For pyrene 
molecule, we also find good agreement with the theoretical result of Basak et 
al obtained using the standard PPP parameters, taking quadruple CI 
basis.\cite{basak-prb92-2015} Even the lowest two-photon gap compares well with their 
results from the standard parameters. \cite{basak-prb98-2018} Triphenylene molecule 
having $C_3$ symmetry shows two two-photon levels below $B^-$ state. Here the 
excitation to the first $B^-$ state is very weak and the second level is 
$\approx$ 0.5eV higher than the 
experimental value. \cite{triphenylene-cpl-1985, triphenylene-rsc-2011}.  In 
other cases also the experimental gaps are nearly $0.5$ eV lower than the 
computed values, which can be attributed to solvent effects. In case of 
benzopyrene, the difference is slightly larger because the gap is obtained 
from the truncated CI basis within DMRG method.\cite{benzopyrene-anal.chem-1976} 
Furthermore, since the molecule is not a quasi-one dimensional system (in 
contrast to picene), the excitation gap is slightly higher than the experimental
gap.

We have given one-photon, two-photon and spin gaps of these molecules 
in Fig. 3. We find that in all the cases except pyrene and picene, the 
lowest triplet state is well above half of the singlet-singlet gap and hence 
none of the molecules are suitable for improving the solar cell efficiency via 
singlet fission. However, for pyrene and picene, since [$E_{S1} - 2 \times E_{T1}$] 
value is almost zero, these molecules can be tuned to have a favourable 
conditions for singlet fission, by substitution. In fact, B-N substituted 
pyrenes are shown to be possible candidate materials for singlet fission by Zeng et al
\cite{bnpyrene-jpcl-2018}. In fluoranthene, pyrene, benzanthracene, 
benzopyrene and picene, the ($2A - T_1$) gap is $\geq$ 1.0 eV. However, 
triphenylene has the lowest $2A - T_1$ gap of 0.5 eV, while the gap is 0.7 eV 
in helicene, and is 0.9 and 1.0 eV in chrysene and benzopyrene, respectively.  
We notice that the optical gap falls off rapidly from anthracene to pentacene
(3.7 eV to 2.92 eV, see Table S1 in supporting information), whereas for 
phenacenes, it is slower (4.3 eV for phenanthrene to 3.82 eV for picene). In the case of oligoacenes, the spin gap
almost reaches zero, whereas for phenacenes, it is quite large,  even in case
of picene (1.95 eV).

\begin{table}[ht!]
\begin{center}
\setlength{\tabcolsep}{1.5pt} 
\begin{tabular}{|l| c |c|}
\hline
Molecule & Optical gap  & singlet-  \\
 & (Expt gap)       & triplet gap \\      
\hline
Fluoranthene   & ~~ 3.45 ~(3.46) \cite{clarence-spectra-1959}~~ &  2.34  \\
Pyrene         & ~~~~ 3.75 ~(3.34) \cite{clarence-spectra-1959,pyrene-fluro-jacs-1993} &  1.86  \\
Benzanthracene & ~~ 3.77 ~(3.23) \cite{clarence-spectra-1959}~~ &  1.96  \\
Chrysene       & ~~ 3.95 ~(3.45) \cite{hussain-spec-chimi-acta-2007}~~ &  2.34  \\
Helicene       & ~~ 4.04 ~(3.49) \cite{palewska}~~ &  2.47  \\
Triphenylene   & ~~~~ 5.11 ~(4.36) \cite{triphenylene-cpl-1985,triphenylene-rsc-2011} &  2.83  \\
Benzopyrene    & ~~ 3.91 ~(3.38) \cite{benzopyrene-anal.chem-1976}~~ &  2.28  \\
Picene         & ~~ 3.82 ~(3.30) \cite{clarence-spectra-1959}~~ &  1.95 \\
\hline
\end{tabular}
\end{center}
\caption{Electronic excitation gap and singlet-triplet gaps in eV for PAH 
molecules. Numbers in the bracket are experimental gaps obtained from 
references shown as superscript. Except benzopyrene and picene other gaps are 
numerically exact. For benzopyrene and picene, the singlet-singlet and singlet-triplet 
gaps are obtained from DMRG studies. }
\label{tabular:sym-basis}

\end{table}

\begin{figure}[h]
\begin{center}
\includegraphics[width=0.8\textwidth]{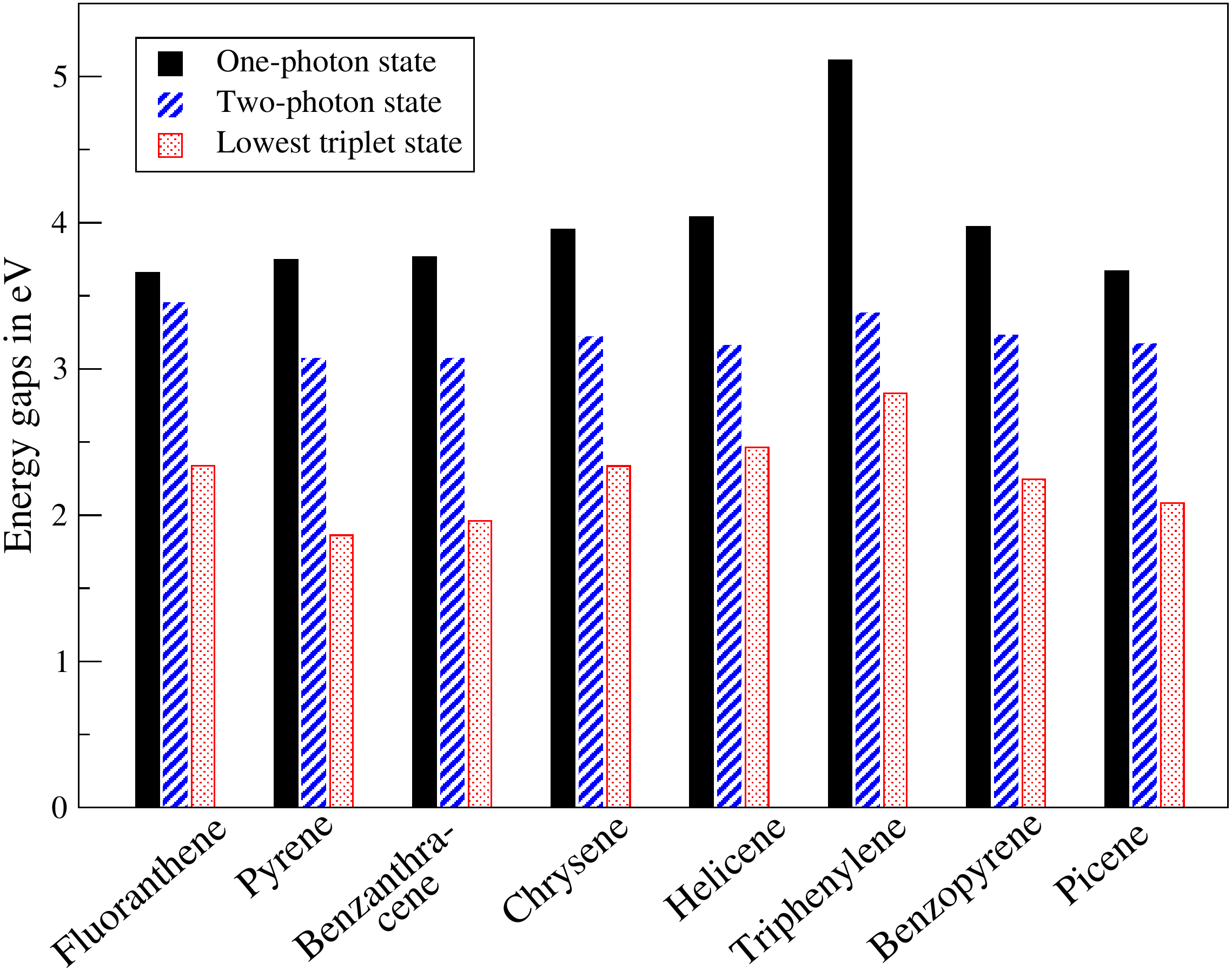}
\end{center}
\caption{Energy gaps in eV, for one-photon state (black filled bar), two-photon 
state (blue hatched bar) and triplet states (red dotted bar) for PAHs molecules.}
\end{figure}

\section{Singlet State Properties}
\subsection{Bond orders} {We have computed the bond orders of the PAH molecules
 in various electronic states. The bond order between sites 'i' and 'j' is 
given by the expectation value of $\hat{p}_{ij} = -\dfrac{1}{2} 
\left( \hat{E}_{ij} + \hat{E}_{ji} \right)$ in the state of interest, 
 where 
$ \hat{E}_{ij} = \sum_{\sigma}c^{\dagger}_{i\sigma} c_{j\sigma}$. The larger 
the bond order, 
the shorter is the bond length and vice versa. Comparison of bond orders in 
a given excited state gives the propensity of a bond to shorten or lengthen 
with respect to the ground state bonds. At equilibrium, the ratio of bond 
length to bond order is expected to be the same for all bonds.}\\

{\indent In Fig. 4, we have given the bond orders in the singlet ground state 
as well as the lowest one-photon excited state for all the PAHs we have 
studied. For comparison, we have given the bond orders and spin densities
of naphthalene, anthracene, phenanthrene and biphenyl molecules in 
supporting information (Fig. S1 and S2). 

\begin{figure}[ht!]
\begin{center}
\hspace{-1.0cm}
\includegraphics[width=4.5cm]{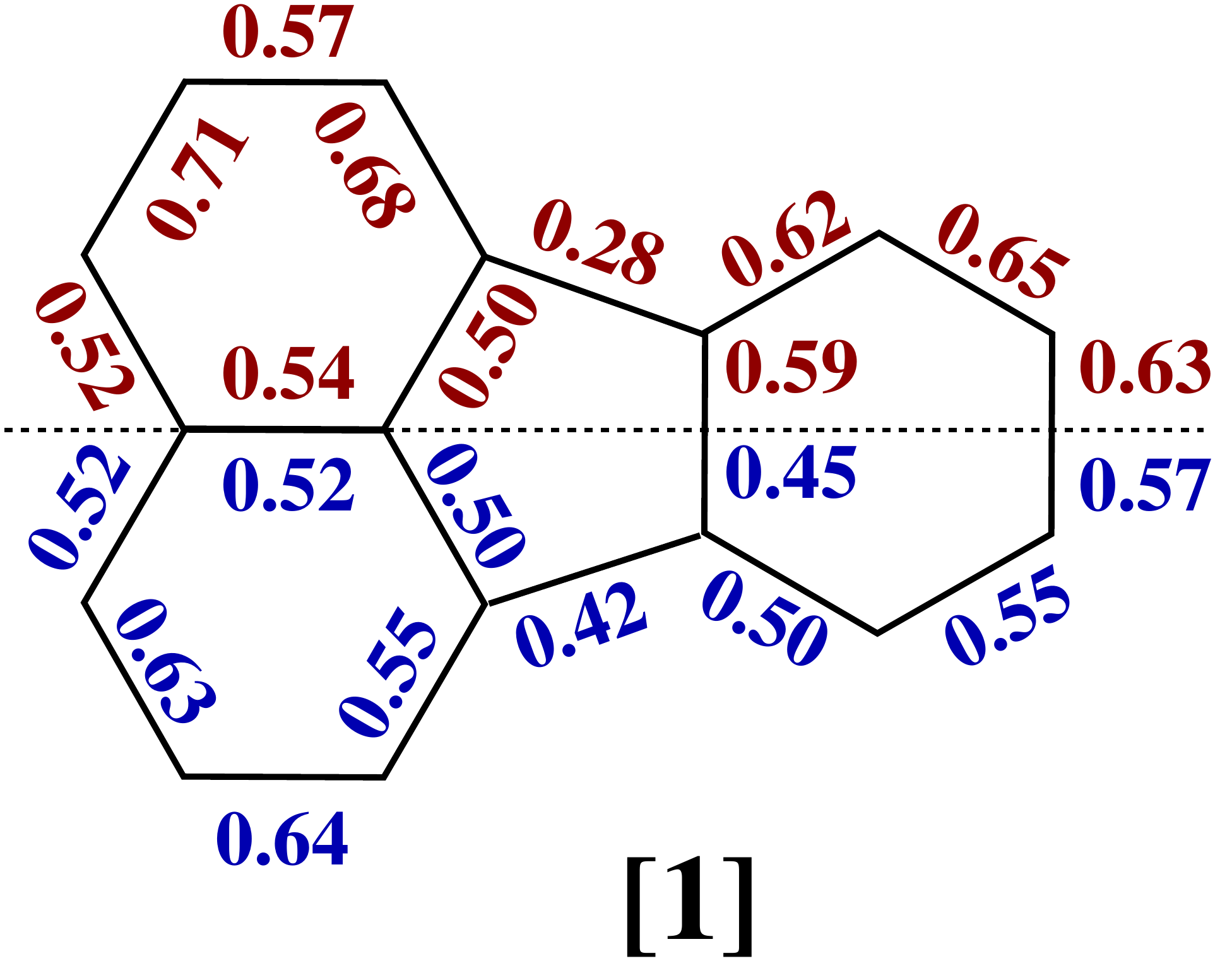}
\hspace{1.0cm}
\includegraphics[width=4cm]{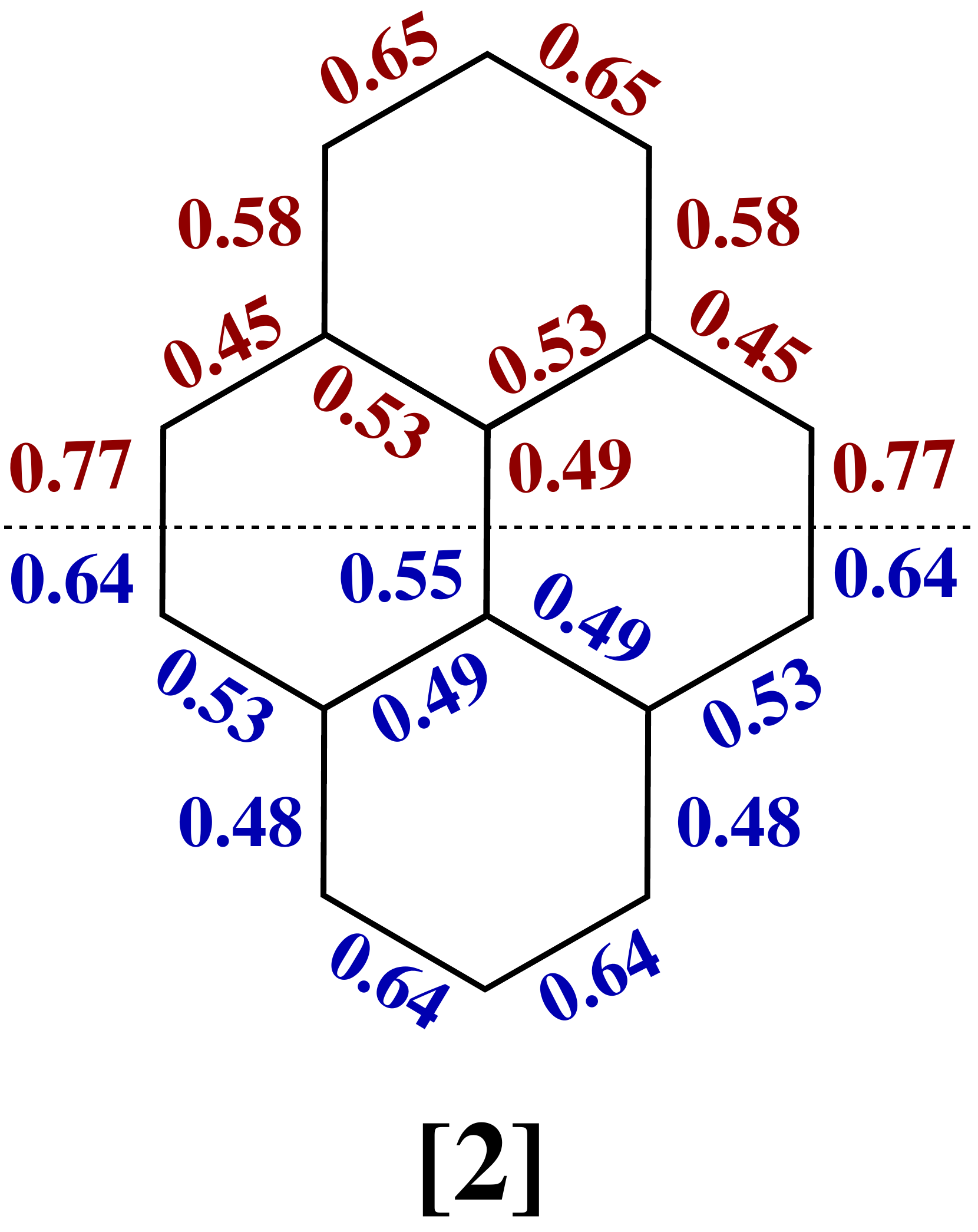}
\hspace{1cm}
\includegraphics[width=6cm]{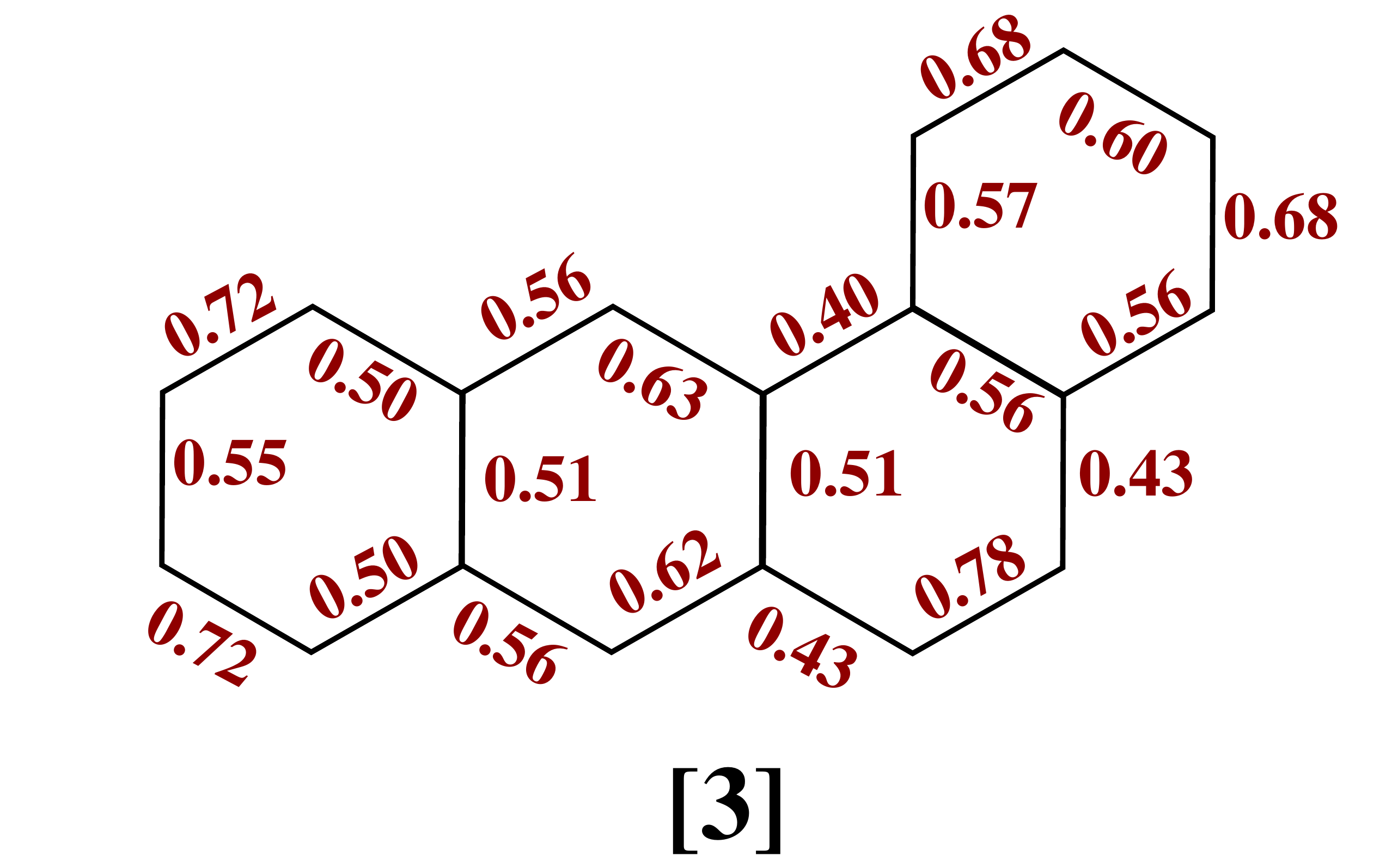} \\
\vspace*{0.5cm}
\hspace*{-2.0cm}
\includegraphics[width=5.5cm]{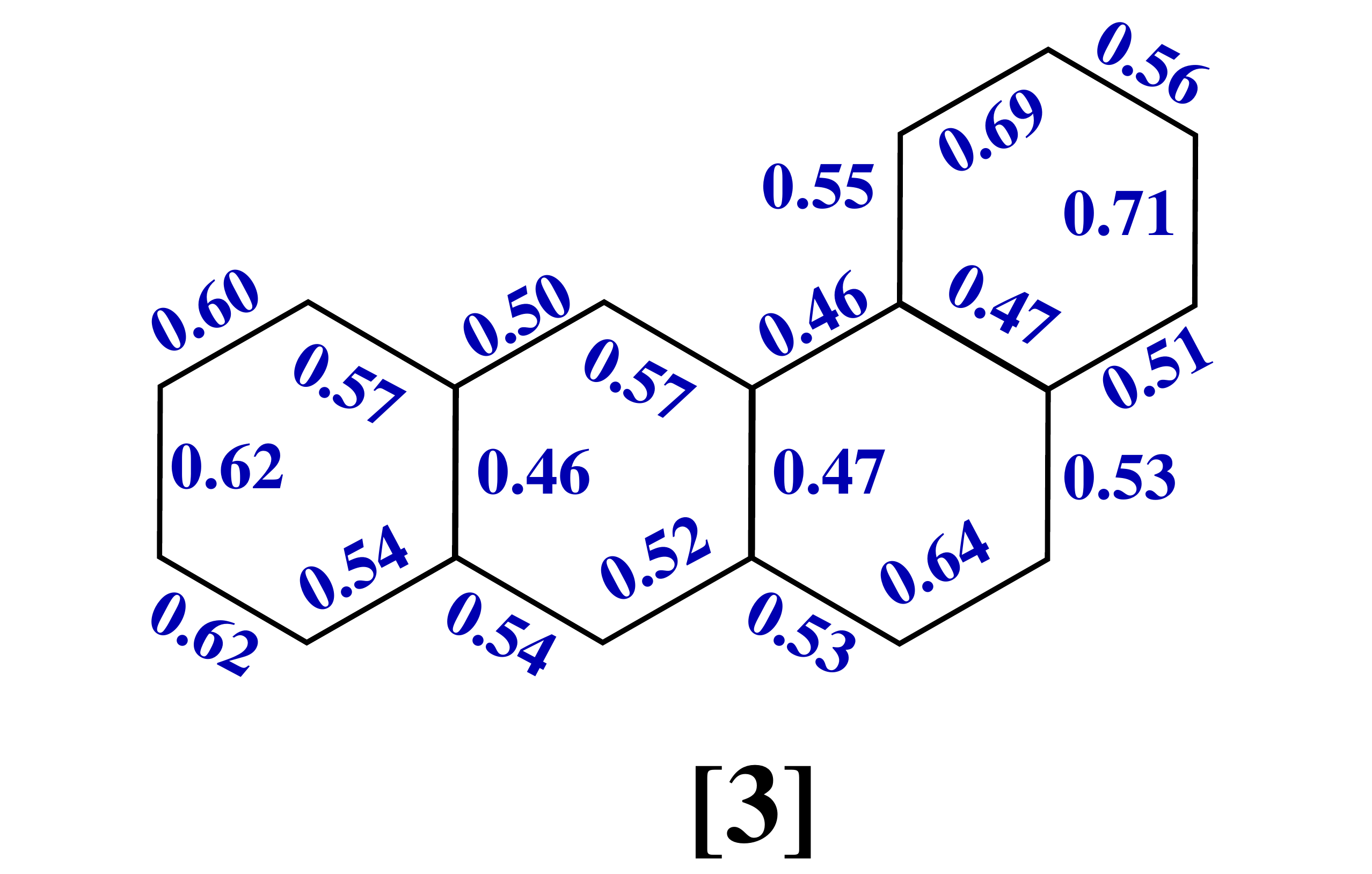}
\hspace{0.3cm}
\includegraphics[width=5.5cm]{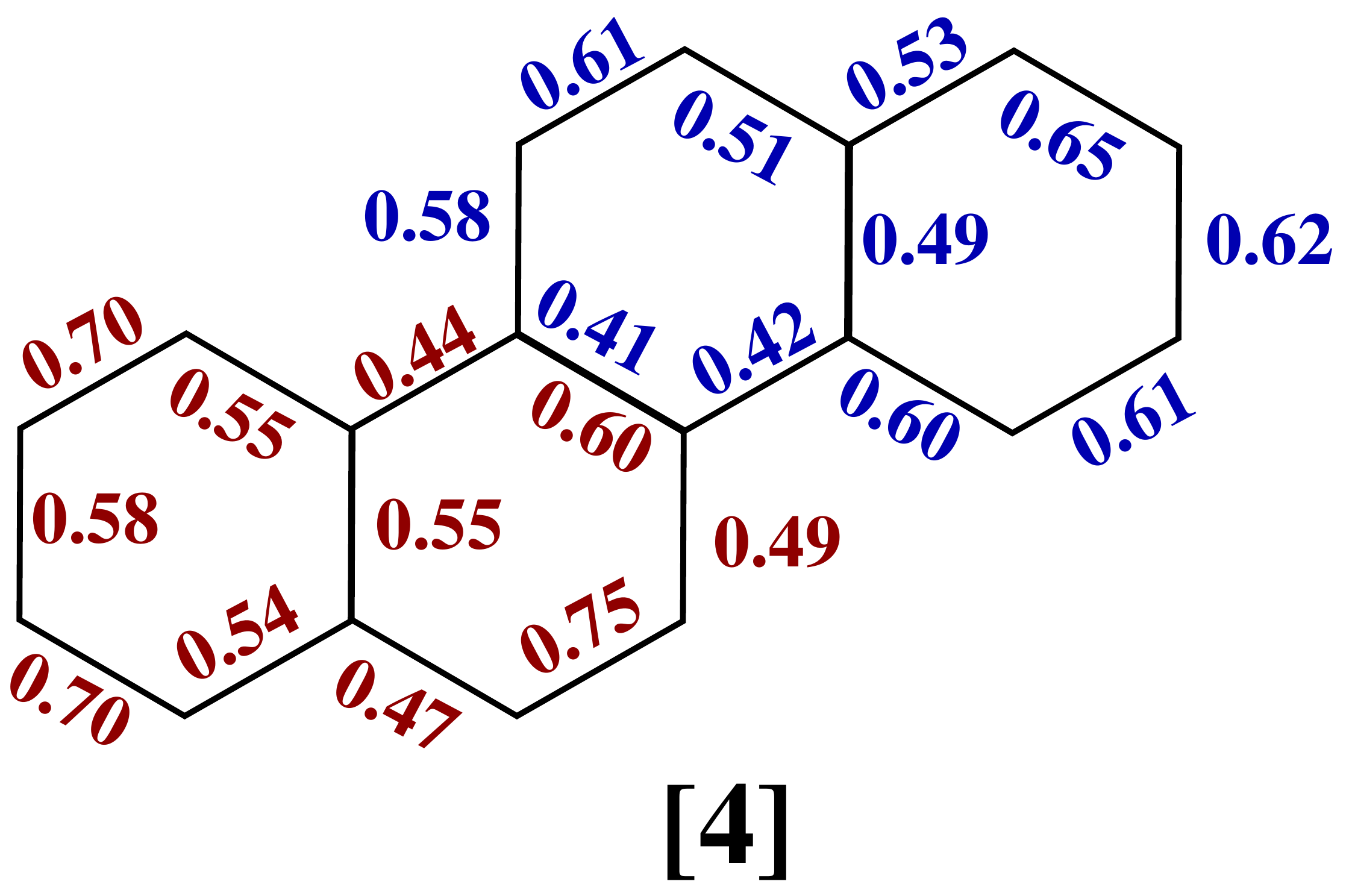}
\hspace{0.4cm}
\includegraphics[width=5.8cm]{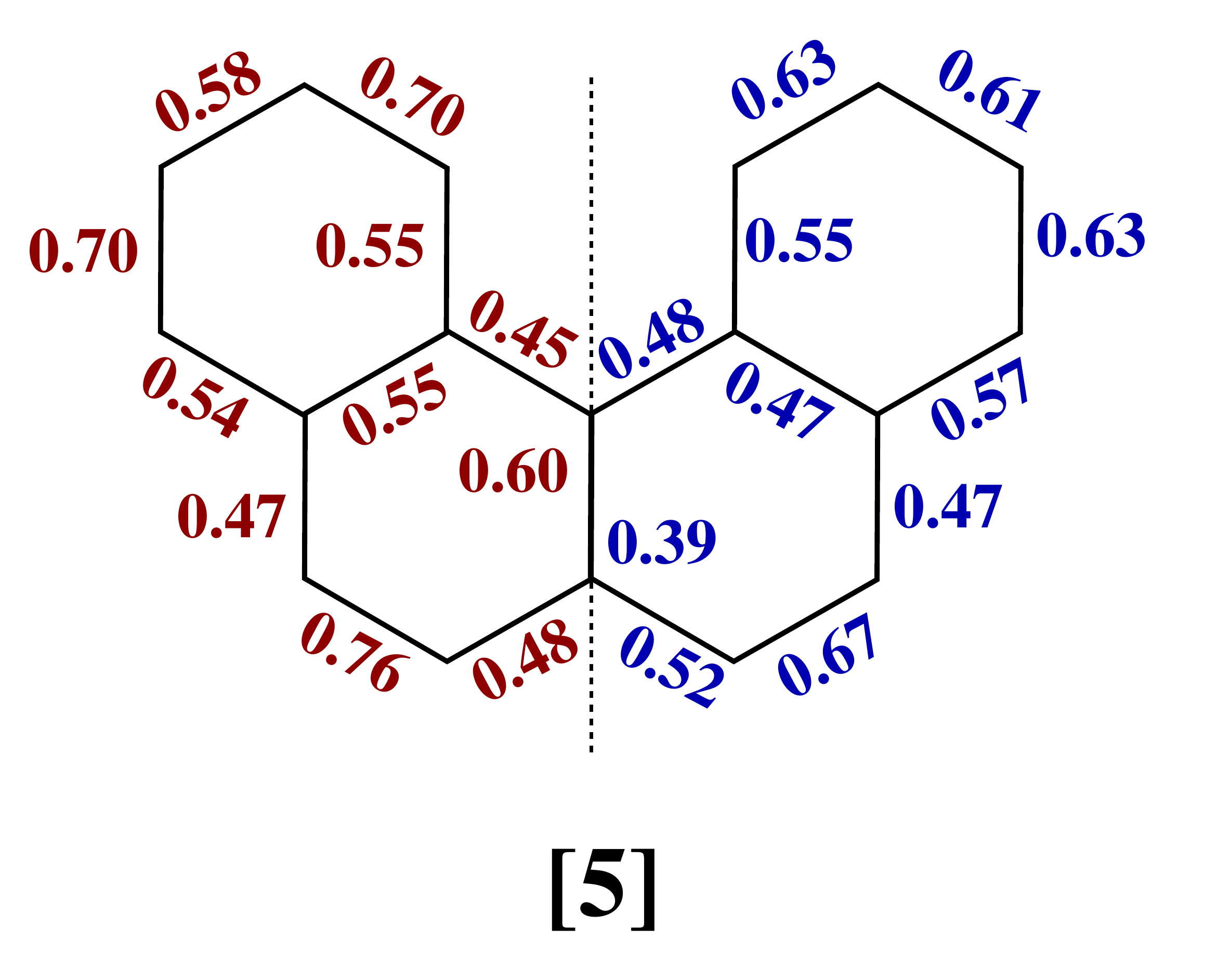} \\
\vspace*{0.5cm}
\hspace*{-1.5cm}
\includegraphics[width=4.0cm,height=5.5cm]{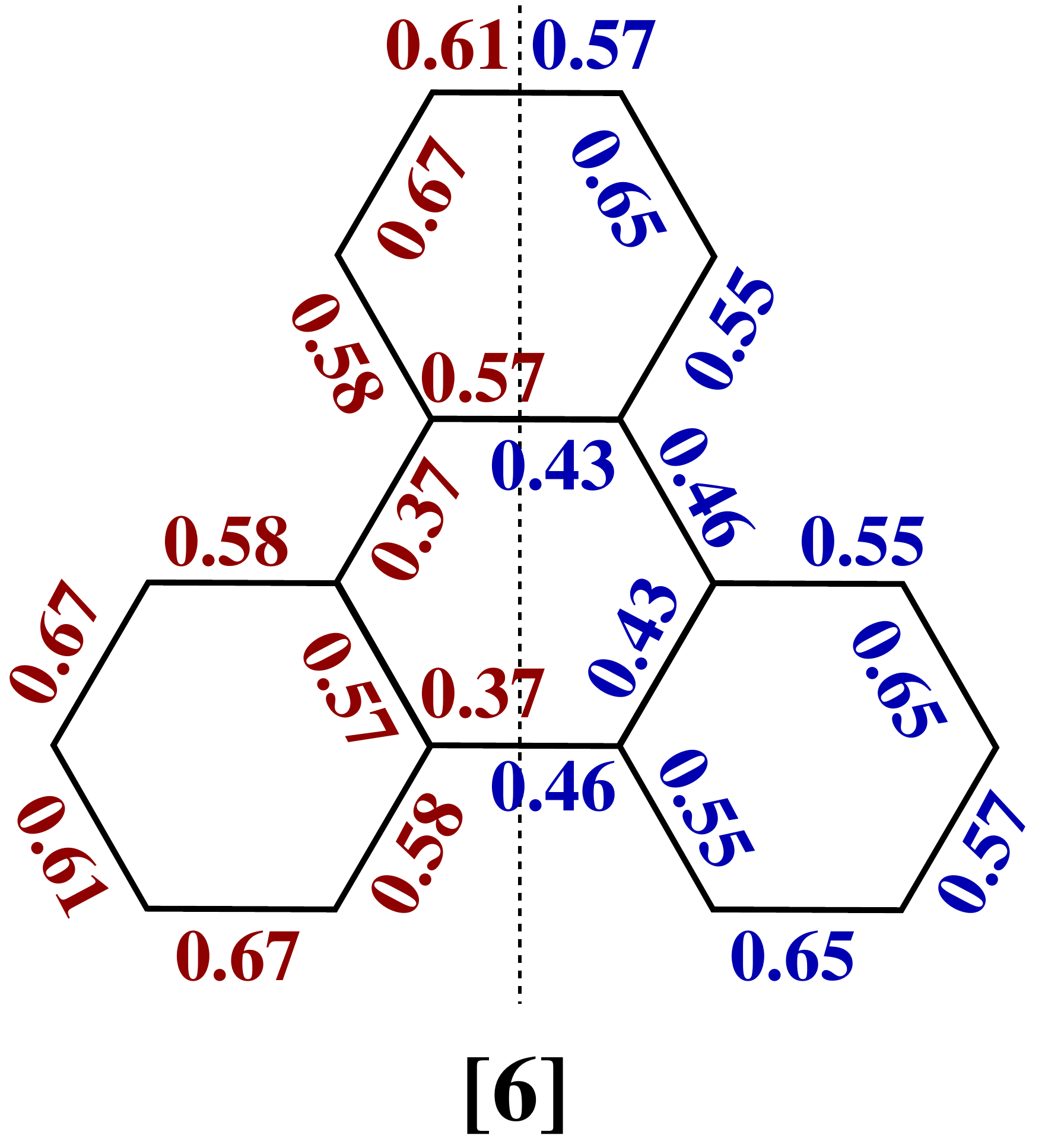}
\hspace{0.3cm}
\includegraphics[width=6.1cm,height=6.2cm]{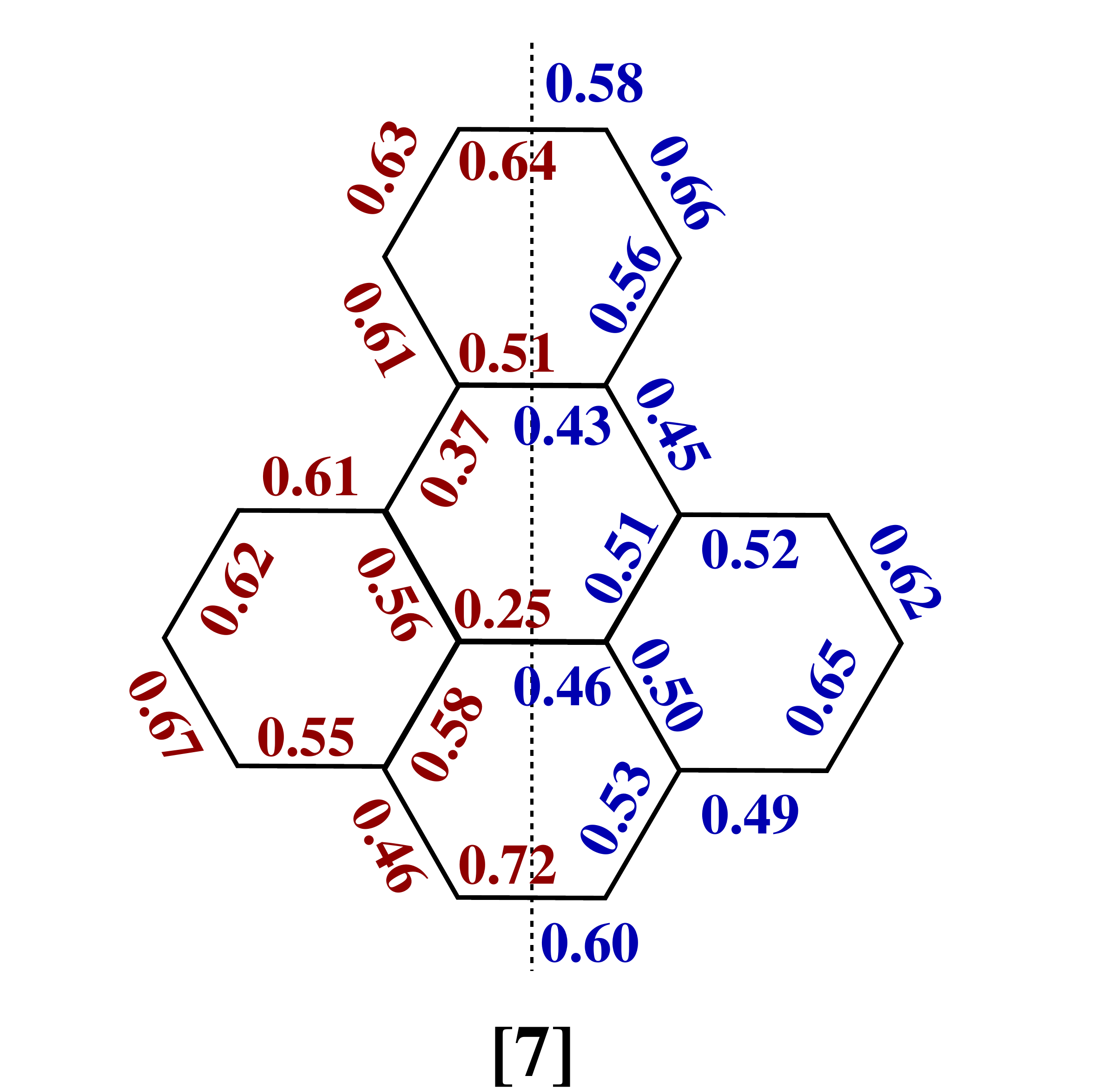} 
\hspace{0.0cm}
\includegraphics[width=6.6cm]{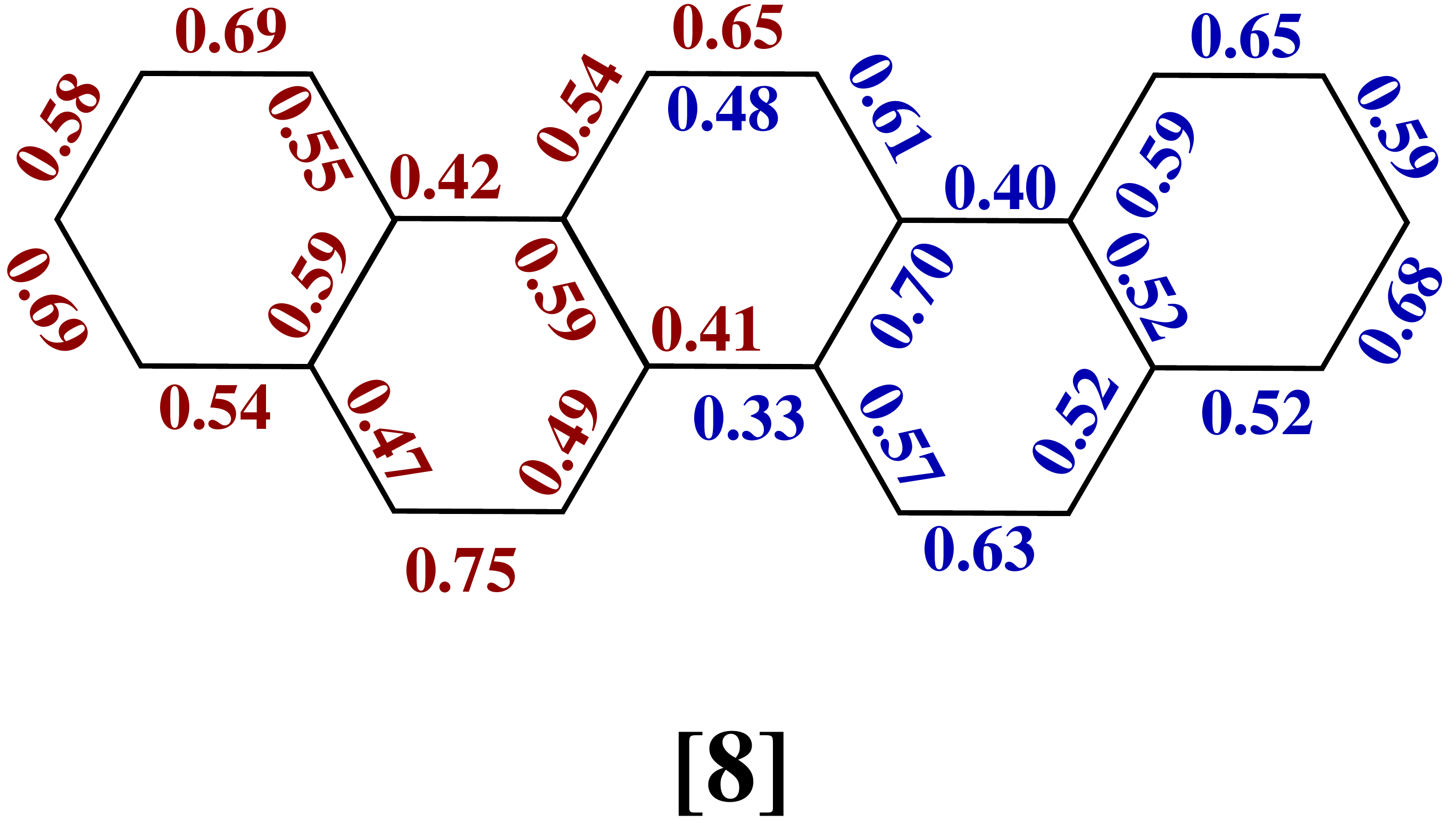}%
\end{center}
\caption{Bond orders in singlet ground state (in red) and in dipole allowed 
state (in blue). The numbering for molecules is as given in Figure 1. \\
}
\label{fig: singlet-BO}
\end{figure}

\subsubsection{Molecules with 16 $\pi$-electrons (Fluoranthene and Pyrene)} 
In the ground state of fluoranthene, we note that the bond order of the bonds 
connecting the naphthalene unit and benzene unit are very small. This 
indicates that the two units are only weakly connected by the 
$\pi-$conjugation. The bond orders in the six membered rings are close to 
those of benzene in the H\"{u}ckel model $\sim$0.66, although bonds which 
are shared between rings have slightly lower bond order. In the optically 
excited state the bond connecting the benzene and naphthalene units become 
much stronger, indicating a contraction of the distance between the two units. 
The other bonds also show slightly smaller but nearly uniform bond orders, 
indicating ring expansion in the excited state. This also shows that 
the Stokes shift in the spectral lines will be large in this system. 
This is in confirmity with large Stokes shift observed experimentally,
by Gusten and Heinrich.
\cite{fluoranthene-jphoto-1982}.

In pyrene, the bond orders in the ground state resemble those of biphenyl 
connected by two ethylenic units on either side, with alternate single and 
double bonds, rather than the bond orders of naphthalene connected by two 
propylenic units at the top and bottom. This is in accordance with the 
most possible resonance structures for pyrene. Compared to biphenyl, the 
central bond becomes stronger in pyrene (see Fig. S1). In the optically excited 
state, all the bond orders become more uniform. This shows that at equilibrium 
geometry, the molecule is not distorted much from the uniform bond lengths 
assumed in the calculations. We find a uniform expansion of rings in 
the optically excited state. We should expect the Stokes shift due to 
relaxation of the geometry in the excited state to be small as opposed to 
what is expected in fluoranthene.

\subsubsection{Molecules with 18 $\pi$-electrons (Benzanthracene, Chrysene, Helicine and Triphenylene)}
Benzanthracene molecule can be viewed as a phenyl ring attached either to  an
anthracene or to a phenanthrene. This is well reproduced  by the bond order 
pattern in the ground state, where one of the two terminal rings have 
anthracenic like and the other two have phenanthrenic like bond orders (see SI 
Figure 1). In benzanthracene, the middle bond of armchair of 
the phenanthrenic moiety is the weakest while the bond opposite to that is 
the strongest. The ground state bond orders of both chrysene and helicene, 
again, resemble those of phenanthrene with a marginal change in bond orders. 
The middle bond of armchair of the phenanthrenic unit is the weakest and 
the bond opposite to that is the strongest, in accordance with the possible 
resonance structures. This is true even for helicene, where there are two 
weaker bonds at the arm chair, connecting the two naphthalenic unit. In both 
molecules, there is a slight enlargement of the rings in the excited state. 
The Stokes shift in these system are not expected to be large. The 
ground state bond orders in triphenylene molecule, correspond to three phenyl 
rings connected by a very weak bond (bond order is 0.37). The bond order 
pattern is reversed upon excitation, with the central ring having a more 
uniform and slightly stronger bonds compared to the ground state geometry. 
Even in this molecule, the excited state geometry has  more uniform bond 
orders. Once again the change in geometry from the ground state to the 
excited state is small, which leads us to expect small Stokes shift. 
Notice that the effect of non-conjugation in triphenylene is reflected in 
the excitation gap. Among all the PAHs we have studied, triphenylene has  
the largest optical gap and the singlet-triplet gap (see Table 2). Among 18 $\pi-$ 
electronic molecules, benzanthracene has the lowest gap while chrysene and 
helicene have almost similar gap ($3.95$ eV and $4.04$ eV, see Table 2) owing
to their similarity in conjugation pattern. \\

\subsubsection{Molecules with 20 (Benzopyrene) and 22 $\pi$-electrons (Picene)}
Benzopyrene molecule has similar bond orders as those of triphenylene with an
extra ethylenic unit connected at lower end of the molecule (see [7] in 
Fig. 4). Picene being a oligomer of phenanthrene, shows the ground state bond 
order similar to phenanthrene with minimal change in the bond orders of 
interior bonds. The change in equilibrium geometry from the ground to the 
excited state is not much, although the bond lengths in the excited states 
are slightly larger. Thus, even in picene, we should not expect large Stokes 
shift in the fluorescence.} 

\section{Triplet State Properties}

\begin{figure}[ht!]
\begin{center}
\hspace{-1.0cm}
\includegraphics[width=4.5cm]{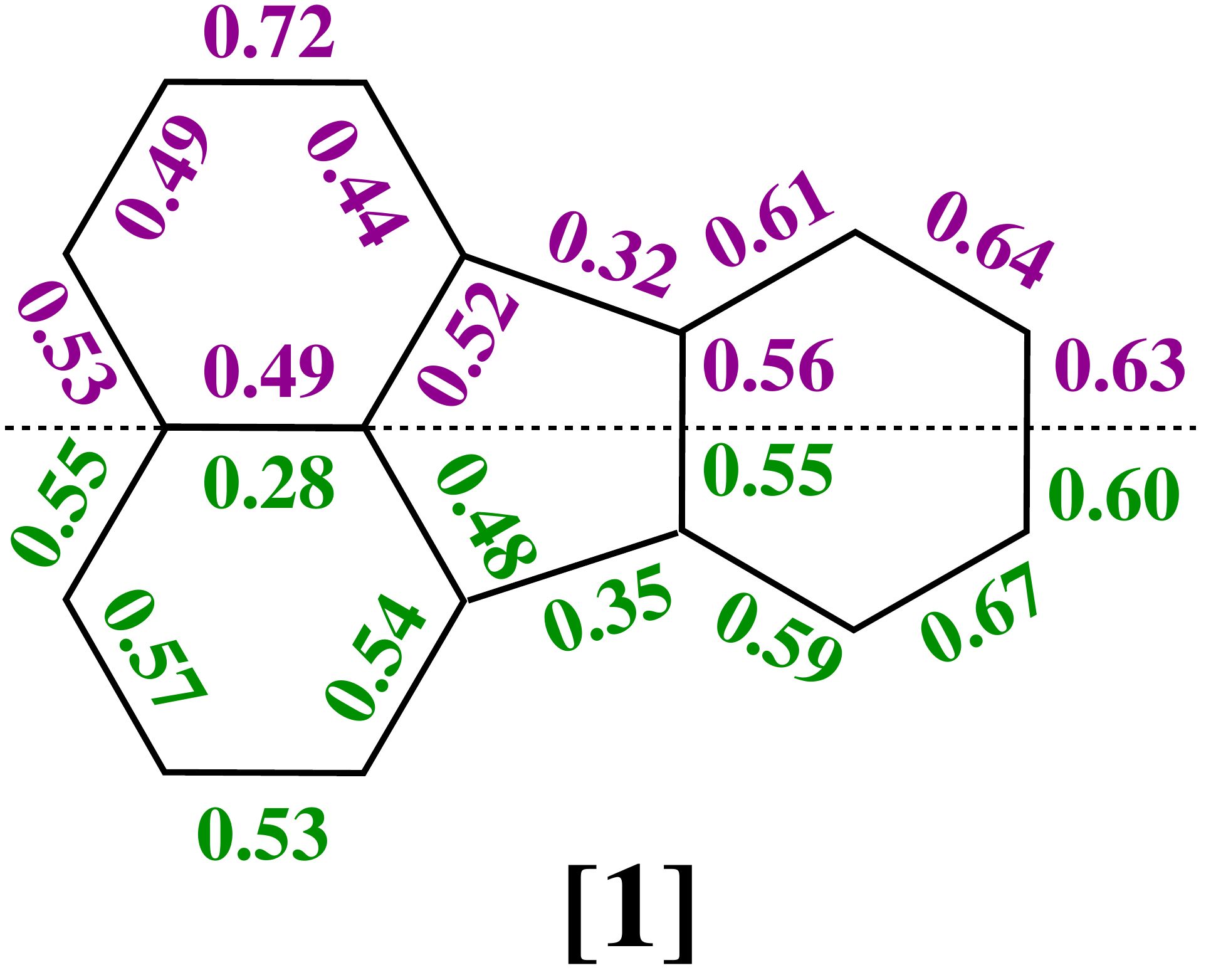}
\hspace{1.0cm}
\includegraphics[width=4cm]{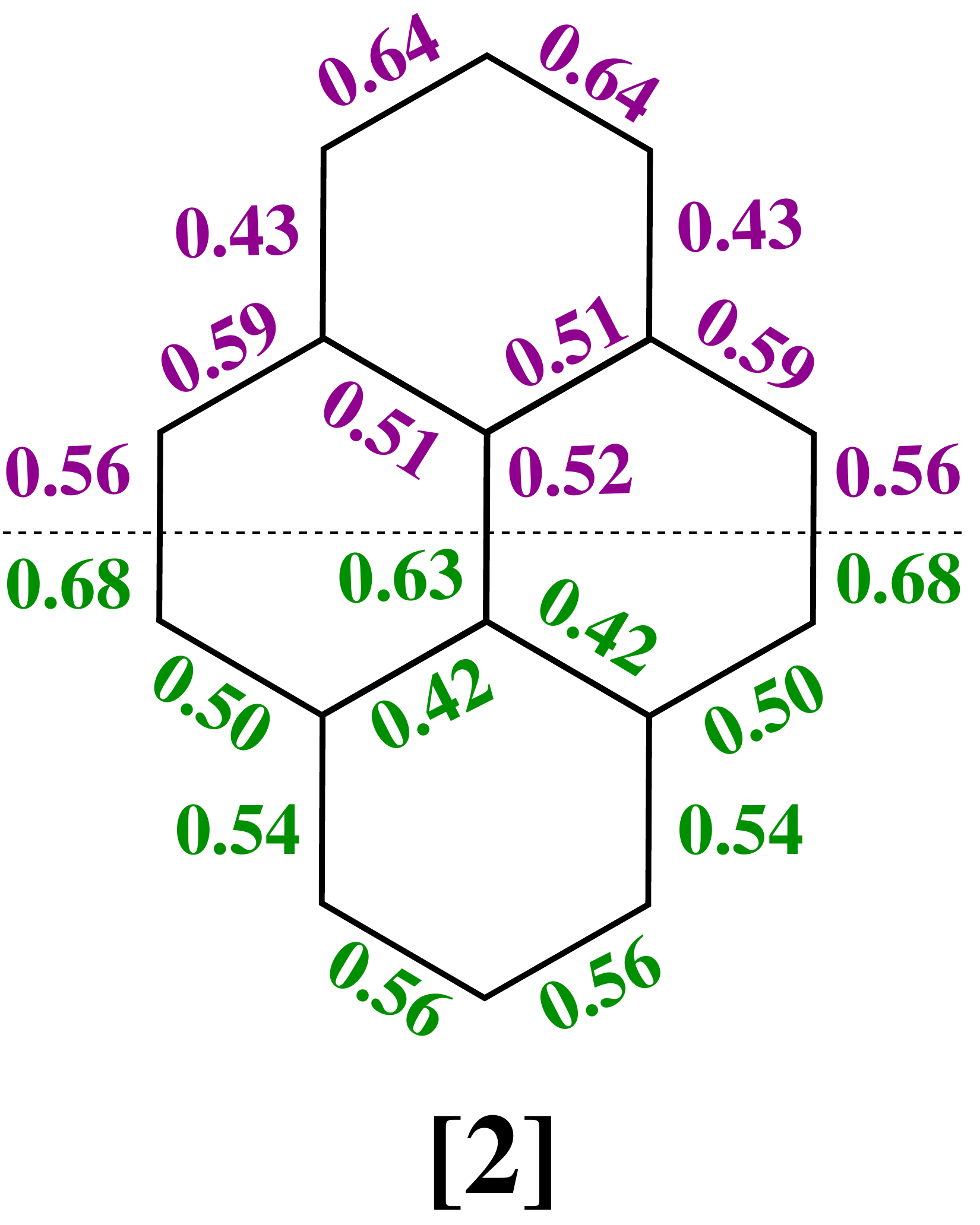}
\hspace{1cm}
\includegraphics[width=6cm]{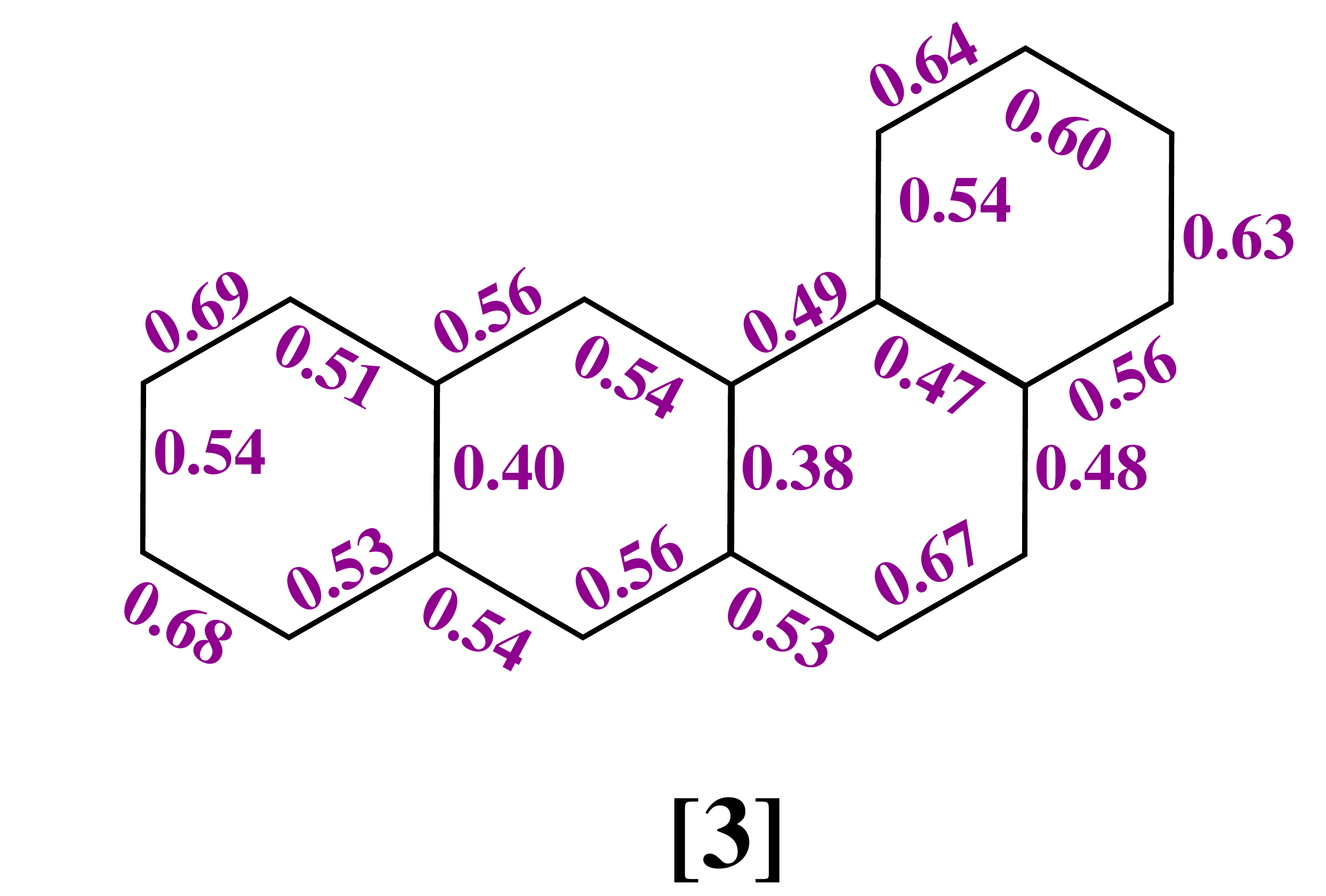} \\
\vspace*{0.5cm}
\hspace*{-2.0cm}
\includegraphics[width=5.8cm]{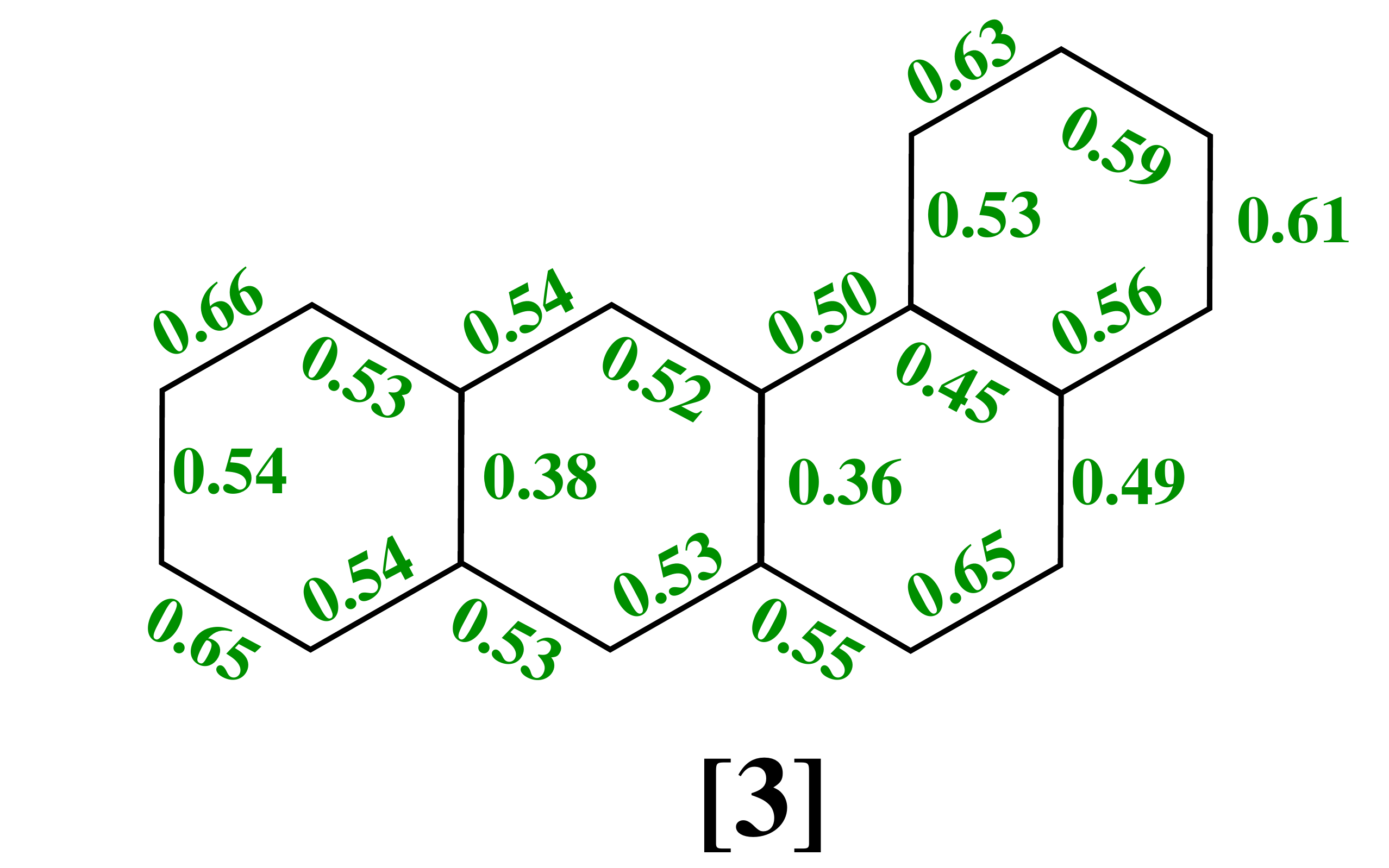}
\hspace{0.3cm}
\includegraphics[width=5.5cm]{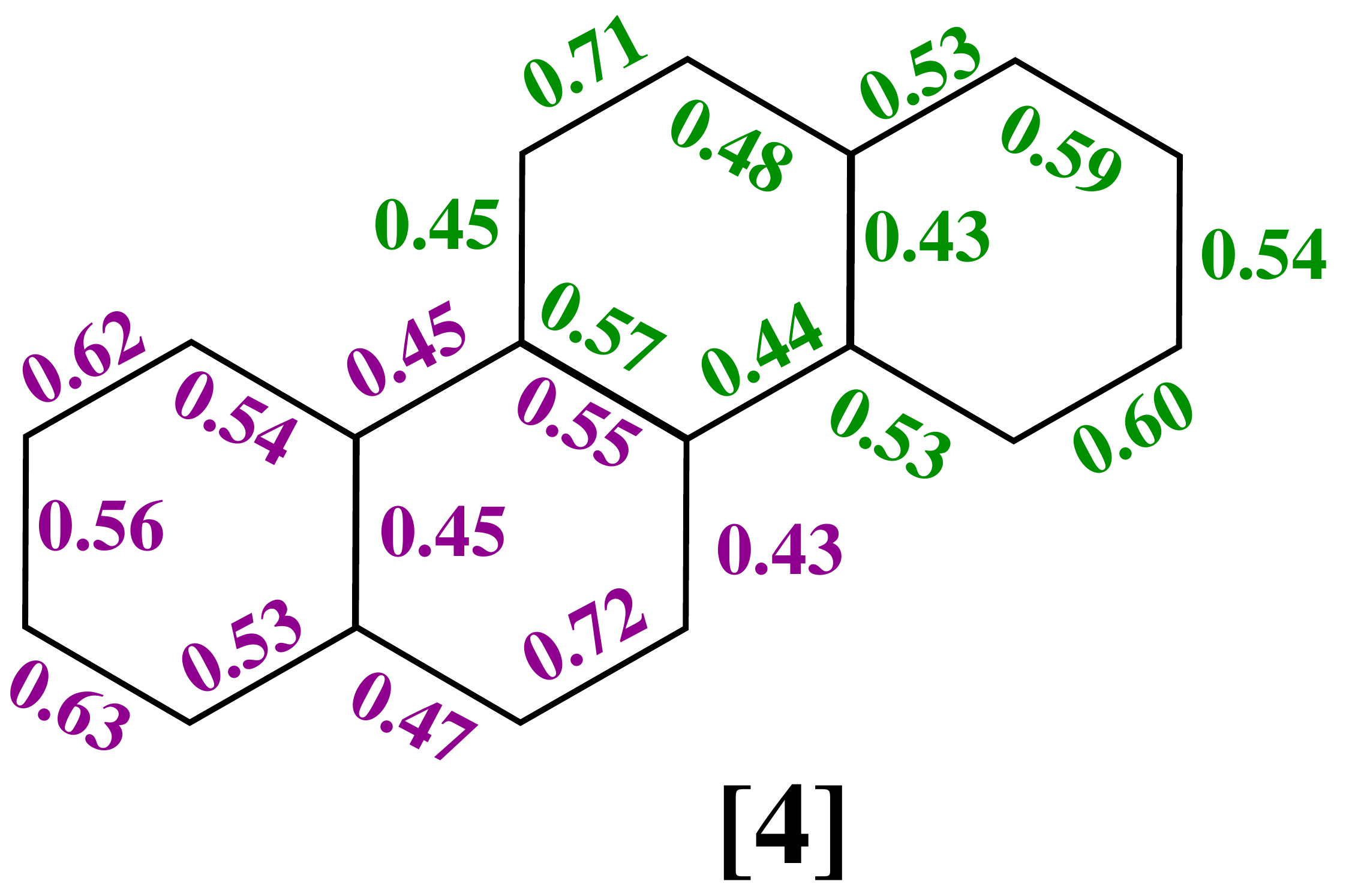}
\hspace{0.3cm}
\includegraphics[width=5.5cm]{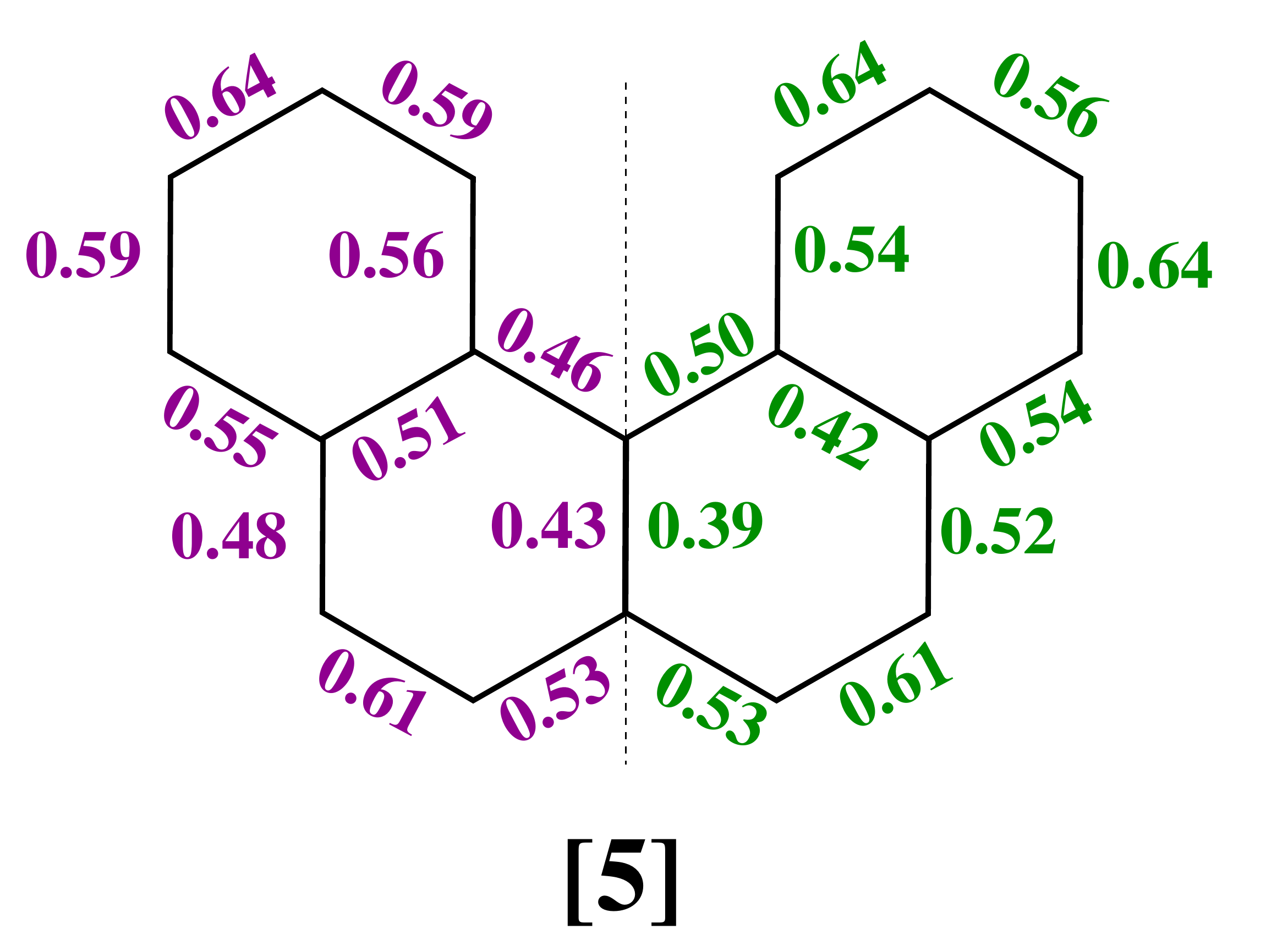} \\
\vspace*{0.5cm}
\hspace*{-1.5cm}
\includegraphics[width=4.5cm,height=5.5cm]{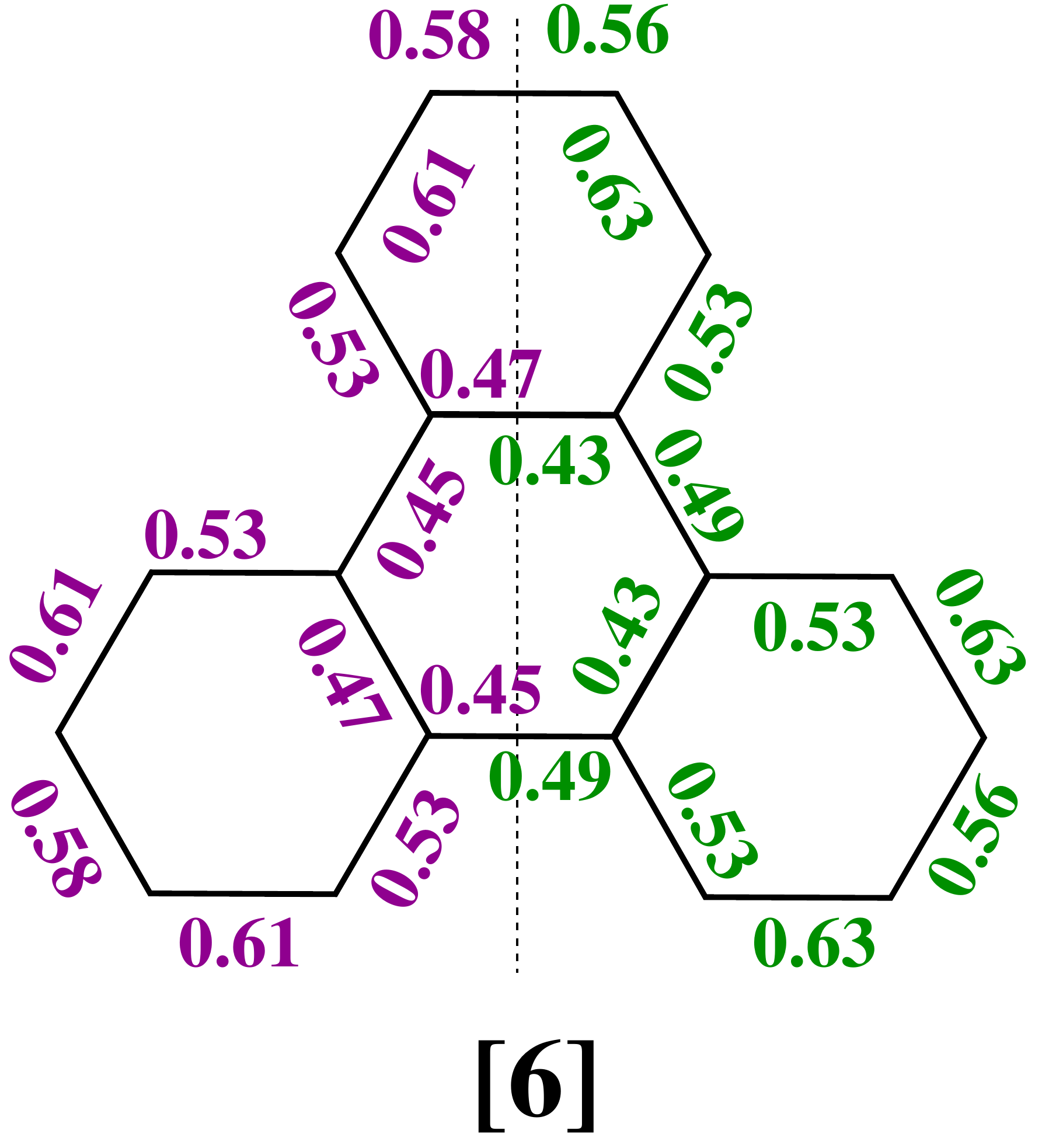}
\hspace{0.5cm}
\includegraphics[width=4.5cm]{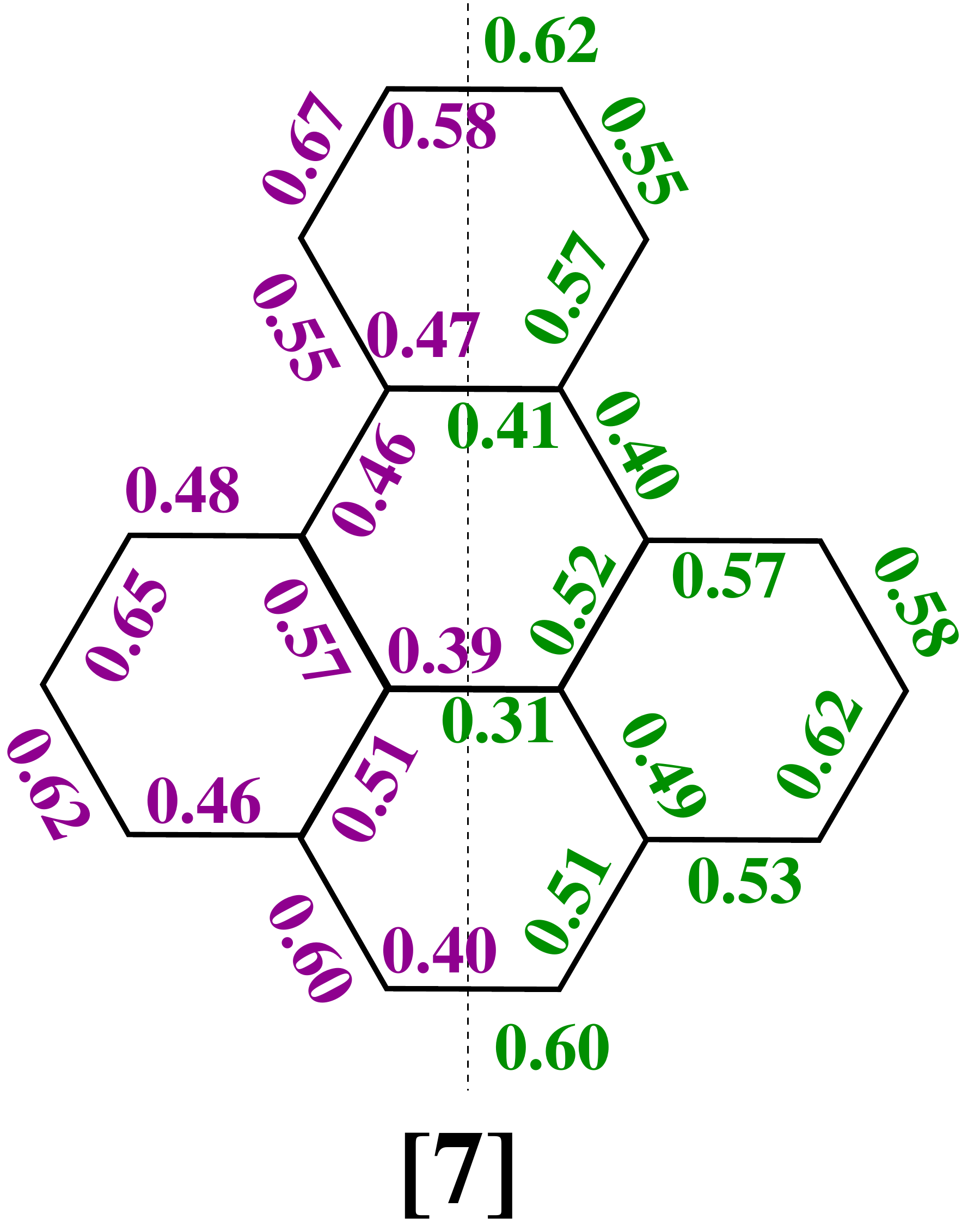} 
\hspace{0.5cm}
\includegraphics[width=7.0cm]{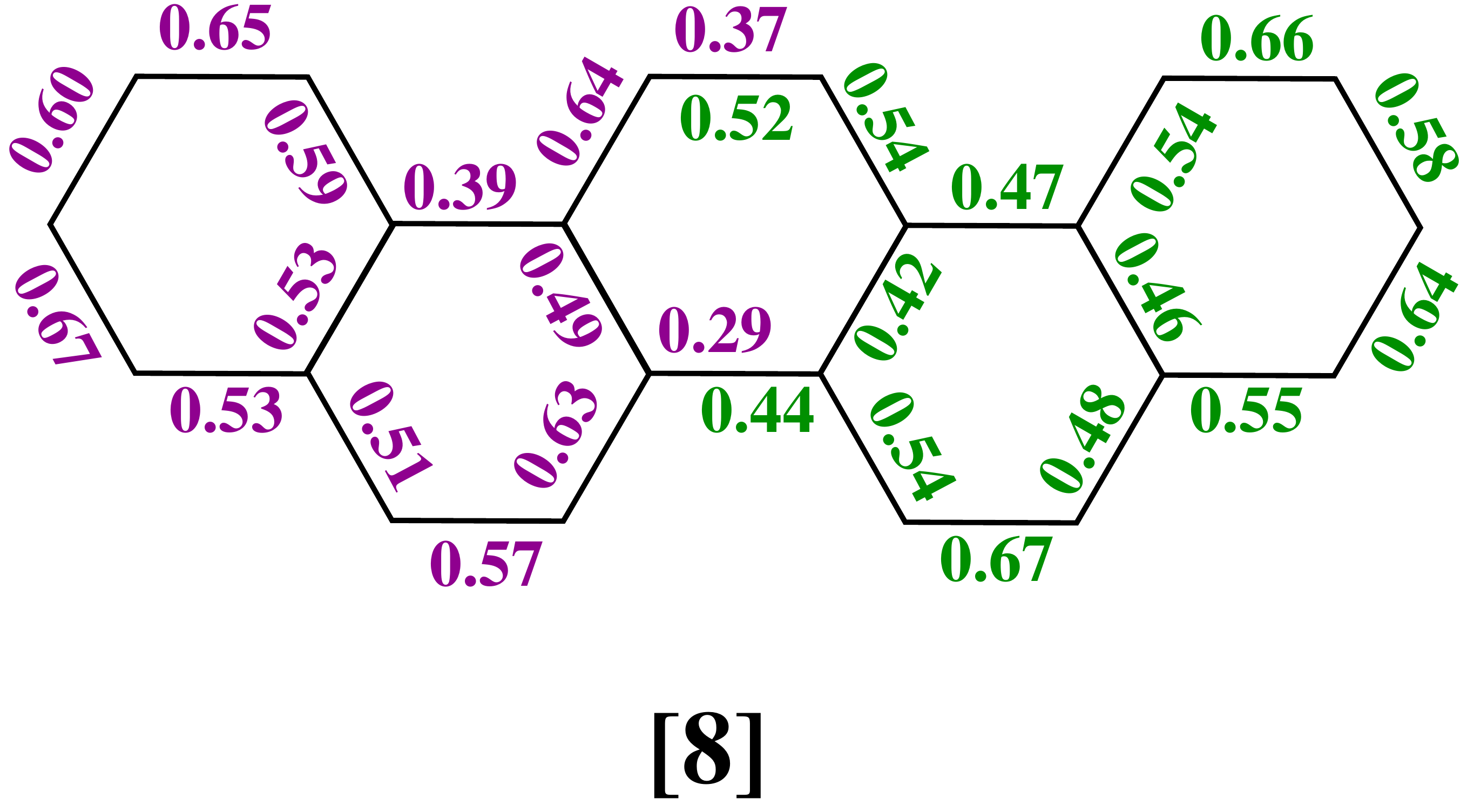}%
\end{center}
\caption{Bond orders in lowest triplet state (in magenta) and in two
photon allowed state (in green). The numbering for molecules is as given in 
Figure 1. \\ }
\label{fig: triplet-BO}
\end{figure}

\subsection{Bond-Orders}{
{Fig. 5} gives bond orders of the PAH molecules in the lowest triplet and the 
lowest two photon states. In fluoranthene (1) the bond common to the two 
benzene rings becomes very weak and elongated while the peripheral bonds, 
except the one that connects the phenyl ring to the five membered ring, all 
have similar bond lengths at equilibrium. However the triplet bond orders 
show more alternation and only the bond connecting the benzene and naphthalene 
units stays elongated. 
In pyrene (2), bond orders in the triplet and two-photon allowed state show an
opposite pattern. i.e. the bond orders are all nearly uniform 
in the naphthalenic unit in the triplet state and in biphenyl unit, they are
alternating. On the other hand, in the two photon state, bonds are uniform in
the biphenylenic unit whereas alternating in the naphthalenic unit. In
benzanthracene (3), both the triplet and the two photon states have similar 
bond order patterns, with nearly uniform peripheral bonds and weaker 
inter-ring bonds. In chrysene as well as in helicene, both the triplet and the 
two photon states show bond alternation, although it is more pronounced in the 
two photon state. In helicene the bond lying at the $C_2$ axis is the weakest 
bond both in the triplet state and two-photon allowed state (2A). In 
triphenylene (6), bond orders in both 2A and triplet states are similar and
nearly uniform along the benzenic unit with a weaker central bond connecting 
the three phenyl rings. In benzopyrene (7), the interior bonds through which
 the $C_2$ axis passes through, are weaker. In picene (8), the outer bonds have 
a marginal bond alternation in both the states, except the middle bond of the
arm-chair of phenanthrenic units ($0.35\pm0.05$). In the triplet state, the 
bonds connecting the two terminal naphthalene units to the central benzene 
ring are very weak ($0.37$ and $0.29$), but only slightly weaker in the case 
of the 2A state ($0.52$ and $0.44$)} compared to other bonds.\\}

\subsection{Spin Densities}
\begin{figure}[ht!]
\begin{center}
\hspace{-1.0cm}
\includegraphics[width=5.5cm]{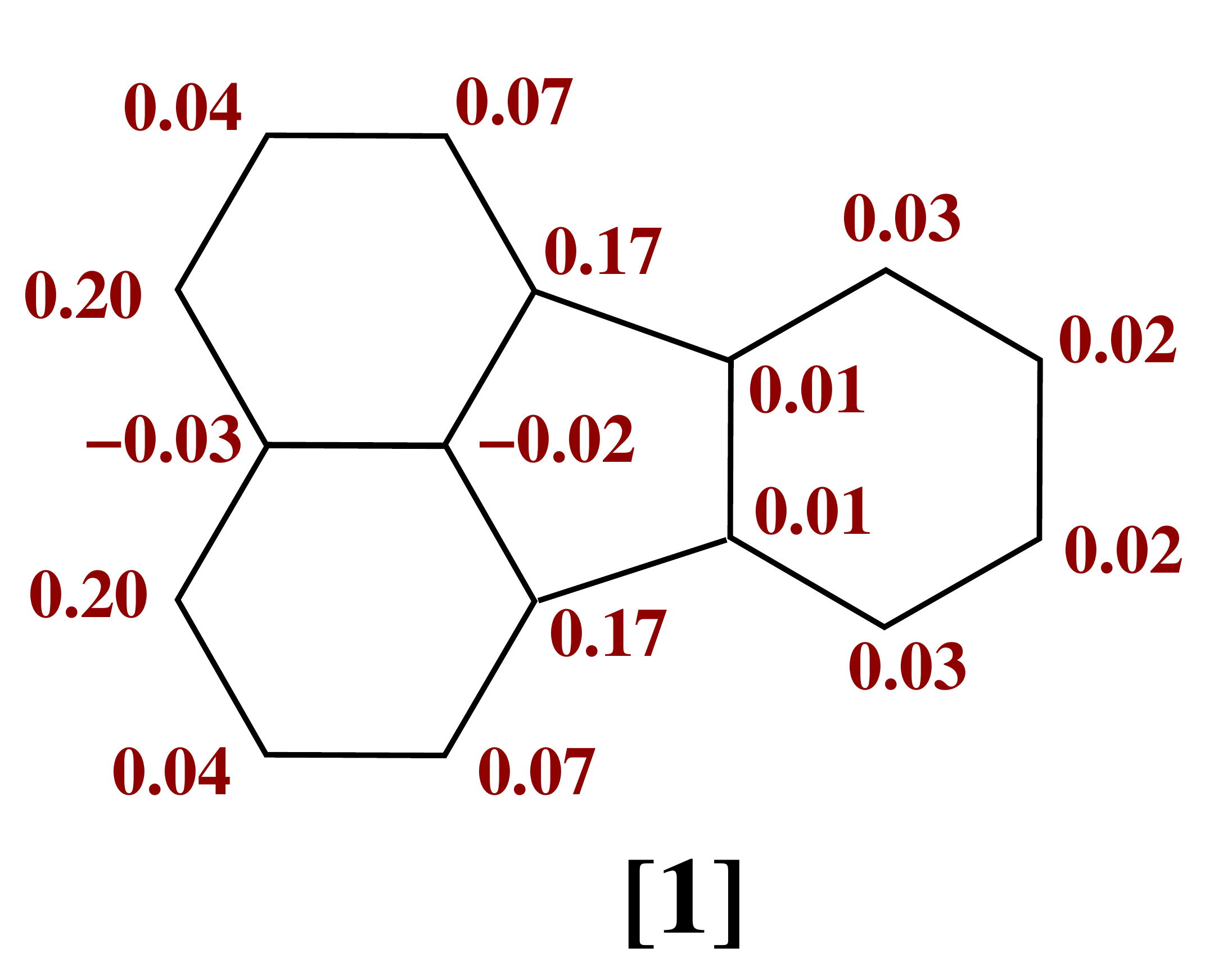}
\hspace{0.5cm}
\includegraphics[width=4cm]{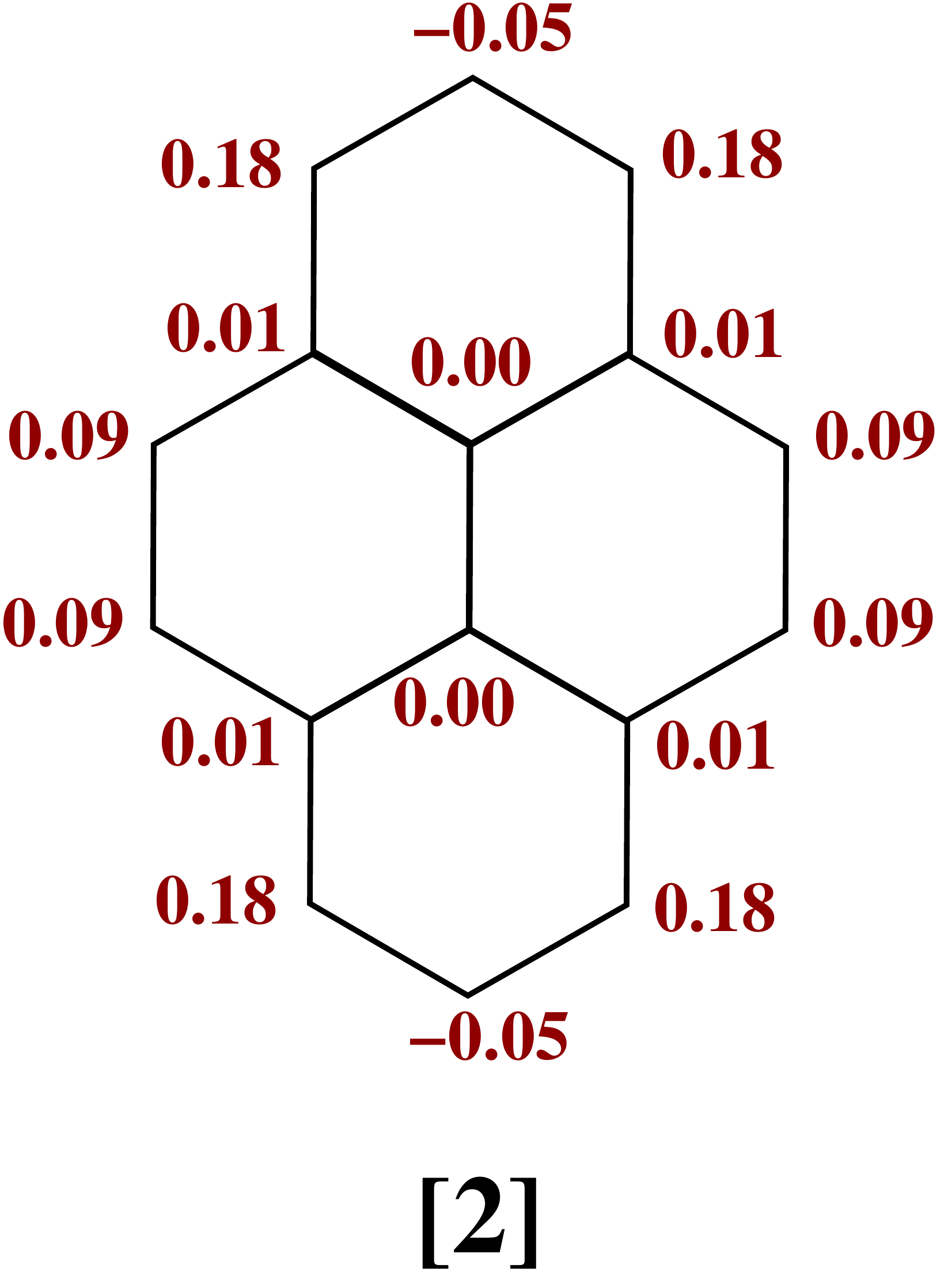}
\hspace{0.5cm}
\includegraphics[width=6cm]{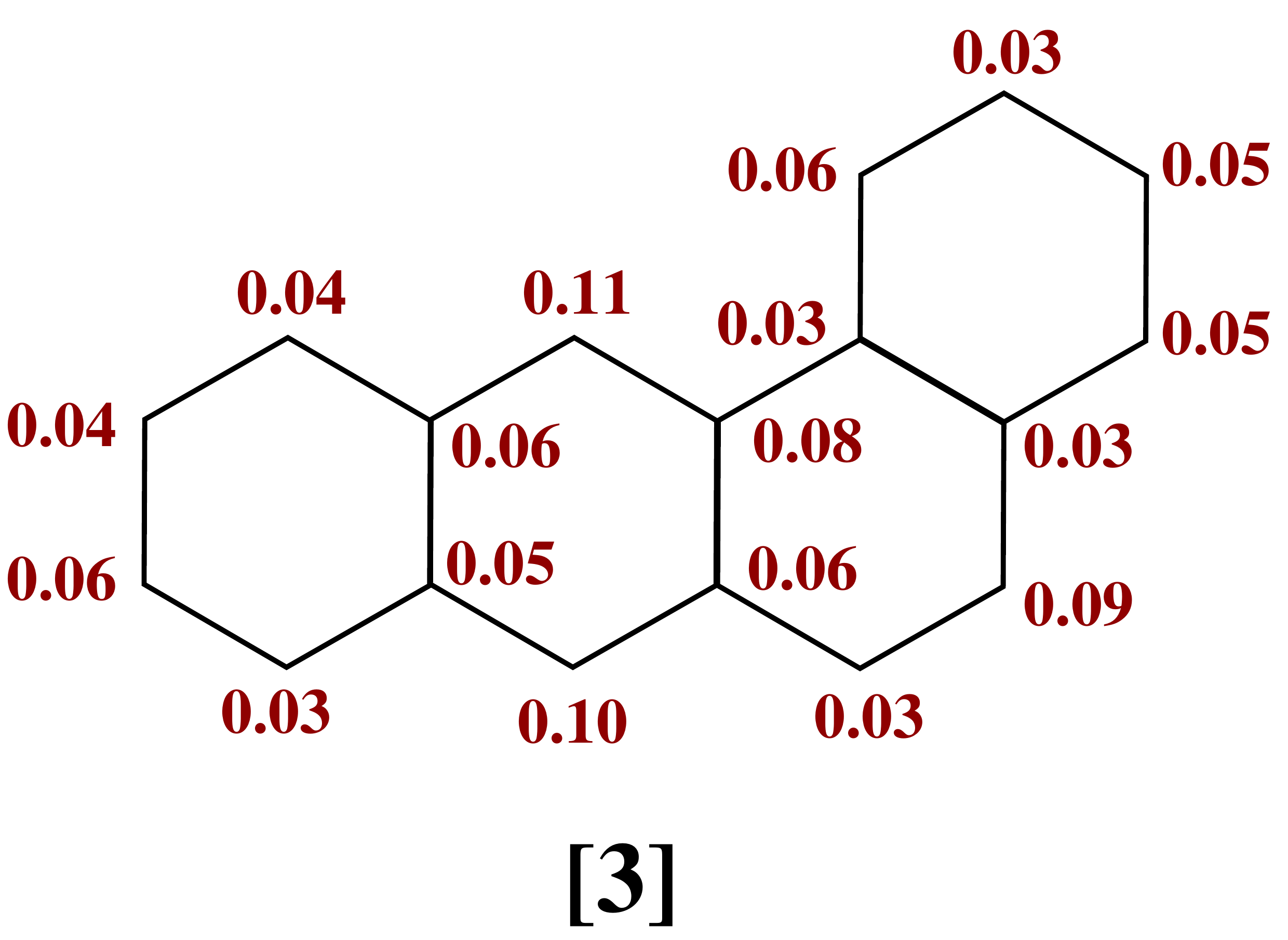} \\
\vspace*{1.0cm}
\hspace{0.0cm}
\includegraphics[width=6.0cm]{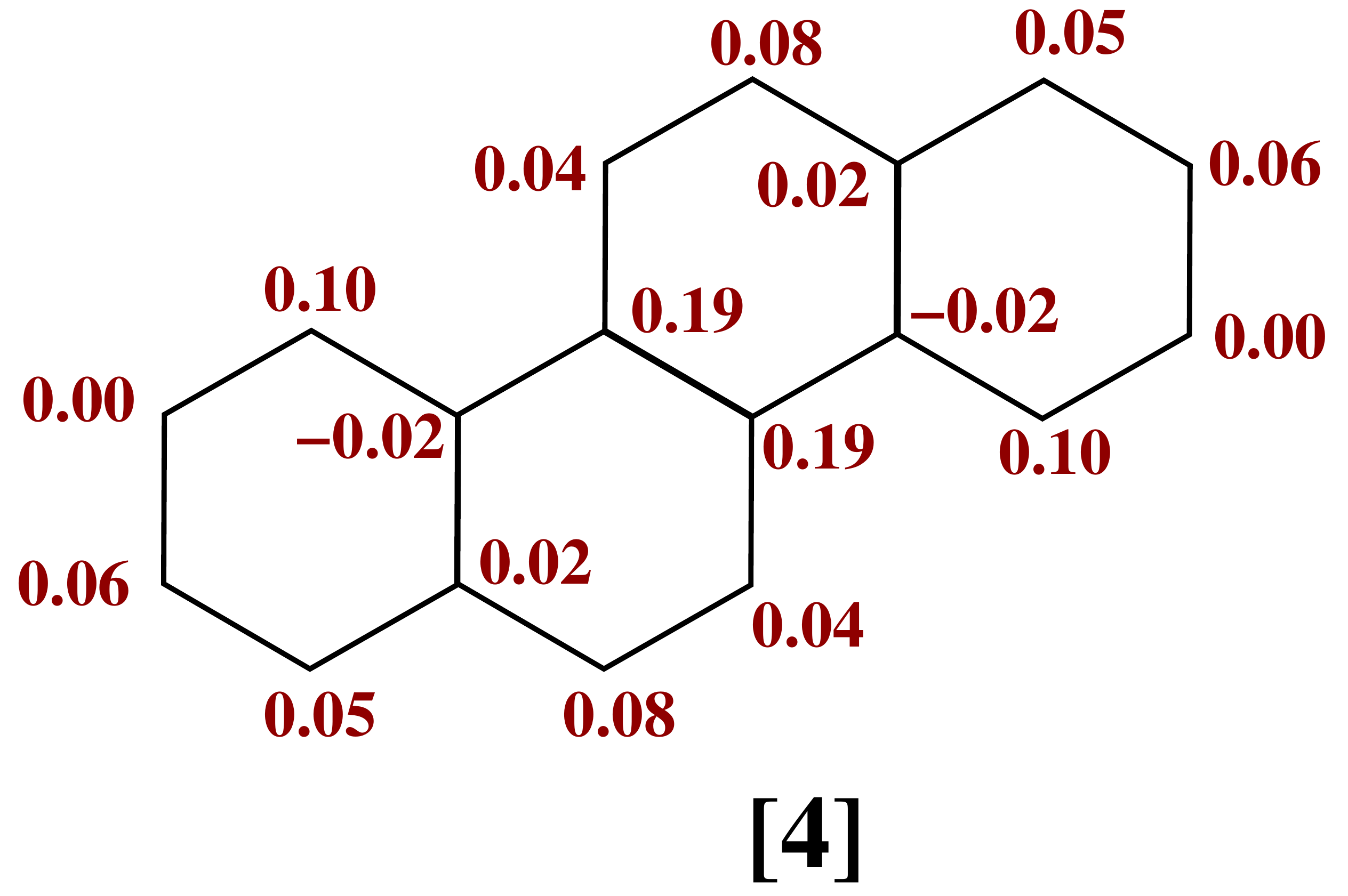}
\hspace{2.3cm}
\includegraphics[width=6.0cm]{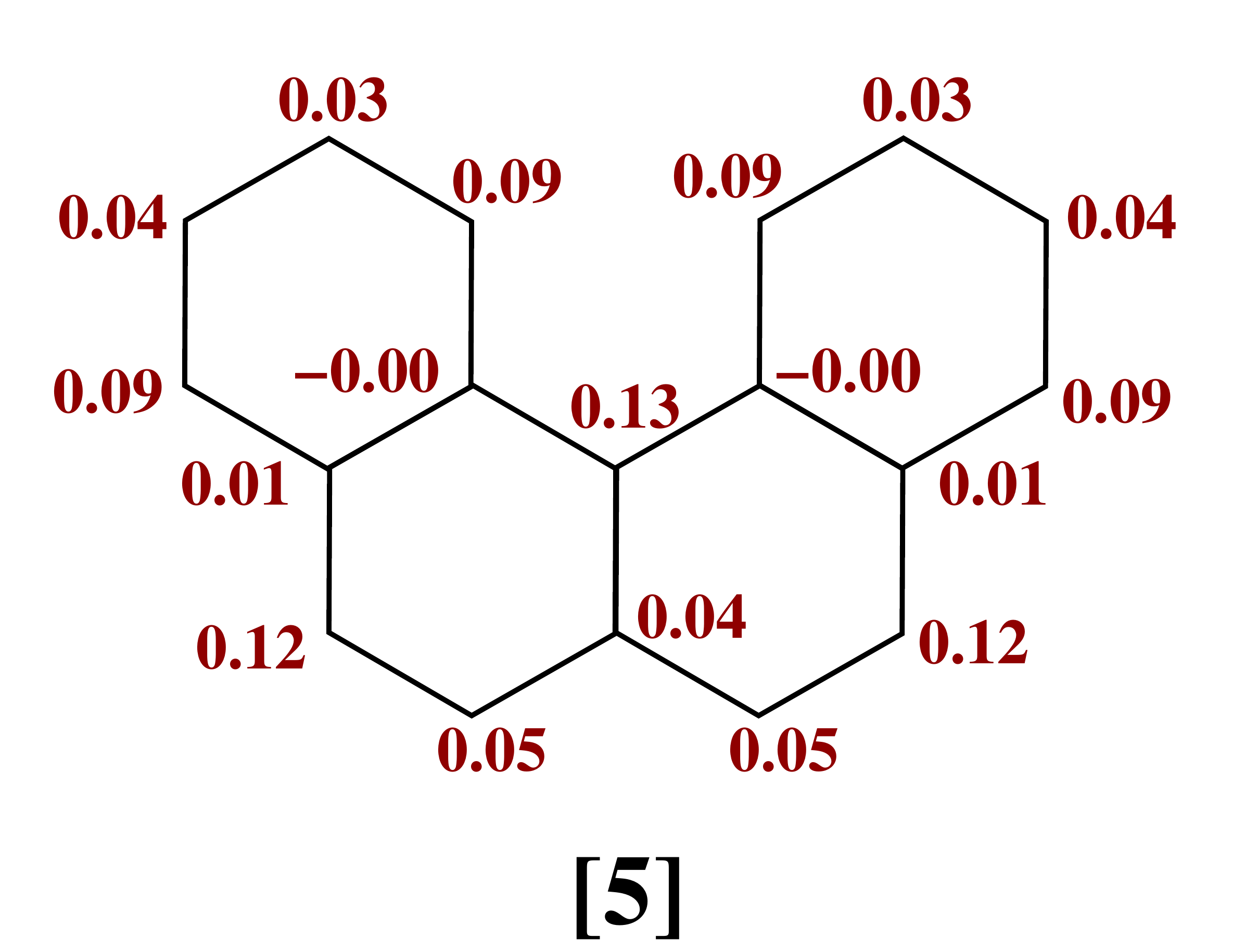} \\
\vspace*{1.0cm}
\hspace*{-2.0cm}
\includegraphics[width=4.5cm,height=5cm]{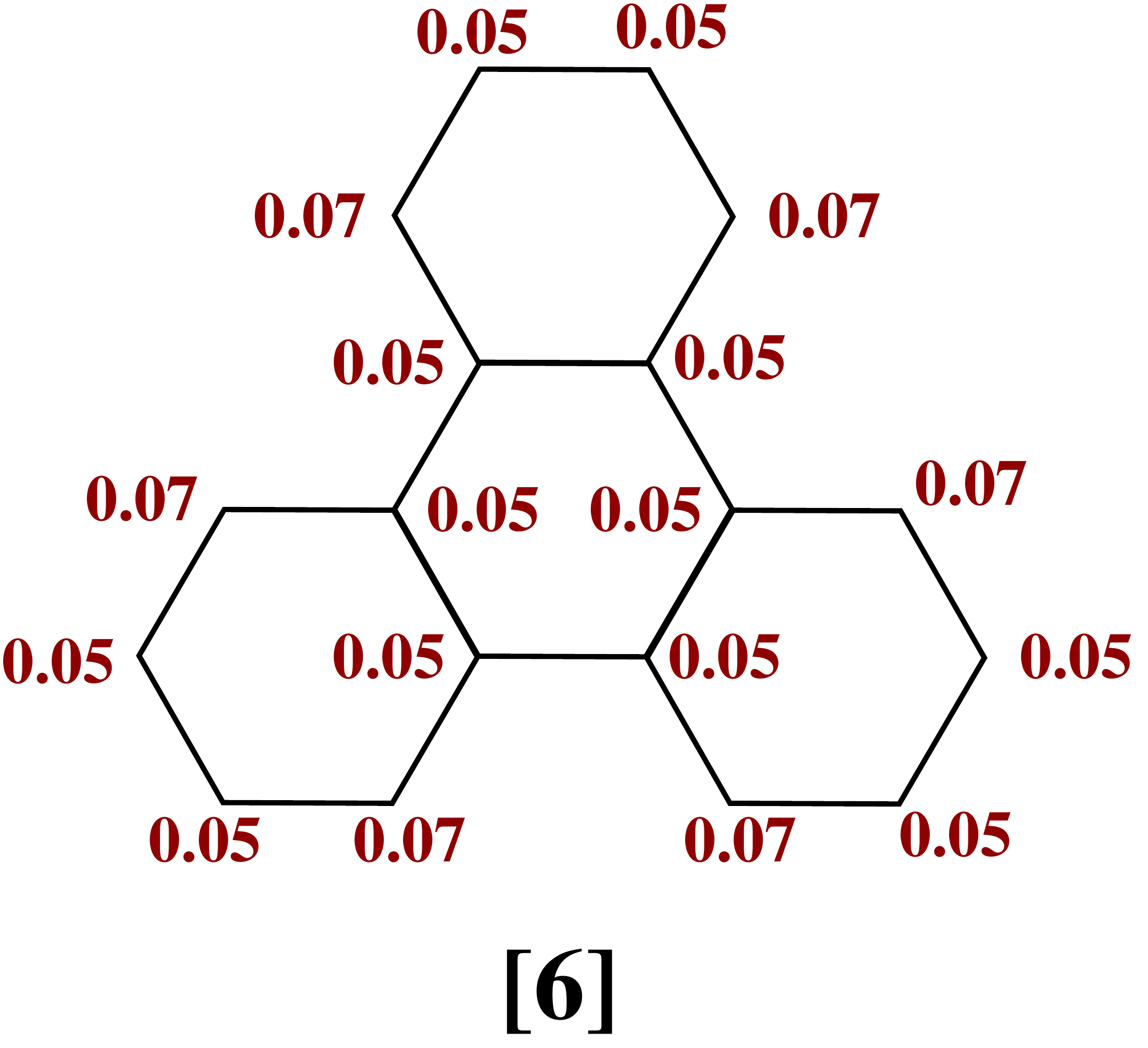} 
\hspace{0.0cm}
\includegraphics[width=6cm]{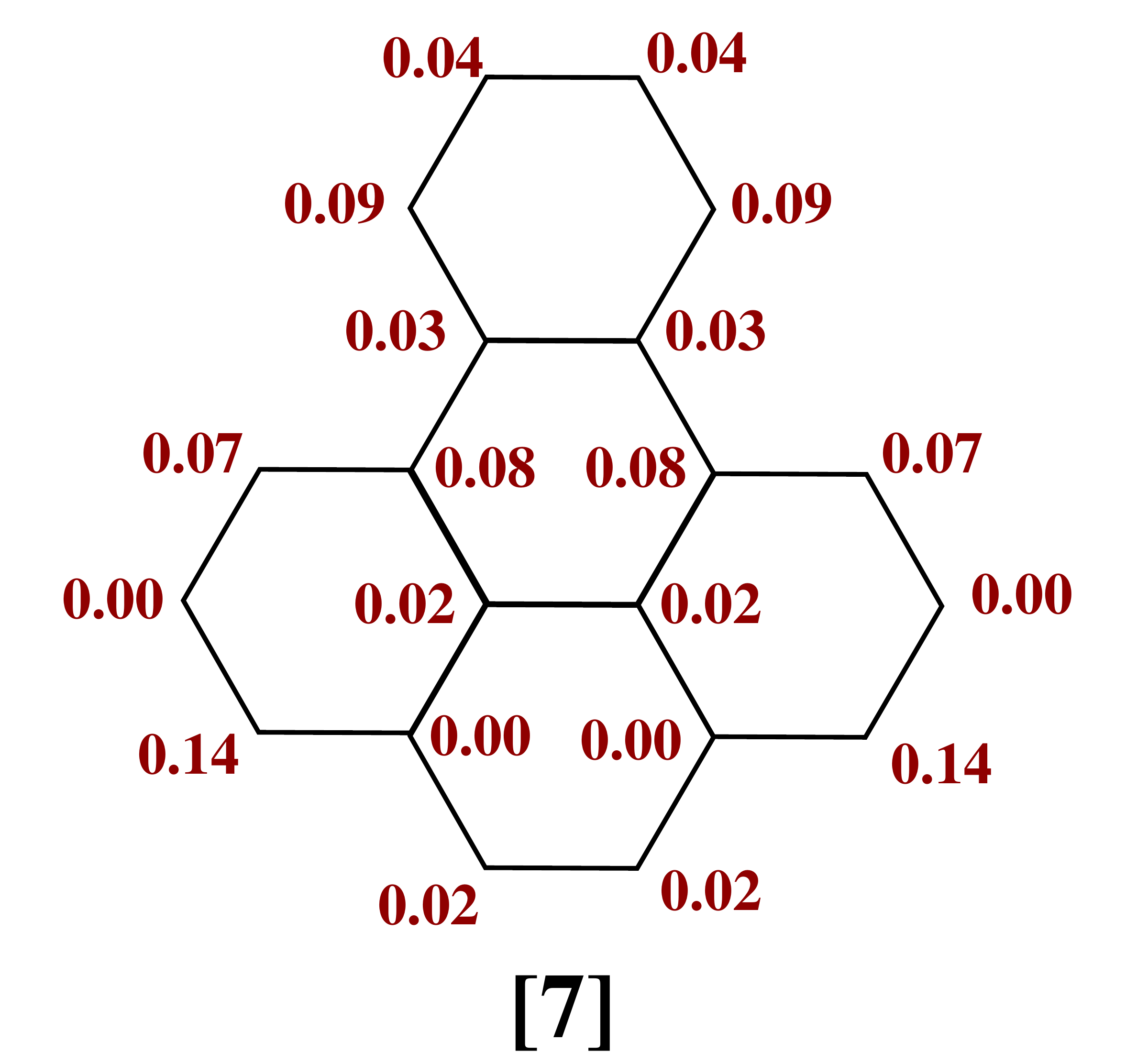} 
\includegraphics[width=7.0cm]{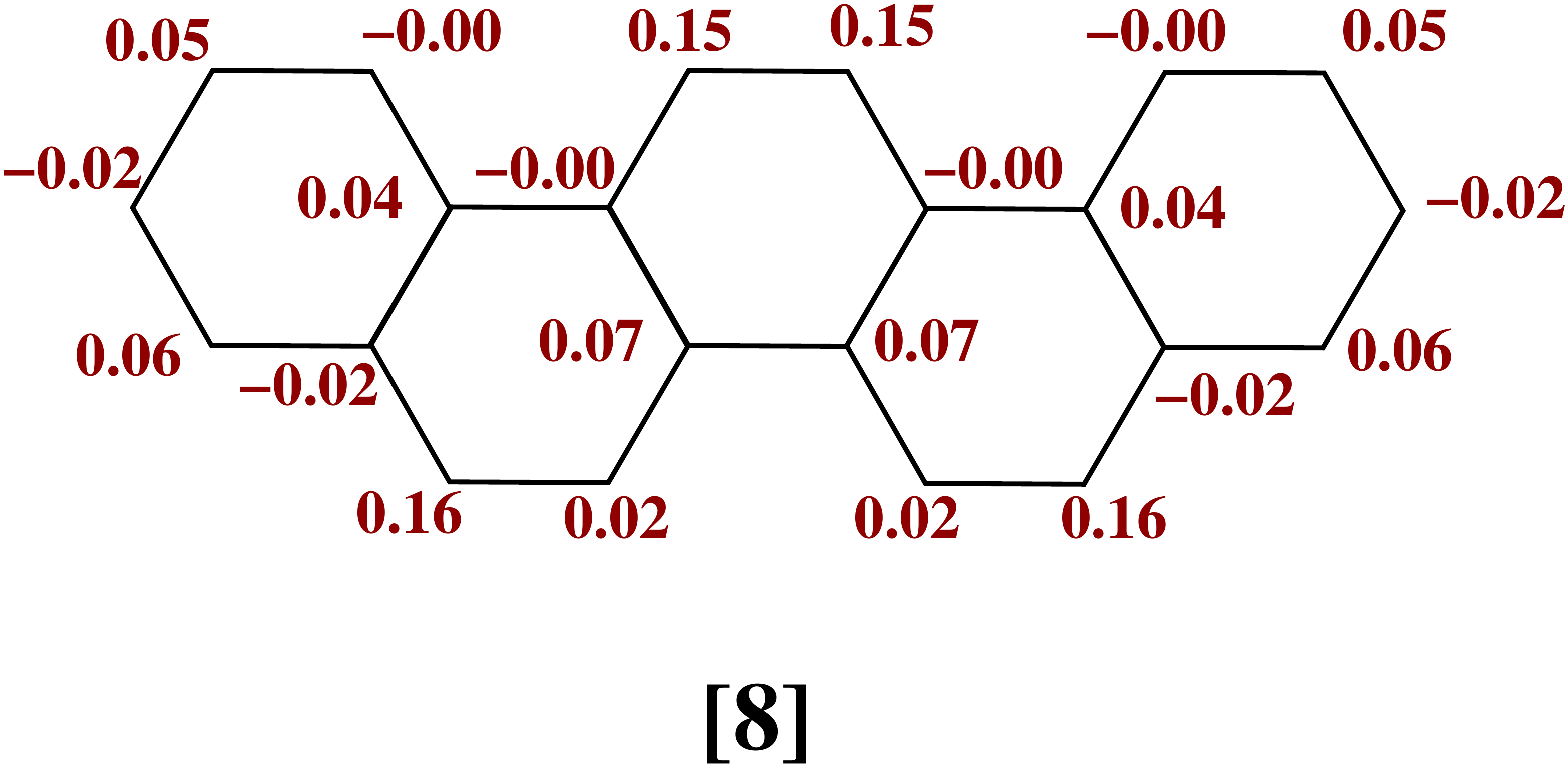}%
\end{center}
\caption{Spin densities in lowest triplet state are given in red color. The 
numbering of molecules are as given in Figure 1.}
\label{fig: triplet-spnden}
\end{figure}

\noindent  Spin density at a site "i" is defined as the difference in charge density 
of up and down spins. i.e.  
$ \rho_{i} = <\psi_k|a^{\dagger}_{i\alpha} a_{i\alpha}-
a^{\dagger}_{i\beta} a_{i\beta}|\psi_k> = 2 <\psi_k|S_z|\psi_k>$.
In Fig. 6 are given the spin densities in the lowest energy triplet 
state of the PAH molecules. We find large positive spin densities at the 
$\alpha$ sites of the naphthalenic units of fluoranthene. There is slightly 
negative spin density at the sites common to the two benzene rings in the 
naphthalene unit. The total spin density is almost localised on the 
napthalenic unit and spin density pattern resembles that of naphthalene 
(see supporting information Fig. S2). On the other hand, while isolated 
benzene has a significant spin density at each site ($0.17$), benzenic unit 
of fluoranthene has negligible spin density ($ < 0.02 \pm 0.01$). In pyrene, 
the largest positive spin densities are found at the $\alpha$ sites ($0.18$) 
of biphenylinic unit followed by the $\beta$ carbon atoms of naphthalenic unit
($0.09$). There is negative spin density at the two peripheral carbon atoms of 
biphenylinic unit ($-0.05$) and all other carbon atoms have negligible spin densities. 

Unlike the bond order pattern of benzanthracene, spin density pattern neither
resembles that of anthracene nor phenanthrenic unit. A uniform spin density 
($0.05\pm0.01$) is seen except at the $\alpha-$ carbon atoms of anthracenic 
moiety ($0.11$). No negative spin density is observed in this molecule. 
Chrysene shows an opposite spin density pattern compared to other molecules. 
Here, the spin density is largest ($0.19$) at the two carbon atoms where 
two naphthelinic units are fused. It also shows negative spin densities at one
of the adjacent sites followed by a large positive spin density ($0.10$ and 
$0.08$) at the $\alpha$ sites of naphthalenic units. On the other hand, in 
helicene almost all the $\alpha$ sites have large positive spin densities
($\sim 0.1$) and $\beta$ sites have smaller spin densities. Triphenylene has a
uniform spin density of $0.05$  and $0.07$ at alternate sites, with the central
benzenic unit having a uniform spin density of $0.05$, owing to its three-fold 
symmetry. 

Benzopyrene has similar spin density pattern as that of triphenylene, with a 
pronounced alternation in spin densities at adjacent sites. The highest 
positive spin density of $0.14$ is seen at the carbon atoms adjacent to the 
vinylenic bridge in benzopyrene. Picene molecule, on the other hand, has 
alternate positive and negative spin densities at the peripheral phenyl rings. 
There is also large positive spin densities ($0.15$) at four alpha carbon 
atoms of the interior naphthalenic units. The carbon atoms of the interior 
arm-chair have spin density of $0.07$.

\section{Summary} {In this paper we have studied the correlated electronic 
states of eight well known polycyclic aromatic hydrocarbons. From energetics, 
we find that in all these molecules, the one-photon state is above the two 
photon state. Hence none of these molecules are expected to be fluorescent. 
The optical gap obtained from our study agree well with the experimental gaps. 
The singlet-triplet gap in all the molecules is also quite large and does 
not satisfy the requirements for singlet fission. The bond orders show that 
the equilibrium geometries of all the states are nearly the same as assumed 
in the calculations. Thus, we expect small Stokes shifts in fluorescence. The 
spin densities in the triplet state are mostly confined to the $\alpha$ sites 
of the naphthalenic units in the molecules. Most molecules show very small 
negative spin densities at a few sites and often all the sites have only 
positive spin densities.}

\noindent
\begin{suppinfo}
We have given the following table and figures:

\noindent
i) A Table for electronic excitation gap and singlet-triplet gaps for PAH
molecules - naphthalene, anthracene and phenanthrene. \\
ii) A figure for bond orders for the lowest singlet and triplet
states for naphthalene, anthracene, phenanthrene and biphenyl.\\
iii) A figure containing spin densities of the lowest triplet states for naphthalene,
anthracene, phenanthrene and biphenyl.
\end{suppinfo}

\begin{acknowledgement} Y. A. P. thanks Disha programme for women
in Science (DST0128), Department of Science and Technology, Government of 
India, for financial support.
\end{acknowledgement}

\end{document}



\begin{table}[ht!]
\begin{center}
\setlength{\tabcolsep}{1.5pt} 
\begin{tabular}{|l| c |c|}
\hline
Molecule & Optical gap  & singlet-  \\
 & (Expt gap)       & triplet gap \\      
\hline
Naphthalene    & ~~ 6.01 ~ (5.63)$^{[1]}$ & 2.52 \\
Anthracene     & ~~ 3.70 ~ (3.31)$^{[2]}$   &  1.71   \\
Phenanthrene   & ~~ 4.28 ~ (3.60)$^{[2]}$  &  2.54  \\
\hline
\end{tabular}
\end{center}
\caption{S1. Electronic excitation gap and singlet-triplet gaps in eV for PAH 
molecules. Numbers in the bracket are experimental gaps obtained from 
references shown as superscript. 
}
\end{table}

\begin{figure}
\includegraphics[width=4.6cm]{naph-bnd.pdf} \hspace{2.5cm}
\includegraphics[width=6cm]{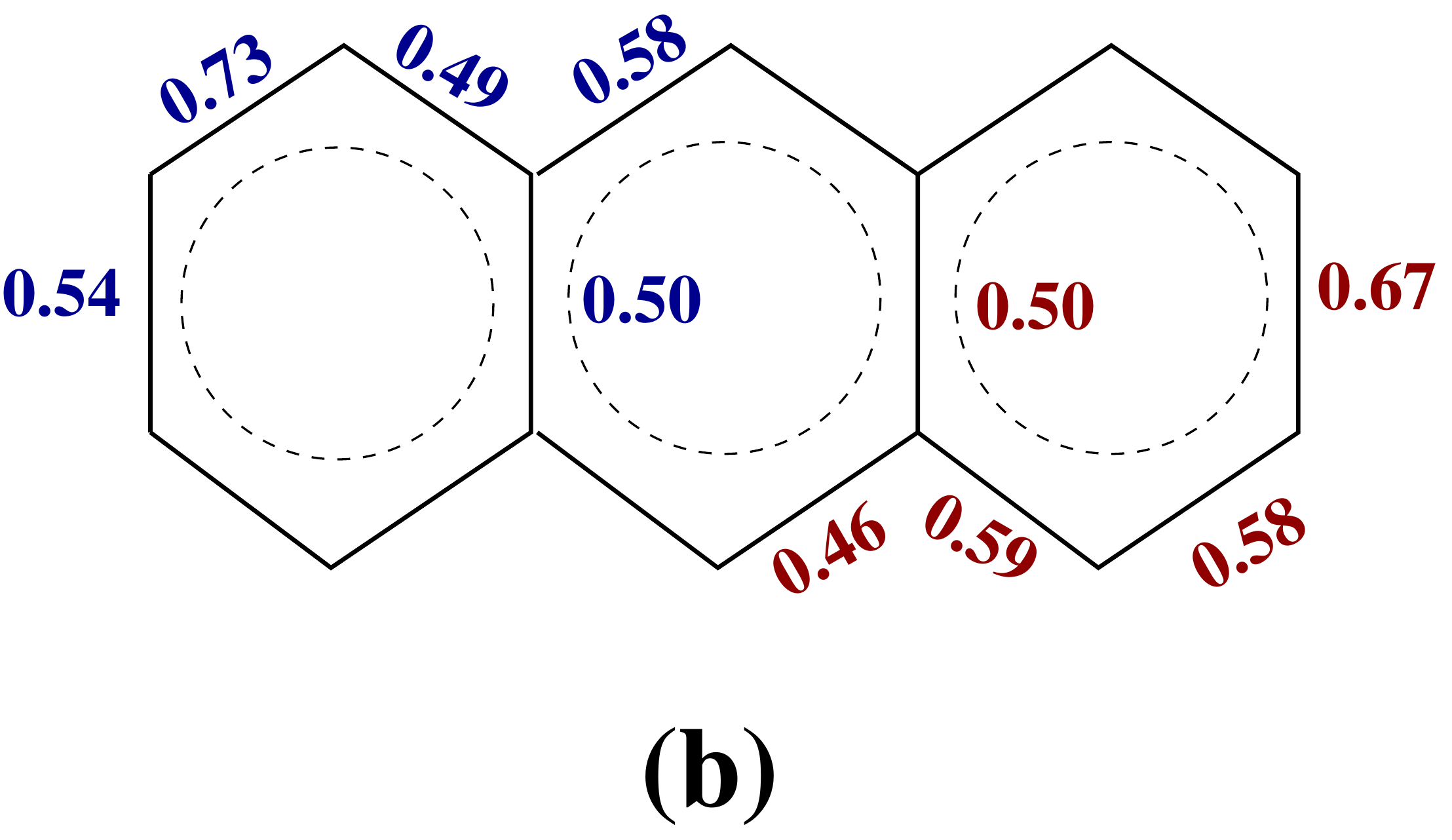} \\
\vspace*{1cm}
\includegraphics[width=6cm,height=3.8cm]{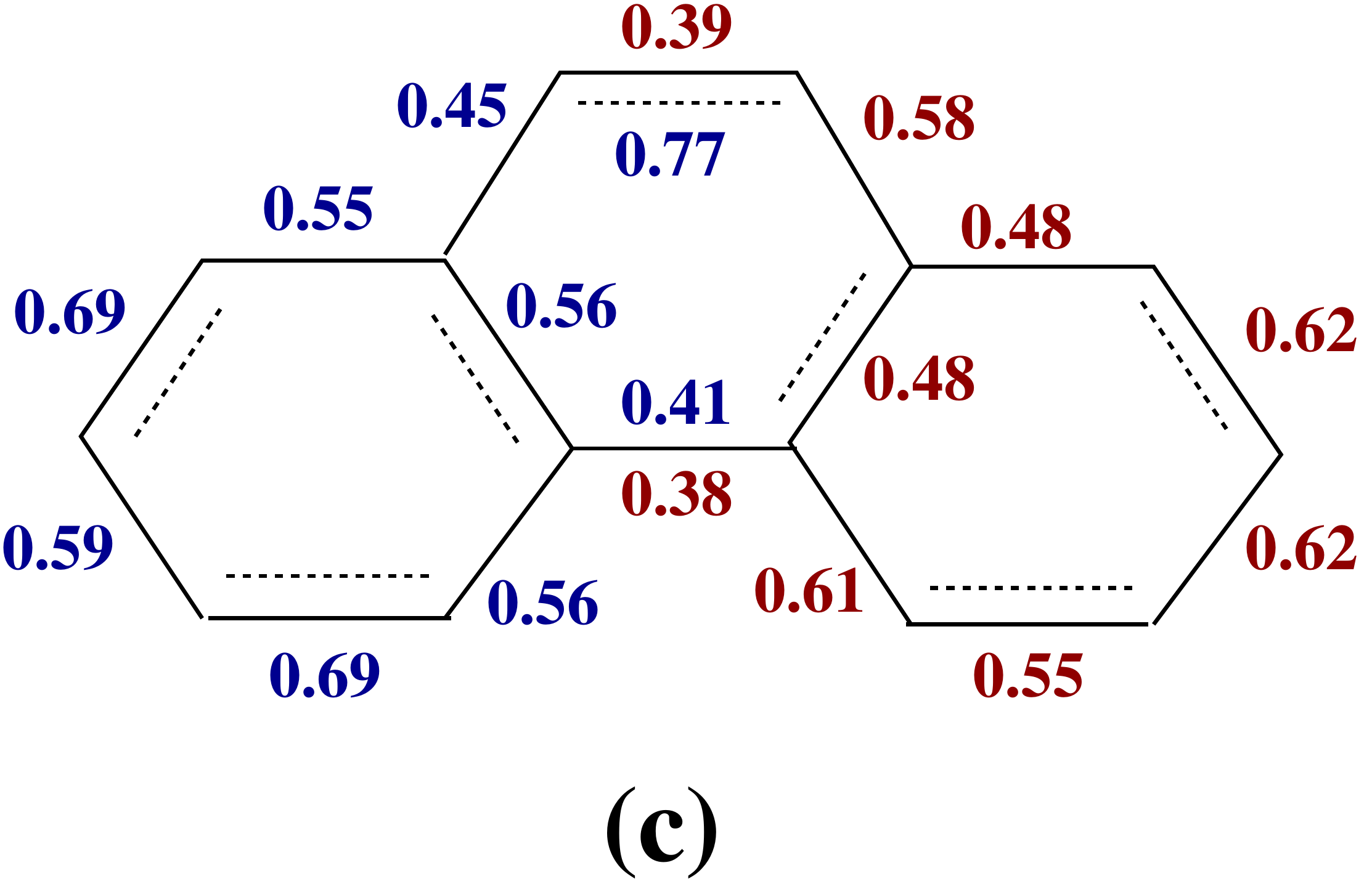} 
\hspace{1.0cm}
\includegraphics[width=6cm,height=2.8cm]{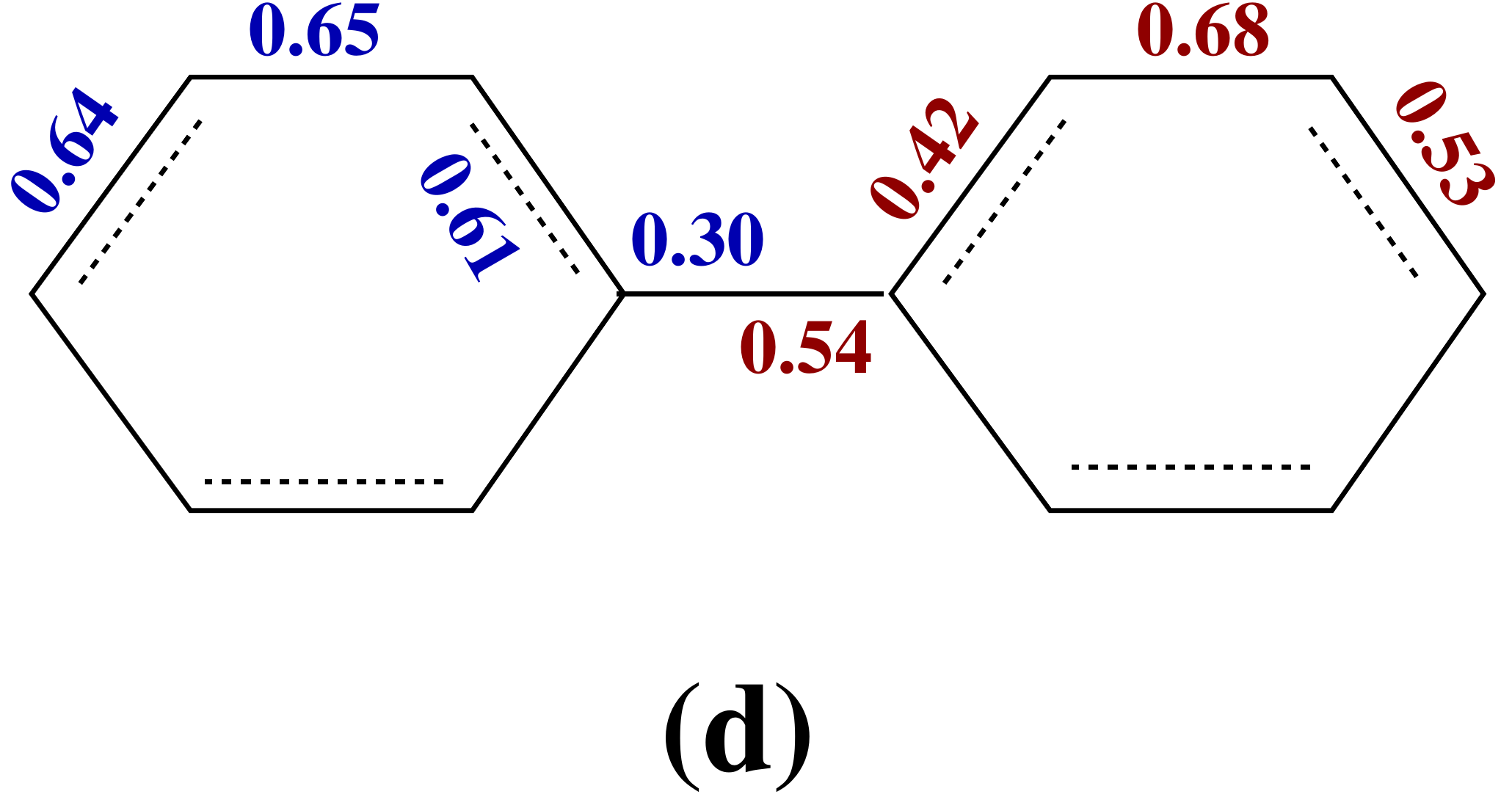} \\
\caption{S1. Bond orders for the lowest singlet (blue in colour) and triplet 
states for (a) naphthalene, (b) anthracene, (c) phenanthrene and 
(d) biphenyl (red in colour).}
\end{figure}

\begin{figure}
\includegraphics[width=4.6cm]{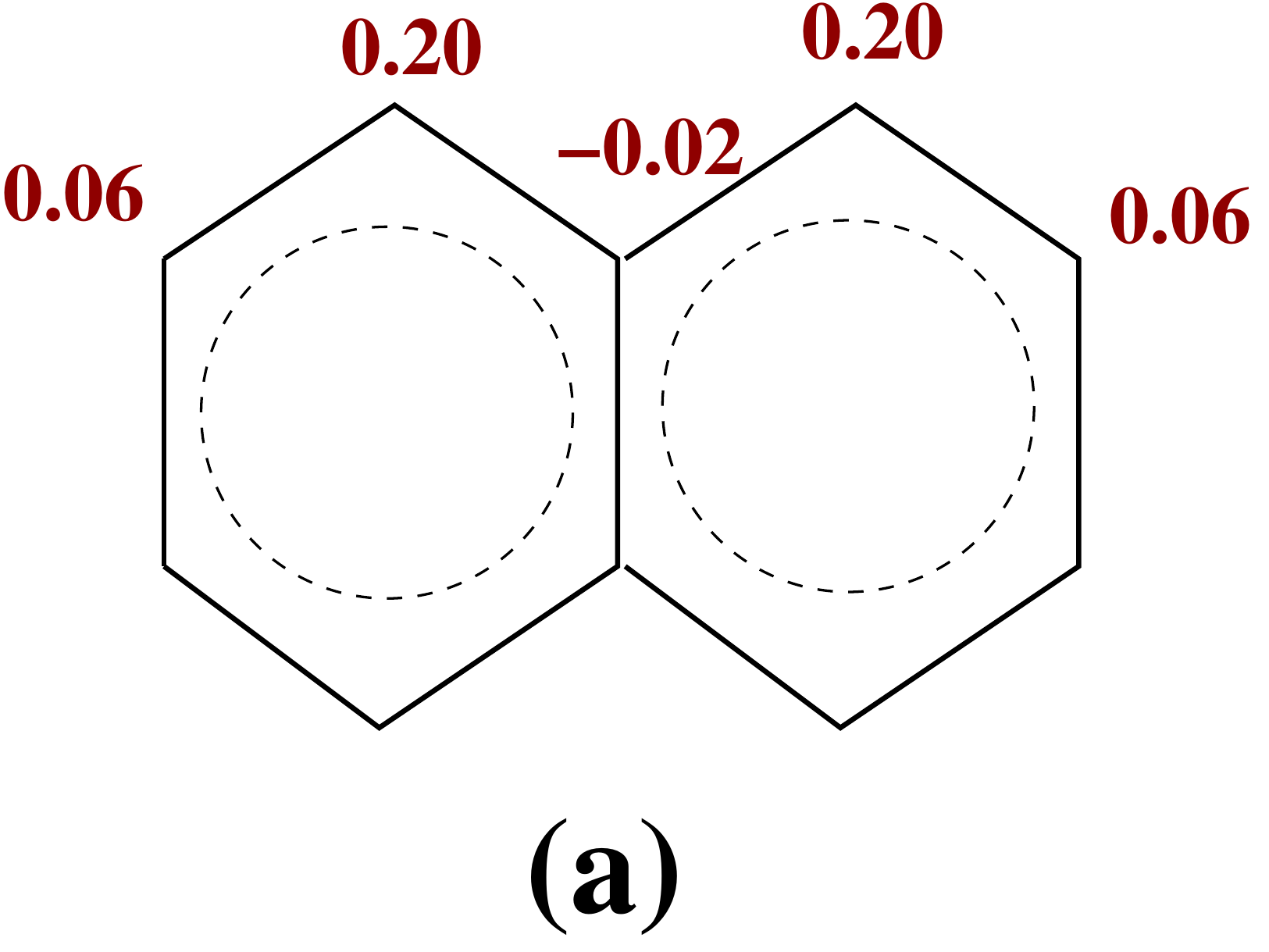} \hspace{2.5cm}
\includegraphics[width=6cm]{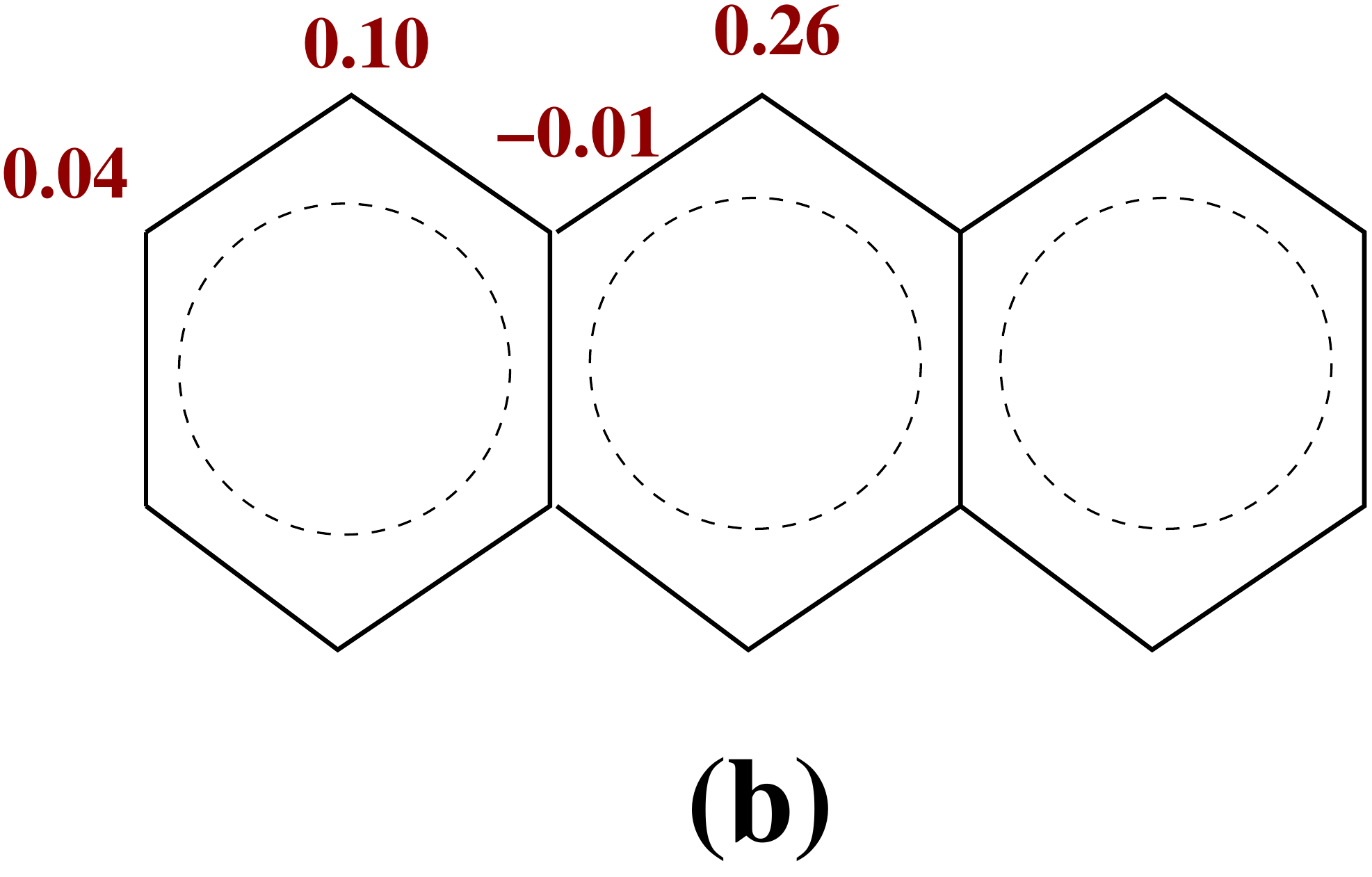} \\
\vspace*{1cm}
\includegraphics[width=6cm,height=3.8cm]{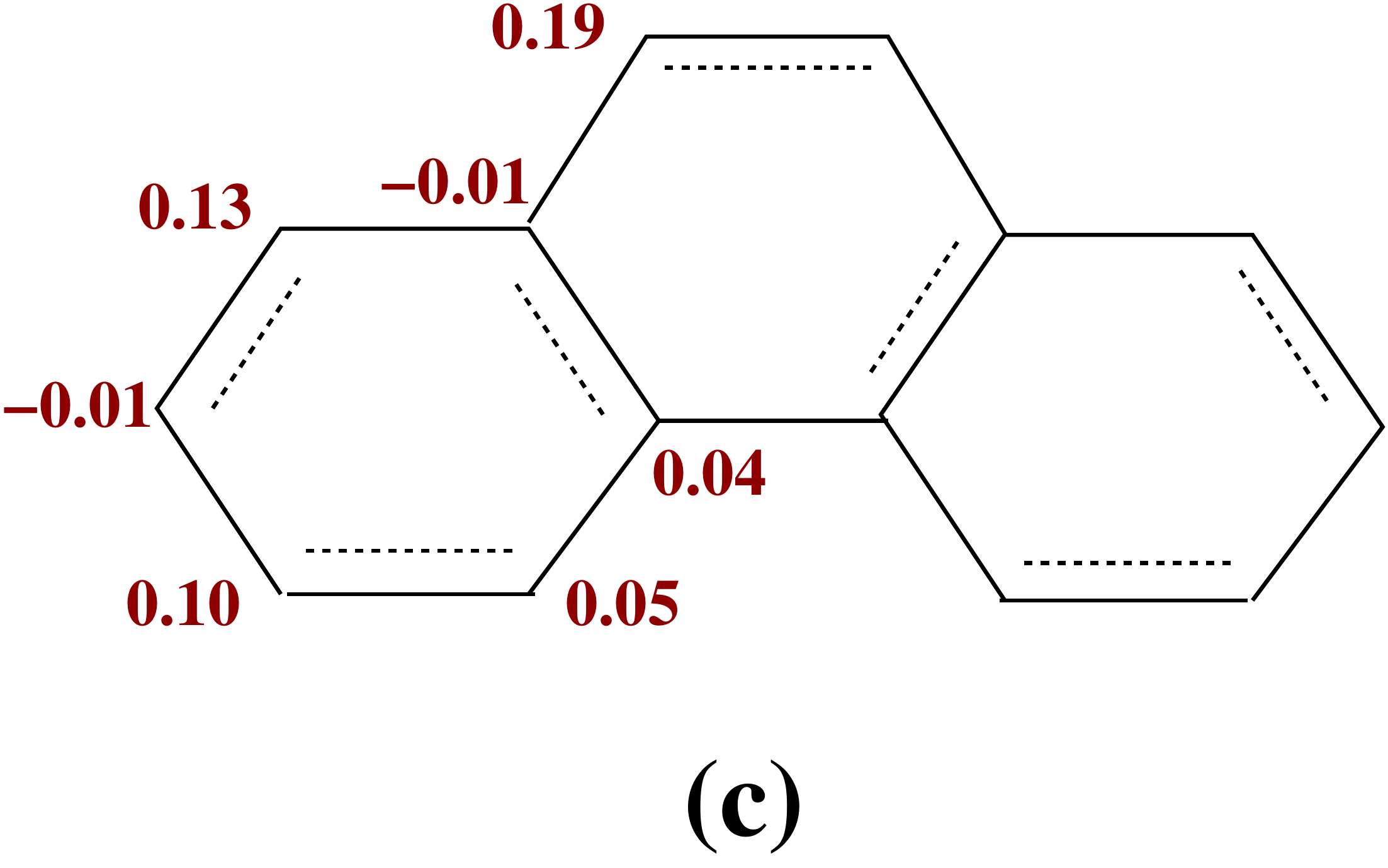} 
\hspace{1.0cm}
\includegraphics[width=6cm,height=2.8cm]{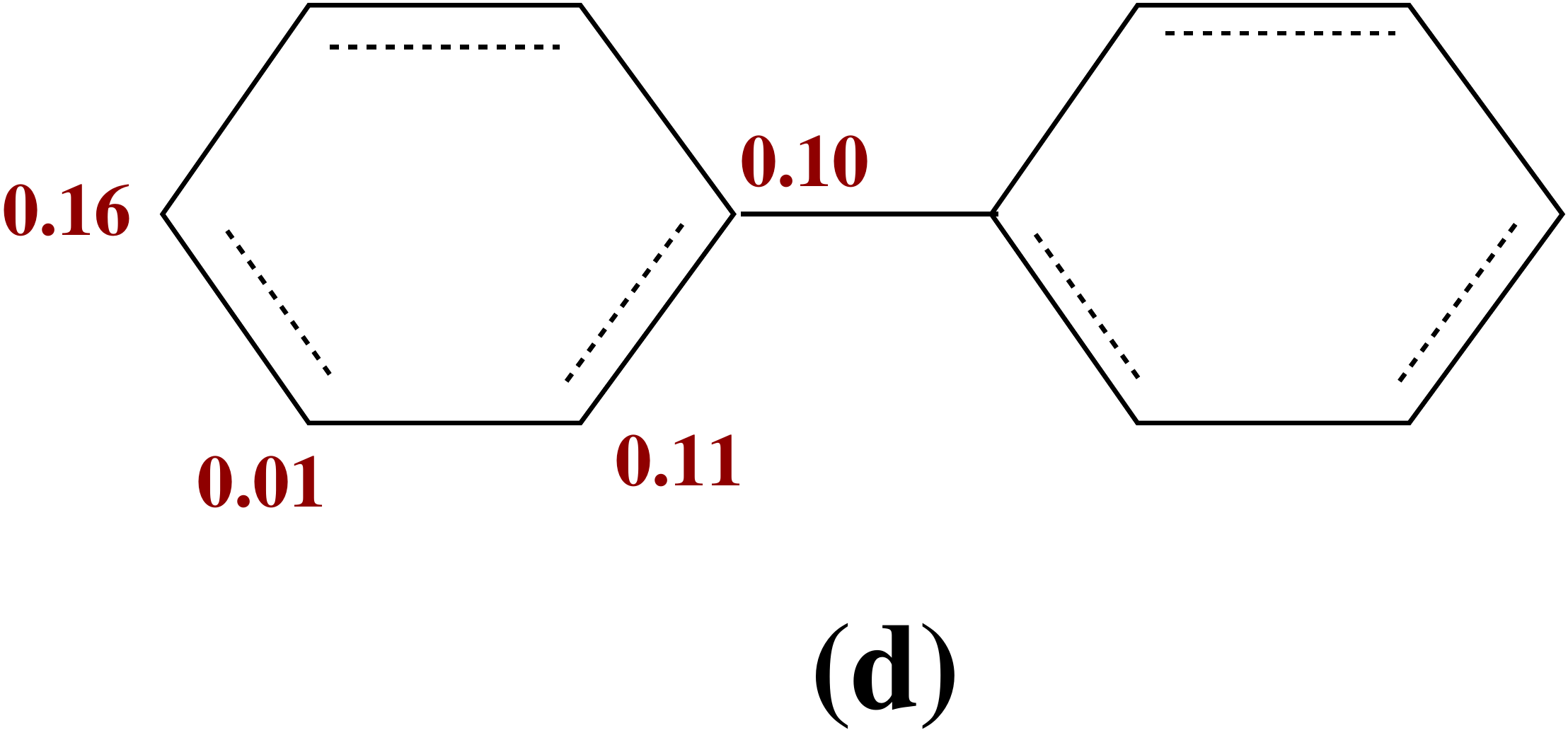} \\
\caption{S2. Spin densities of the lowest triplet states for  (a) naphthalene, 
(b) anthracene, (c) phenanthrene and (d) biphenyl.}
\end{figure}

\pagebreak
{\bf References} 
\begin{enumerate}
\item  Birks, J. B.; Christophorou, L. G.; Huebner, R. H.;
Excited Electronic States of Benzene and Naphthalene,
{\it Nature}, {\bf 1968}, {\it 217}, 809-812.


\item Clarence K. Jr. The Longest Wavelength Band in the Electronic Spectra
of Polycyclic Aromatic Hydrocarbons for Analytical Use, {\it Appl. Spectr.},
{\bf 1959}, {\it 13}, 15-25.

\end{enumerate}